\documentclass[preprint,12pt]{emulateapj}
\usepackage{epsfig}
\usepackage{natbib}
\usepackage{amssymb}
\usepackage{amsbsy}
\usepackage{natbib}
\usepackage[mathcal]{euscript}
\bibpunct{(}{)}{,}{a}{}{,}
\begin{document}
 
\def\simlt{\vcenter{\hbox{$<$}\offinterlineskip\hbox{$\sim$}}}
\def\simgt{\vcenter{\hbox{$>$}\offinterlineskip\hbox{$\sim$}}}
\def\etal{et al.\ }

\title{CSI~2264: Characterizing Accretion-Burst
           Dominated Light Curves for Young Stars in NGC~2264\footnotemark[*]}
\footnotetext[*]{Based on data from the {\em Spitzer} and {\em CoRoT} missions, as well as the Canada France Hawaii Telescope (CFHT) MegaCam CCD, and 
the European Southern Observatory Very Large Telescope, Paranal Chile, under program 088.C-0239. The {\em CoRoT} space mission was developed and is 
operated by the French space agency CNES, with particpiation of ESA's RSSD and Science Programmes, Austria, Belgium, Brazil, Germany, and Spain. 
MegaCam is a joint project of CFHT and CEA/DAPNIA, which is operated by the National Research Council (NRC) of Canada, the Institute National des 
Sciences de l'Univers of the Centre National de la Recherche Scientifique of France, and the University of Hawaii.}
\author{John Stauffer\altaffilmark{1}, Ann Marie Cody\altaffilmark{1}, 
Annie Baglin\altaffilmark{2}, Silvia Alencar\altaffilmark{3}, 
Luisa Rebull\altaffilmark{1}, Lynne A. Hillenbrand\altaffilmark{4},  
Laura Venuti\altaffilmark{5}, 
Neal J. Turner\altaffilmark{6}, John Carpenter\altaffilmark{4},
Peter Plavchan\altaffilmark{7}, Krzysztof Findeisen\altaffilmark{4}, 
Sean Carey\altaffilmark{1}, 
Susan Terebey\altaffilmark{8}, Mar\'ia Morales-Calder\'on\altaffilmark{9},  
Jerome Bouvier\altaffilmark{5}, 
Giusi Micela\altaffilmark{10}, Ettore Flaccomio\altaffilmark{10}, 
Inseok Song\altaffilmark{11}, Rob Gutermuth\altaffilmark{12},  
Lee Hartmann\altaffilmark{13}, Nuria Calvet\altaffilmark{13}, 
Barbara Whitney\altaffilmark{14}, David Barrado\altaffilmark{9}, 
Frederick J. Vrba\altaffilmark{15}, Kevin Covey\altaffilmark{16}, 
William Herbst\altaffilmark{17}, Gabor Furesz\altaffilmark{18}, 
Suzanne Aigrain\altaffilmark{19}, Fabio Favata\altaffilmark{20} }
\altaffiltext{1}{Spitzer Science Center, California Institute of
Technology, Pasadena, CA 91125, USA}
\altaffiltext{2}{LESIA, Observatoire de Paris-Meudon, 5 place Jules 
Janssen, 92195, Meudon, France}
\altaffiltext{3}{Departamento de F\'{\i}sica -- ICEx -- UFMG, 
Av. Ant\^onio Carlos, 6627, 30270-901, Belo Horizonte, MG, Brazil}
\altaffiltext{4}{Astronomy Department, California Institute of
Technology, Pasadena, CA 91125, USA}
\altaffiltext{5}{UJF-Grenoble 1 / CNRS-INSU, Institut de Plan\'etologie 
et d'Astrophysique de Grenoble (IPAG) UMR 5274, Grenoble, F-38041, France}
\altaffiltext{6}{Jet Propulsion Laboratory, California Institute
of Technology, Pasadena, CA 91109, USA}
\altaffiltext{7}{Infrared Processing and Analysis Center, California Institute of
Technology, Pasadena, CA 91125, USA}
\altaffiltext{8}{Department of Physics and Astronomy, 5151 State University
Drive, California State  University at Los Angeles, Los Angeles, CA 90032}
\altaffiltext{9}{Centro de Astrobiolog\'ia, Dpto. de
Astrof\'isica, INTA-CSIC, PO BOX 78, E-28691, ESAC Campus, Villanueva de
la Ca\~nada, Madrid, Spain}
\altaffiltext{10}{INAF - Osservatorio Astronomico di Palermo, Piazza 
del Parlamento 1, 90134, Palermo, Italy}
\altaffiltext{11}{Department of Physics and Astronomy, The University 
of Georgia, Athens, GA 30602-2451, USA}
\altaffiltext{12}{Five College Astronomy Department, Smith College, 
Northampton, MA 01063, USA}
\altaffiltext{13}{Department of Astronomy, University of Michigan, 
500 Church Street, Ann Arbor, MI 48105, USA}
\altaffiltext{14}{Astronomy Department, University of Wisconsin-
Madison, 475 N. Charter St., Madison, WI 53706, USA}
\altaffiltext{15}{U.S. Naval Observatory, Flagstaff Station, 10391 
West Naval Observatory Road, Flagstaff, AZ 86001, USA}
\altaffiltext{16}{Lowell Observatory, 1400 West Mars Hill 
Road, Flagstaff, AZ 86001, USA}
\altaffiltext{17}{Astronomy Department, Wesleyan University, 
Middletown, CT 06459, USA}
\altaffiltext{18}{Harvard-Smithsonian Center for 
Astrophysics, Cambridge, MA 02138, USA}
\altaffiltext{19}{Sub-department of Astrophysics, Departmentof Physics,
University of Oxford, Oxford OX1 3RH}
\altaffiltext{20}{European Space Agency, 8-10 rue Mario Nikis, 
F-75738 Paris Cedex 15, France}
\email{stauffer@ipac.caltech.edu}

\begin{abstract}
Based on more than four weeks of continuous high cadence photometric
monitoring of several hundred members of the young cluster NGC~2264
with two space telescopes, NASA's {\em Spitzer} and the CNES {\em
CoRoT}\ (Convection, Rotation, and planetary Transits), we provide
high quality, multi-wavelength light curves for young stellar objects
(YSOs) whose optical variability is dominated by short duration flux
bursts, which we infer are due to enhanced mass accretion rates. These
light curves show many brief -- several hour to one day --
brightenings at optical and near-infrared (IR) wavelengths with
amplitudes generally in the range 5-50\% of the quiescent value. 
Typically, a dozen or more of these bursts occur in a thirty day
period.  We demonstrate that stars exhibiting this type of variability
have large ultraviolet (UV) excesses and dominate the portion of the
$u-g$ vs.\ $g-r$ color-color diagram with the largest UV excesses. 
These stars also have large H$\alpha$ equivalent widths, and either
centrally peaked, lumpy H$\alpha$ emission profiles or profiles with
blue-shifted absorption dips associated with disk or stellar winds.  
Light curves of this type have been predicted for stars whose
accretion is dominated by Rayleigh-Taylor
instabilities at the boundary between their magnetosphere and inner
circumstellar disk, or where magneto-rotational instabilities modulate
the accretion rate from the inner disk.  Amongst the stars with the
largest UV excesses or largest H$\alpha$ equivalent widths, light
curves with this type of variability greatly outnumber light curves
with relatively smooth sinusoidal variations associated with
long-lived hot spots.  We provide quantitative statistics for the
average duration and strength of the accretion bursts and for the
fraction of the accretion luminosity associated with these bursts.
\end{abstract}


\section{Introduction}

Time series photometry obtained with space based telescopes -- {\em
MOST} (Microvariability and Oscillations of STars; Walker et al.\
2003), {\em CoRoT} (Convection, Rotation, and planetary Transits;
Baglin et al.\ 2009) and Kepler -- have recently been used to derive
very high quality light curves for normal low mass field stars.  
Those data have provided a wealth of new information on, for example,
the size evolution and stability of cold spots on main sequence stars
(Silva-Valio \& Lanza 2011; Walkowicz et al. 2013),  the ages of main
sequence stars from asteroseismology (Silva Aguirre et al. 2013), and
the frequency of occurence of planets (Fressin et al. 2013).  These
results required the very high quality data achievable from space
because the variability amplitudes for these old stars are quite
small, and in many cases, the needed cadence and duration to
successfully address these topics cannot be readily obtained with
ground-based data.

Very young, pre-main sequence (PMS) stars are in principle much more 
favorable targets for time series photometry.  In the first place, the
amplitudes of variability for young stellar objects (YSOs) are often
much larger, with amplitudes up to a magnitude or more.  In addition,
there are many more physical processes that can cause variability in
YSOs (variable extinction, variable accretion, flares, very large cold
spots, relatively rapid rotation, etc.).  The different mechanisms
have differing timescales, ranging from hours to weeks.  They also
have different spectral signatures, giving impetus to multi-wavelength
and spectral monitoring.  The rich variety of phenomena in YSOs have
motivated ground-based observing campaigns to characterize their
variability properties at least for the past half century.

For the past two decades, the primary reference for the
characterization and interpretation of YSO light curves has been
Herbst et al.\ (1994; H94). H94 sorted YSO light curves into four
basic types. Class I light curves are periodic and roughly sinusoidal
in shape, and result from the rotational modulation of cold,
non-axisymmetrically distributed starspots; stars with Class I light
curves are typically Weak-Lined T Tauri stars (WTTS), though some
Classical T Tauri stars (CTTS) may also show light curves of this
type. Class II light curves are thought to be the result of variations
in the veiling continuum of CTTS, and to reflect changes in the size,
shape or effective temperature of hot spots on their surfaces due to
variations in the accretion rate from their circumstellar disks.  Most
Class IIs vary irregularly in time, with no obvious periodicity, but
with characteristic timescales less than a few days.   However, a
subset of the Class IIs -- designated as Class IIp -- are periodic,
presumably due to a dominant hot spot group whose lifetime is long
compared to the star's rotation period.  Class III light curves show
(generally) irregularly spaced flux dips on an otherwise relatively
slowly varying quiescent brightness maximum; these stars generally
show little or no veiling in their spectra.  Stars with Class III
light curves in the H94 sample were all relatively high mass PMS stars
with spectral types earlier than K2.  H94 was least certain about the
physical mechanism for the Class III light curves, but advocated
variable extinction by circumstellar material as the most likely
cause.

Perhaps the most interesting of these light curve types is the Class
II (and IIp) group, because their properties should provide direct
insight into the accretion process. Recent advances in computational
astrophysics have made it possible to begin to construct physically
realistic 3D magnetohydrodynamic (MHD) models of accretion disks
(Kulkarni and Romanova 2008 = KR08; Romanova et al.\ 2012 = R12;
McKinney et al.\ 2012; Cemeljic et al.\ 2013).   For stars with
relatively strong magnetic dipoles, these models predict that the
accreting gas flows along field lines and impacts the star at nearly
free-fall velocities at relatively high latitude, resulting in the
stable, long-lived hot spots and predicting light curves that are
periodic and approximately sinusoidal (see Figure 7a of KR08). If one
of these ``funnel flows" passes through our line of sight, the
infalling warm gas could be the source for variability in the inverse
P Cygni profiles often seen in the emission line profiles of classical
T Tauri stars (Kurosawa \& Romanova 2013).  Stars with this type of
accretion could plausibly be associated with the Class IIp light
curves of H94 (as originally suggested by Bertout et al.\ 1988).

Other theoretical models predict less well-behaved accretion
variability, plausibly better matched to the Class II light curves of
H94. Some 3D MHD model simulations from KR08 exhibit Rayleigh-Taylor
(RT) instabilities at the disk-magnetosphere boundary, resulting in
many short-lived tongues of matter penetrating between the stellar
magnetic field lines.  Typically, a few such tongues reach the surface
of the star at any one time, producing short-lived hot spots at lower
latitudes than for the funnel flows and relatively chaotic looking
light curves (see Figure 7b of KR08).  Other 3D MHD models show that
in some cases magneto-rotational instabilities (MRI) in the inner
circumstellar disk of a YSO can lead to highly variable accretion
rates and predicted light curve shapes similar to those arising from
the RT instabilities (see Figure 6 of R12).

Quantitative comparison between these theoretical model light curves
and those of actual YSOs, particularly for the Class IIs, is  more
difficult than it might appear.   That is because the amplitudes
predicted for the instability-driven accretion bursts are often only
5-10\%, and their typical durations are of order a day.  Typical
ground-based campaigns with one or two epochs per night can detect
this type of variability, but does not provide the cadence, regularity
or photometric accuracy to accurately measure such features. Higher
cadence observations are needed in order to clearly characterize the
physical mechanisms. Do the light curves of the majority of real CTTS,
in fact, look anything like the model predictions?  If so, what types
of quantitative information can be drawn from these light curves --
for example, what are the relative frequency of periodic, sinusoidal
light curves as would result from stable funnel flows at high
latitudes, versus aperiodic, chaotic light curves resulting from RT or
MRI driven accretion?  Do the aperiodic, chaotic light curves dominate
at high accretion rates, as predicted by KR08?  Are the amplitudes and
durations of the observed accretion bursts roughly in accord with the
theoretical predictions?

The first steps towards clarifying how YSOs actually accrete using
high cadence, space-based monitoring have in fact been published
recently.  Alencar et al. (2010) provided {\em CoRoT} monitoring for
the YSOs of NGC 2264, and suggested that a few of their CTTS light
curves were likely due to stochastic accretion (see their Fig.\ 1e). 
Rucinski et al.\ (2008) used MOST to obtain a well-sampled light curve
for TW Hya, which also showed numerous brief flux excesses that they
associated with a variable accretion rate. A small sample of other
bright YSOs was also observed with MOST (Siwak et al. 2011; Cody et al
2013); for the YSOs with disks in this sample, there was some short
timescale variability which could not be attributed to long-lived
spots or flares, and which could be evidence for variable accretion.
However, none of these studies were accompanied by simultaneous
spectroscopic or multi-wavelength time-series data, which limited the
ability to provide conclusive interpretations to the light curves.

We have recently conducted an observational campaign for the
star-forming region NGC~2264 designed to have the cadence, duration,
precision, wavelength coverage and sensitivity to resolve many of the
long-standing issues for characterizing YSO variability.  That program
is called the Coordinated Synoptic Investigation of NGC~2264, or CSI~2264.
This campaign was built around 30+ days of continuous, simultaneous
monitoring by three space telescopes -- {\em CoRoT} (Baglin et al.\ 2009),
MOST (Walker et al.\ 2003), and {\em Spitzer} (Werner et al.\ 2004).   Here,
we report one result from that campaign: the empirical identification
of a set of YSOs whose light curves are best explained as being
dominated by short-timescale variations in the star's accretion rate.
We demonstrate that these stars include the most actively accreting
members of NGC~2264.

\section{Data Used for This Paper}

Cody et al.\ (2014) provide a detailed description of all of the 
optical and infrared (IR) observational data we have obtained for the
young stars in NGC~2264 and our basic data reduction procedures.  That
paper also provides an overview of the light curve morphologies we
find for the CTTS in the cluster.   Readers
are referred to that paper for specific details concerning  the
observations.   Here we provide just a brief synopsis of the
observations in order to provide the context necessary for our
discussion.   

{\em Spitzer}'s Infrared Array Camera, IRAC (Fazio et al.\ 2004) was
used as part of the CSI~2264 program (program ID 80040, PI: J.\
Stauffer) to observe NGC~2264 nearly continuously from December 3,
2011 to January 1, 2012.  In the first week of that period,
approximately four days were spent in ``staring mode" where the
telescope was pointed to a position near NGC~2264-IRS-1 (i.e., near
the center of the Spokes cluster; Teixeira et al.\ 2006) and repeated
frames were taken at that position for blocks of 19-26 hours without
interruption.  The integration time for each exposure was 4.4
seconds.  The two IRAC fields of view are separated on the sky by
about seven arcminutes, so we collect these light curves for one set
of 283 stars in Channel 1 (3.6 microns), and for a disjoint set of 249
stars in Channel 2 (4.5 microns).   The relative photometric accuracy
for these light curves approaches 1 mmag for stars that are bright but
not saturated.  For the remaining observing time within the CSI~2264
program, {\em Spitzer} observations of  NGC~2264 were done in mapping
mode, where a rectangular region centered near $(\alpha,\delta)\sim$06
40 45.0,+09 40 (2000) and about 45$\arcmin$ by 40$\arcmin$ in size was
observed about twelve times per day, with a total integration time of
about 40 seconds per point on the sky.  For one or two of these maps
each day, the data were taken in high-dynamic range (HDR) mode,
resulting also in data with 1.6 second total integration time.  These
data provide typical 1-$\sigma$ uncertainties of about 0.01 mag for
sufficiently bright stars. Most stars have light curves in both
channels, but stars near the edge of the map may have data in only one
channel.

The {\em CoRoT} camera has a field of view in excess of one square
degree, which is larger than the physical extent of NGC~2264. However,
only stars included within an input catalog have their photometry
downlinked to the ground, and these stars must satisfy brightness,
crowding and nebular background constraints.  In total,  about 490
probable cluster members  and more than a thousand likely field stars
were included in the input catalog for the December 2011 observing
campaign.  Data were obtained from MJD 55896.8 through MJD 55935.5,
with only a brief interruption around MJD 55917 due to  a telescope
pointing problem.  For most stars, the photometry is provided at a
cadence of 512 seconds; for a small subset of the stars, data are
provided at 32 second cadence.   While there is no hard limit, stars
fainter than R = 16.5 were generally excluded from the target list in
order to allow reasonably good signal-to-noise for the {\em CoRoT}
light curves.

In principle, {\em CoRoT} can provide multi-color light curves for
some of the observed stars.  However, for our purposes, we have only
utilized the ``white light'' data.   {\em CoRoT}'s bandpass is
essentially a broad $V + R$ filter.  The {\em CoRoT} data are
sky-subtracted aperture photometry which have been run through a
pipeline designed to mitigate artifacts that are present in the
downlinked data (Samadi et al.\ 2006).  In some cases, jumps in the
photometry are still present even after this pipeline has been run. 
When possible, we have removed these jumps manually; in other cases
where the jumps are less obvious, we have left the data ``as is.'' 
The flux RMS over 512 seconds in the light curves at $R$ = 12 is
typically about 0.002, increasing to about  0.015 at $R$ = 16.

A 20 day {\em CoRoT} ``short run'' for NGC~2264 was also obtained in
March 2008 (Alencar et al.\ 2010).  Light curves for many of the CTTS
members were also obtained in that campaign, and we use those data
where relevant.

We obtained high-resolution spectra of most of the NGC~2264 members
using either Hectochelle on the Multiple Mirror Telescope (MMT;
Szentgyorgyi et al.\ 2011) or the FLAMES multi-object spectrograph on
the Very Large Telescope (VLT; Pasquini et al.\ 2002).   The
Hectochelle spectra have been discussed previously in Furesz et al.\
(2006); the observations were obtained in 2004 and 2005, have a
resolution of about 34,000, and cover the spectral range 6460-6650 \AA.  
In general, no sky background has been
subtracted from these profiles, which can leave narrow emission cores
from the \ion{H}{2} region at the cluster rest velocity in the stellar
spectra.  The FLAMES spectra were obtained
with UT2 at the VLT, using the HR15N grating, producing a resolution
of about 17,000, and covering the spectral range 6375-6815 \AA.  
Only six of the stars that are the focus of  this
paper have FLAMES spectra, but for those stars we have up to 22
spectra, with usually five or six of them obtained during the time
period when {\em CoRoT} was obtaining photometry of the cluster (and the
others being obtained within about 60 days following the {\em CoRoT}
observations).  Spectroscopic rotational velocities ($v \sin i$) for
many of the NGC~2264 members are available from either Baxter et al.\
(2009)\footnotemark[1]\footnotetext[1]{Values for $v\sin i$ for just 97 stars were published
in Baxter et al.\ (2009); however, those authors derived $v \sin i$
for a larger set of stars from the same MMT Furesz et al.\ spectra,
and those additional $v \sin i$ values were made available to us by K.
Covey.} or from our analysis of the VLT/FLAMES spectra.

We also use a variety of broad and narrow band photometry of NGC~2264
members in order to measure the ultraviolet (UV) excesses of the YSOs
and thereby identify those that are actively accreting matter from
their circumstellar disks, and to measure their spectral energy
distributions (SEDs) and thereby characterize the IR excesses from
their disks.  The UV excess determinations are made using $ugri$
photometry obtained in 2010 and 2012 with the Canada-France-Hawaii
Telescope (CFHT) Megacam (Venuti et al.\ 2014). The broad-band
photometry used for the SED fits comes from a variety of sources, but
is primarily reported in either Rebull et al.\ (2002) or Sung et al.\
(2009).

\section{Identification and Initial Characterization of YSOs with 
       Burst-Dominated Optical Light Curves}

\subsection{Identification of Burst-Dominated Light Curves}

Cold spots on the photosphere, clumps or warps in the inner disk that
occult our line of sight, and companions that transit our line of
sight to the primaries all produce light curves with periodic signals.
These types of variables have been found in abundance by previous
ground-based surveys of star-forming  regions (Rebull et al.\ 2002;
Lamm et al.\ 2004 = L04; Herbst et al.\ 1994 and 2000; Stassun et
al.\ 1999;  Grankin et al.\ 2007; etc.).  As long as the duration of
the monitoring period is long enough, and the period is sufficiently
stable, one can make up for the relative sparseness of the sampling 
and the interruptions by weather in these ground surveys by folding
the light curve at the identified period.  We also find these types of
variables in our data, and we provide an initial summary of their
prevalence and characteristics in Cody et al.\ (2014) and Affer et
al.\ (2014). Rotation periods derived from the 2008 {\em CoRoT}
NGC~2264 observations are reported in Affer et al.\ (2013).

Types of variability that are not periodic (e.g., the Class II
variables of H94) are much harder to identify
and characterize using ground-based photometric monitoring data.  The
most extensive ground-based photometric monitoring campaign
conducted for the YSOs of NGC~2264 was that of L04.  Those authors
obtained 88 epochs of $I$ band photometry for about 600
cluster members, plus additional candidate members and field stars,  
during an approximately 40 day campaign. Roughly 400 of these stars,
primarily weak-lined TTauri stars (WTTS), were identified as periodic
variables and 180 stars, primarily CTTS, as irregular variables.  L04
surmised that in most cases the irregular variability could be
attributed to ``variable mass accretion resulting in hot spots which
are not stable in brightness, size and location over a few rotation
periods."  However, the small number of epochs and irregular cadence
did not allow further elucidation of the physical nature of the
variability.  

\begin{figure*}
\begin{center}
\epsscale{0.80}
\plottwo{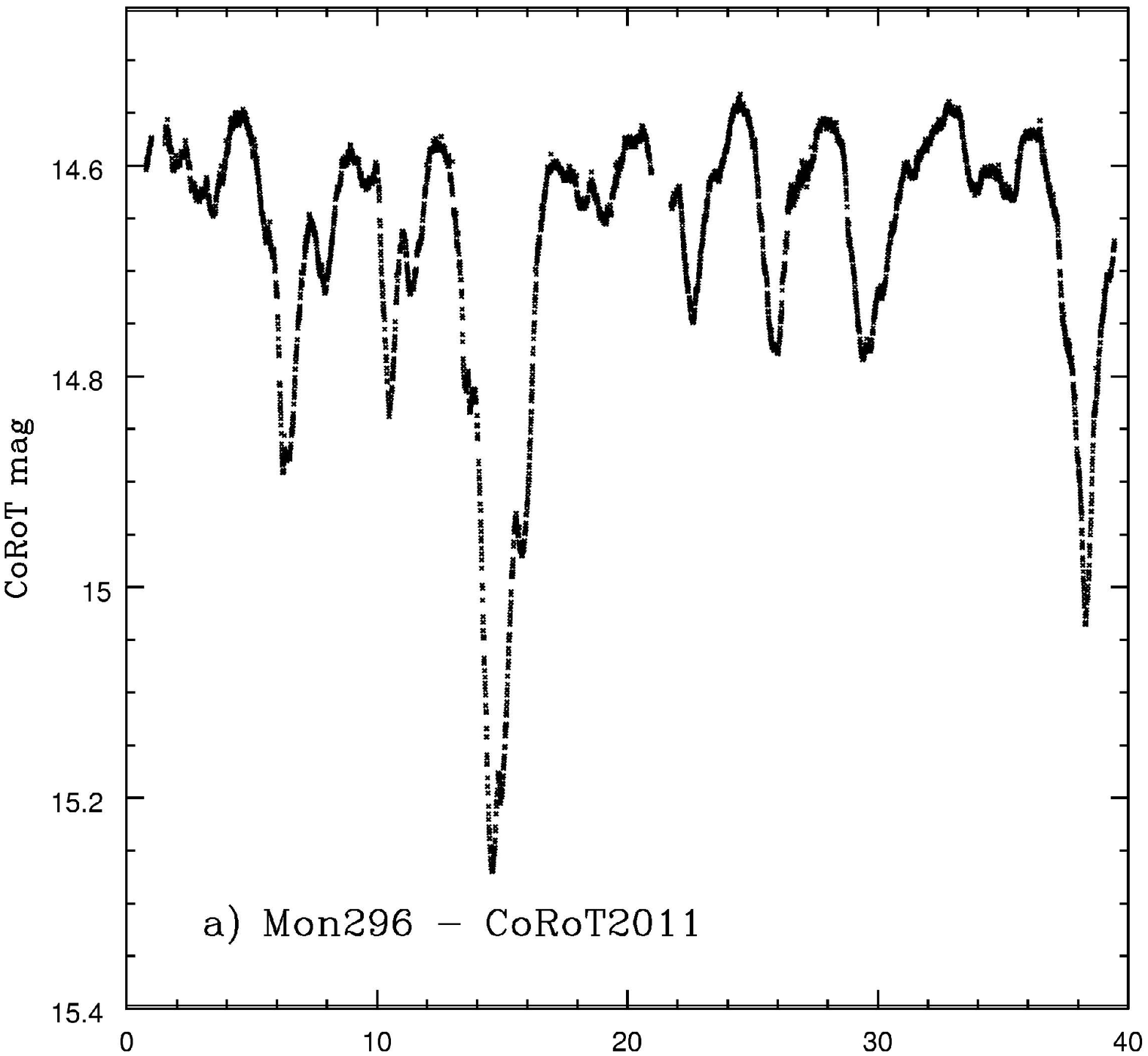}{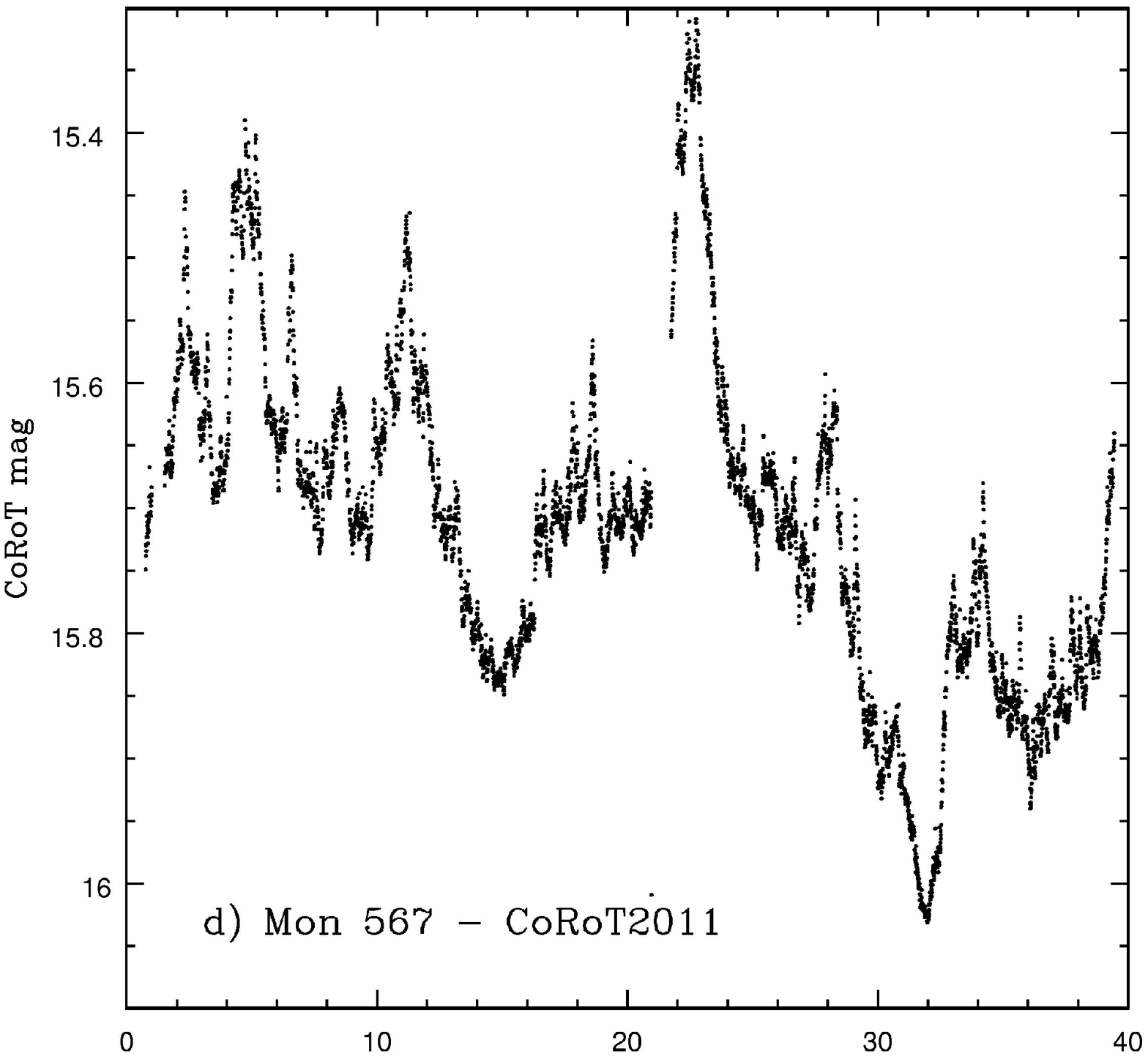}
\vspace{-0.cm}
\plottwo{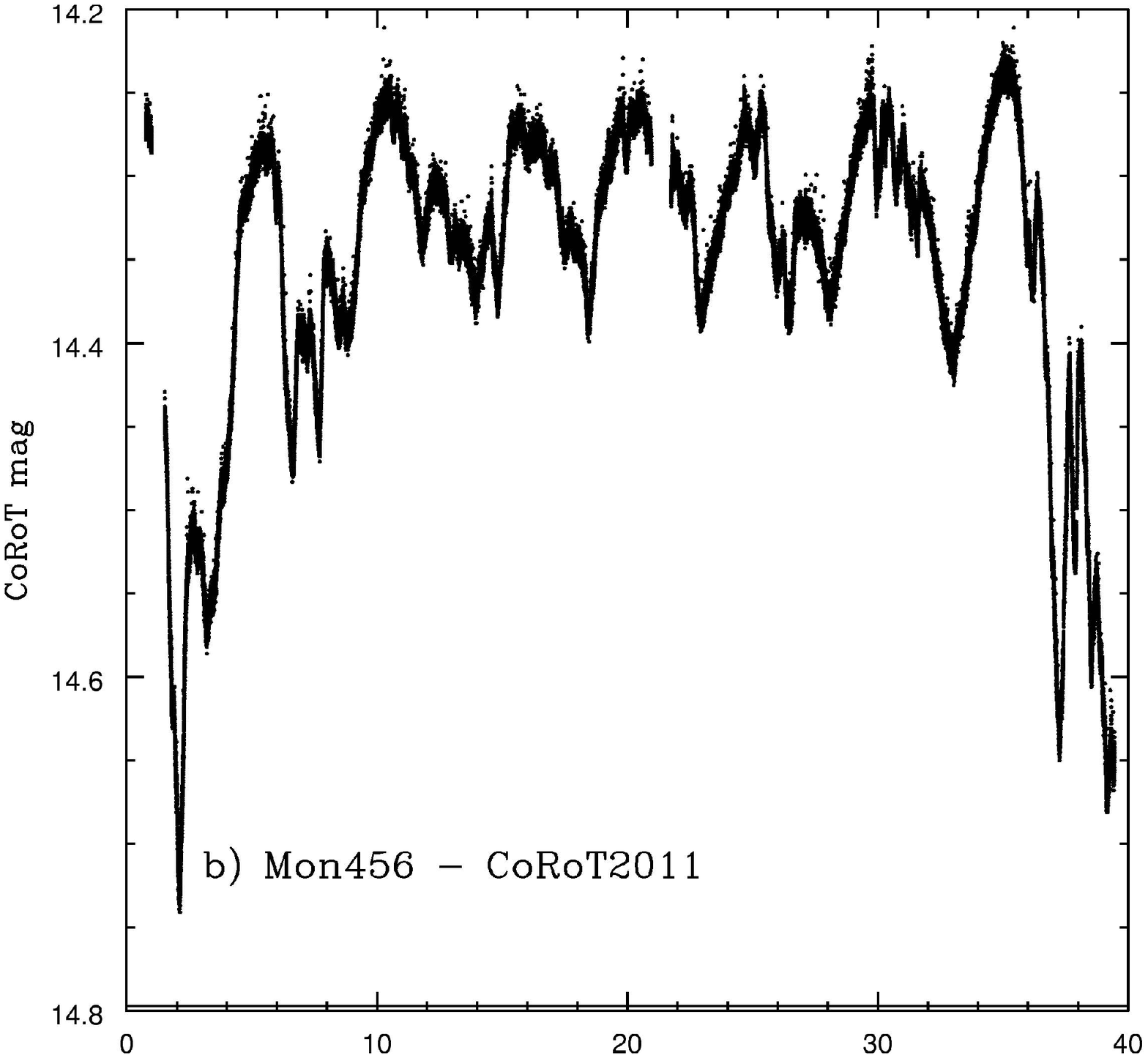}{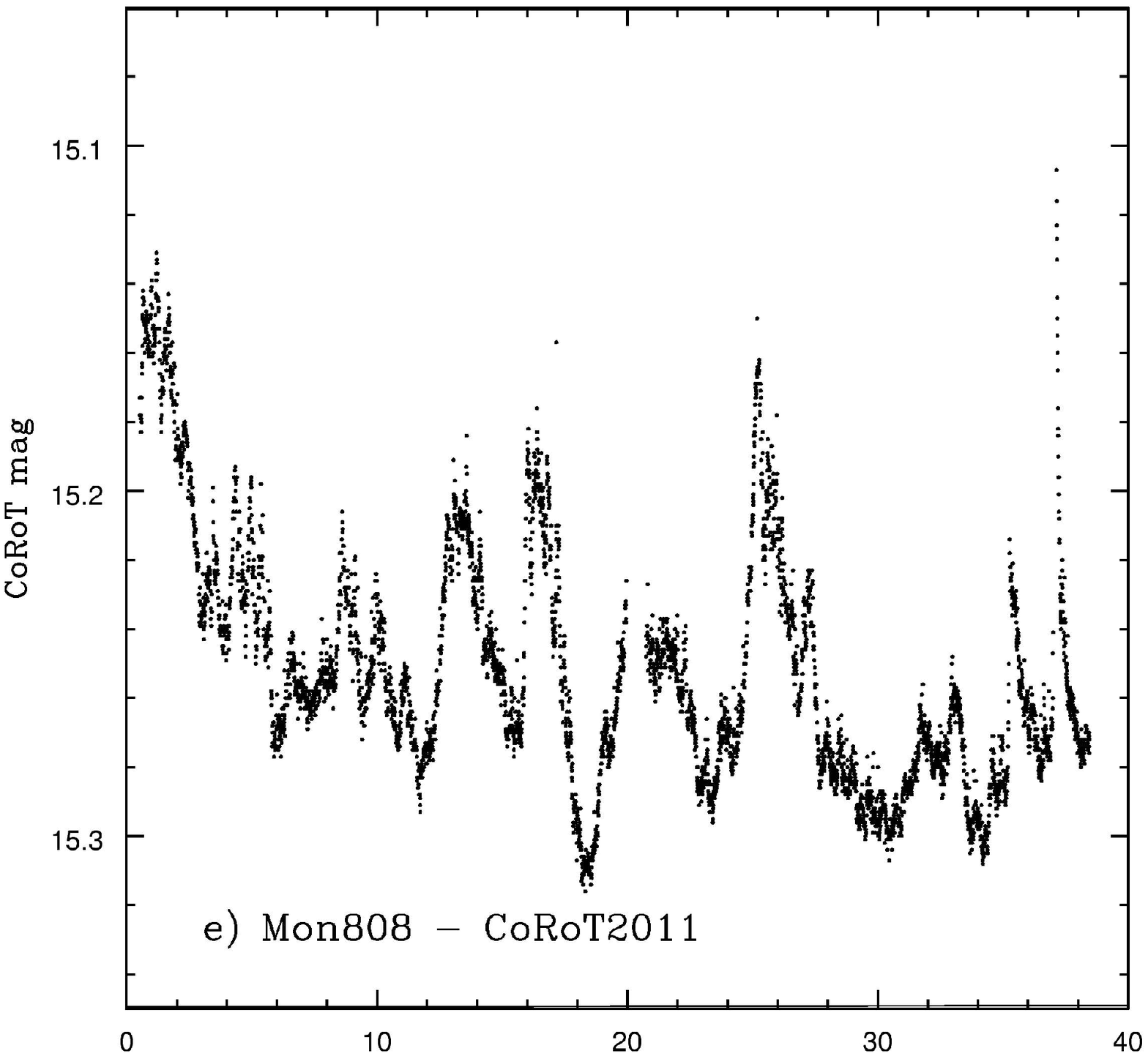}
\vspace{-0.cm}
\plottwo{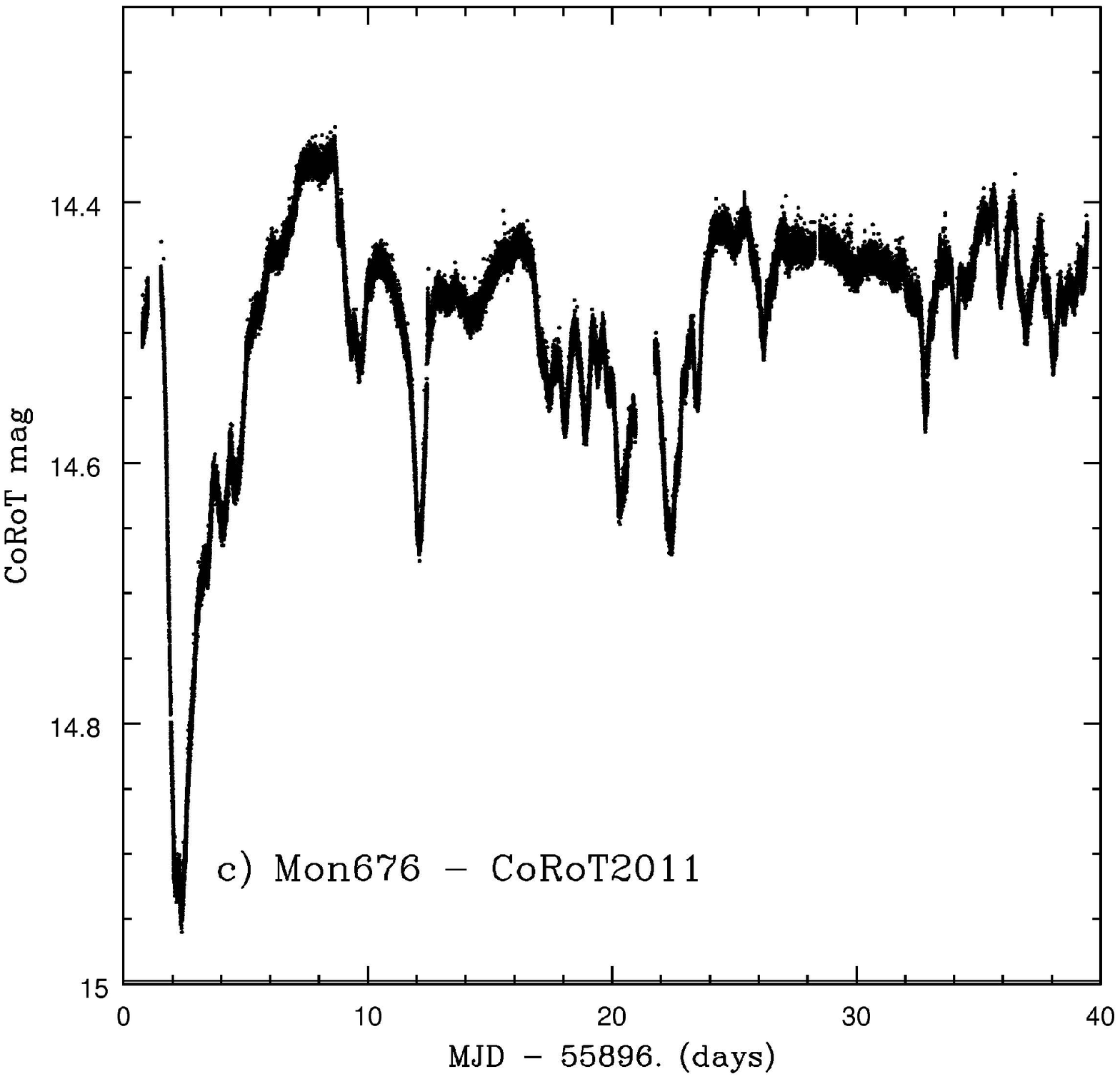}{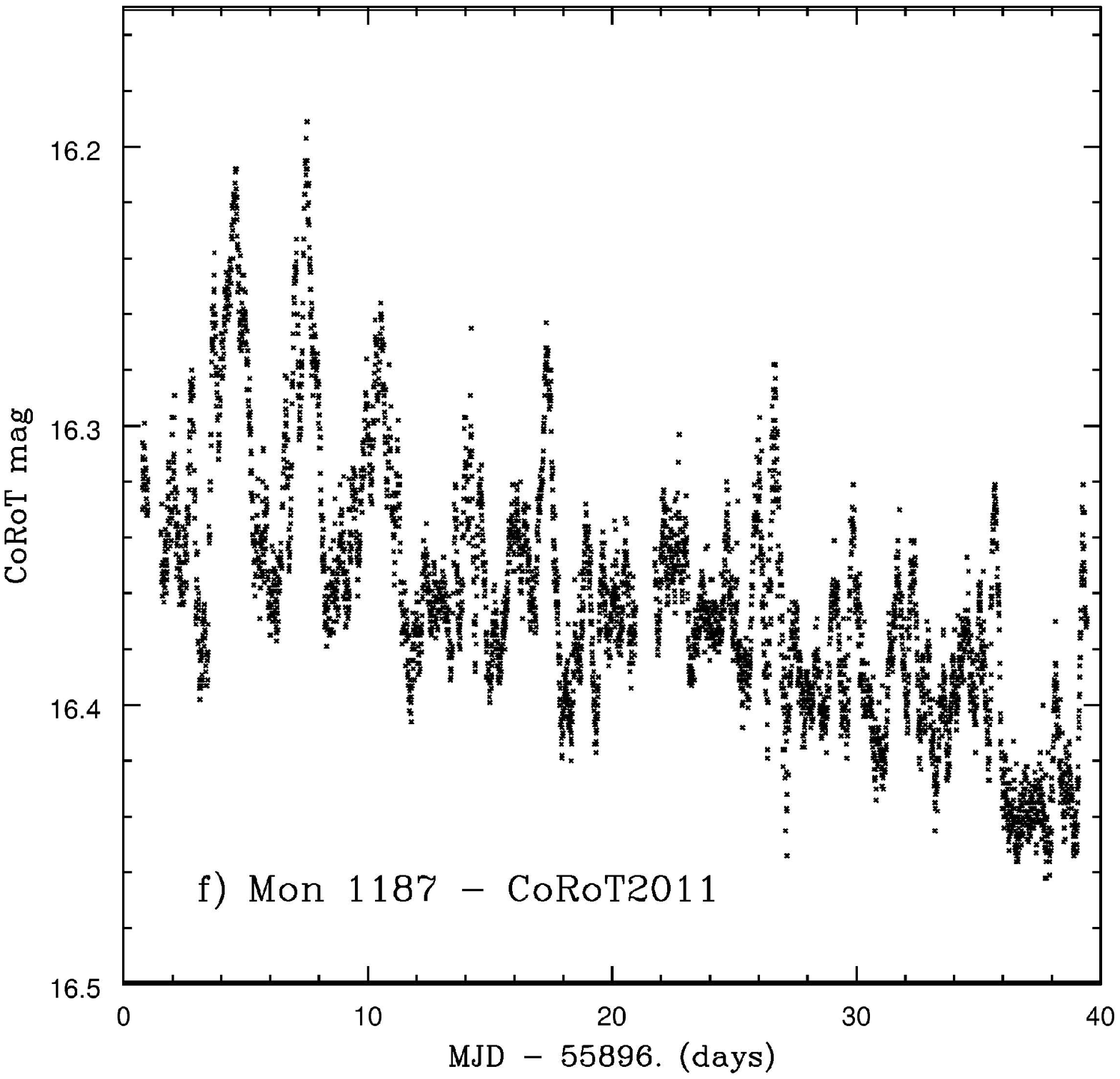}
\end{center}
\caption{{\em CoRoT} light curves for six CTTS members identified by
L04 as  irregular variables.  We identify the first three
(Mon-000296=panel a, Mon-000456=panel b and Mon-000676=panel c) as
having light curves dominated by variable extinction events. The next
three stars (Mon-000567=panel d, Mon-000808=panel e, and
Mon-001187=panel f) have light curves dominated by short duration flux
bursts. Each light curve covers roughly 40 days of time; the $y$-axis
is in {\em CoRoT} counts, where the {\em CoRoT} bandpass corresponds
roughly to a very broad $V +R$ band filter.  For two of the stars
(Mon-000456 and Mon-000676) the cadence is roughly every 30 seconds;
for the four other stars, the cadence is about 7.5 minutes.
\label{fig:sixctts}}
\end{figure*}

We have examined our data, and find that only 39 of the L04
irregular variables have {\em CoRoT} light curves; the overlap is not
better primarily because the L04 survey went considerably deeper
than the {\em CoRoT} data.   Twenty of the thirty-nine stars in common
with L04 stars have light curves that are heterogeneous in their
properties and where more than one physical mechanism may contribute
significantly to the light curve shape. However, for the other half,
the light curves fall into two dominant light curve types.  Ten of the
stars have light curves characterized as having a more or less stable
light maximum interspersed with well-defined flux dips.   Based on our
previous experience with these types of light curves (Bouvier et al.\
2007; Alencar et al.\ 2010), we believe that the flux dips in these
stars are due to periods of enhanced extinction, as for the Class III
variables of H94; in some cases, the flux dips occur approximately
periodically and may be due to an inner disk warp passing through our
line of sight (Terquem \& Papaloizou 2000). Nine other stars, however,
have light curves that appear roughly similar to each other but are
dominated by short duration flux bursts, which we believe are the
stars with variable mass accretion which L04 had hoped to identify.
These nine light curves appear reasonably similar to those predicted
by KR08 or Romanova et al.\ (2012)  for instability-driven accretion. 
Figure~\ref{fig:sixctts} shows {\em CoRoT} light curves for three
stars from each of these two classes.

\begin{deluxetable*}{lccccccccc}
\tabletypesize{\scriptsize}
\tablecolumns{10}
\tablewidth{0pt}
\tablecaption{Basic Information for YSOs with Burst-Dominated {\em CoRoT}
Light Curves\label{tab:basicinformation}}
\tablehead{
\colhead{Mon ID\tablenotemark{a}}  & \colhead{2MASS ID} &
\colhead{{\em CoRoT} 2008\tablenotemark{b}} & \colhead{{\em CoRoT} 2011\tablenotemark{b}} &
\colhead{SpT\tablenotemark{c}} & \colhead{H$\alpha$ EW
(\AA)\tablenotemark{c}} & \colhead{H$\alpha$ Type} &
\colhead{$v \sin i$ (km s$^{-1}$)\tablenotemark{d}} & \colhead{Veiling}\tablenotemark{e} &
\colhead{FR(4.5)\tablenotemark{f}} 
}
\startdata
Mon-000007 &  06415304+0958028 & 223994721 & 223994721 & K7 & 11.5 & I & 80.8 & \nodata & 3.62 \\
Mon-000011 & 06411725+0954323 & 223985009 & 223985009 & K7 & 58.3 & \nodata & \nodata & \nodata & 5.14 \\
Mon-000117 & 06405413+0948434 & \nodata & 602095753 & M2.5 & 353. & I  & \nodata & \nodata & 2.42 \\
Mon-000185 & 06413876+0932117 &  500007249 & 616919566 & K4 & 58.6 & I & 9.6 & \nodata & 3.68 \\
Mon-000260* & 06411099+0935556 & 500007727 & \nodata & K7 & 61.5 & III-B & 17.5 & 1.99 & 6.60  \\
Mon-000341* & 06405426+0949203 & 500007473 & 616849439 & M0.5 & 161. & I & 16.2 & 0.74 & 4.22  \\
Mon-000406 & 06405968+0928438 & \nodata & 616943998 & \nodata & 46.1 & \nodata & \nodata & \nodata & 3.78   \\
Mon-000412 & 06404711+0932401 & \nodata & 616919737 & M1 & 30.7 & I & 8.9 & \nodata & 2.56  \\
Mon-000469 &  06404114+0933578 & \nodata & 602083890 & K7 & 236.5 & III-B & \nodata & \nodata & 4.40  \\
Mon-000474 & 06410682+0927322 & \nodata & 603396438 & G & 104.7 & I & 17.7 & \nodata & 4.97  \\
Mon-000510* & 06410429+0924521 & 500007335 & 602079845 & M0 & 101.8 & I & 18. & 0.30 & 3.67  \\
Mon-000567 & 06405639+0935533 & \nodata & 616919752 & K3 & 84.1 & III-B & \nodata & \nodata & 5.75  \\
Mon-000808 & 06405159+0928445 & \nodata & 603396401 & K4 & 50.2 & I & 9.0 & \nodata & 2.38  \\
Mon-000860 &  06415492+0942527 & 223995308 & \nodata & M2.5 & 261.0 & II-B & \nodata & \nodata & 0.98   \\
Mon-000877 & 06411678+0927301 & \nodata & 616943883 & K4 & 91.4 & II-B & 13.3 & \nodata & 2.64   \\
Mon-000919 & 06411329+0931503 & \nodata & 616919654 & M4 & 79.9 & I & 2.6 & \nodata & 2.52  \\
Mon-000945* & 06404989+0936494 & 223977953 & 223977953 & K4 & 66.3 & I & 16. & 0.67 & 2.97  \\
Mon-000996* & 06404131+0951023 & 500007315 & 616849542 & K7 & 24.5 & I & 15.5 & 0.61 & 2.60   \\
Mon-001022* & 06403911+0950586 & 500007252 & 616849543 & K4 & 46.5 & III-B & 15. & 0.64 & 5.23  \\
Mon-001174 & 06401370+0956305 & 400007614 & 616826810 & M2 & 130.3 & III-B & 10.1 & \nodata & 5.59  \\
Mon-001187 & 06401417+0934283 & \nodata & 602083884 & \nodata & 8.0 & \nodata & 15.5 & \nodata & 2.31   \\
Mon-001217 & 06403665+0952032 & \nodata & 616849540 & K4 & 87.0 & III-B & \nodata & \nodata & 6.88   \\
Mon-001573 &  06401258+1005404 & 223968439 & 223968439 & \nodata & 26.0 & I & \nodata & \nodata & 2.53  \\
\enddata
\tablenotetext{a}{The Mon IDs are our internal naming scheme for stars in
the field of NGC~2264 -- see Cody et al.\ (2014).   An asterisk
following the Mon ID indicates that we have VLT FLAMES spectra for
this star from December 2011.}
\tablenotetext{b}{CoRoT identification numbers.  These uniquely identify the light curve
for each CoRoT short run.}
\tablenotetext{c}{See Cody et al.\ (2014) for the sources of the
spectral type and H$\alpha$ equivalent width (EW) data. All of these
are in emission.}
\tablenotetext{d}{Data from MMT spectra except for Mon 510, 945 and 1022, which are from
the VLT/FLAMES spectra.}
\tablenotetext{e}{Veiling estimate at a wavelength of 6600\AA\ derived from the VLT spectra.} 
\tablenotetext{f}{FR(4.5) is an estimate of
the ratio of the flux from the disk to the flux from the stellar
photosphere at 4.5 $\mu$m.  Because the SED models do not include veiling, these values
are likely lower limits.}
\end{deluxetable*}

In order to determine how common light curves of the type shown in
Figure~\ref{fig:sixctts}d-f are, two of us (JRS and AMC) have closely
examined the {\em CoRoT} light curves of all $\sim$550 probable
NGC~2264 members and the $\sim$1250 {\em Spitzer} light curves for
cluster members.   The {\em CoRoT} data provide light curves for 197
CTTS likely to have ongoing  accretion, with nearly 1000 hours of
monitoring per star. Within these light curves, we find dozens of YSOs
with some evidence of short duration flux bursts like those shown in
Figure~\ref{fig:sixctts}d-f, though in some cases the flux bursts are
superposed with other types of variability, often of larger
amplitude.   To restrict ourselves to stars whose light curves are
dominated by flux bursts, we require the light curves to have: (a) a
relatively flat or only slowly varying ``continuum"; (b) an intrinsic
noise level in the light curve less than 1\%; and (c) presence in the
light curve of at least half a dozen narrow (one hour to one day),
approximately symmetric (unlike flares - see \S3.2 and \S6.1) sharply
peaked flux ``bursts", with at least one of the bursts having an
amplitude greater than 5\%  of the continuum level.    We have
identified a set of 23 YSOs whose 2008 and/or 2011 {\em CoRoT} light
curve variability satisfies these criteria. Hereafter, we refer to 
light curves with these characteristics as {\bf burst-dominated light
curves}\footnotemark[2]\footnotetext[2]{While this paper and Cody et al.\ (2014) are being submitted
together and are closely linked, they were written in parallel and
evolved somewhat independently. The two papers used slightly
different sets of data  - most importantly, Stauffer et al.\ included
stars with only 2008 CoRoT light curves whereas such stars were
excluded from Cody et al. - and slightly different criteria for defining
light curve classes.  This resulted in somewhat
different sets of stars belonging to each class.  In
particular, the burst-dominated class in Table 1 of this paper
includes 23 stars; 19 of those are also listed as burst-dominated
in Table 3 of Cody et al.  Of the remaining four, one (Mon-000185) was
included here based on its 2008 CoRoT light curve and
hence was not in the parent sample for the Cody et al.\ paper.  The other three
were all classified as having ``stochastic" light curves in Cody et al.  Inclusion or
exclusion of these stars from the burst-dominated class would not
appreciably change the conclusions in either paper.}. 
Table~\ref{tab:basicinformation} provides the list of the
YSOs displaying such light curves and some of their spectroscopic
properties. Table~\ref{tab:basicphotom} provides single-epoch optical,
near-IR and {\em Spitzer} photometry for these stars.  The {\em CoRoT}
light curves for all of the stars in Table~\ref{tab:basicinformation}
are provided in the Appendix (Figure~\ref{fig:allcorotlcs}).  For the
stars of Table~\ref{tab:basicinformation} where we also have IRAC
light curves, Figure~\ref{fig:corot2011} of the Appendix overplots the
{\em CoRoT} and {\em Spitzer} light curves.

In the next several sub-sections, we discuss qualitative aspects
of these light curves.   The remaining sections will be used to provide
more quantitative analysis of these stars and their light curves.

\begin{deluxetable*}{ccccccccccccc}
\tabletypesize{\scriptsize}
\tablecolumns{10}
\tablewidth{0pt}
\tablecaption{Photometric Data for the Stars of
Table~\ref{tab:basicinformation}\label{tab:basicphotom}}
\tablehead{
\colhead{Mon ID}  & \colhead{$u$} & \colhead{$g$} & \colhead{$r$} &
\colhead{$i$} & \colhead{$J$} & \colhead{$H$} & \colhead{$K_s$} &
\colhead{[3.6]} &
\colhead{[4.5]} & \colhead{[5.8]} & \colhead{[8.0]} & \colhead{[24]}
}
\startdata
Mon-000007 & 15.850 & 14.807 & 14.674 & 14.192 & 12.220 & 11.529 & 11.175 & 10.136 & 9.785 & 9.340 & 8.692 & 6.17 \\
Mon-000011 & 18.548 & 16.808 & 15.531 & 14.535 & 12.833 & 12.045 & 11.552 & 10.486 & 10.017 & 9.649 & 8.852 & 5.541 \\
Mon-000117 & 18.562 & 18.326 & 16.832 & 16.358 & 14.162 & 13.430 & 13.070 & 12.469 & 12.136 & 12.038 & \nodata & \nodata \\
Mon-000185 & 17.175 & 15.647 & 14.593 & 14.198 & 12.875 & 12.112 & 11.723 & 10.861 & 10.581 & 10.347 & 9.775 & 6.22 \\
Mon-000260 & 19.983 & 18.550 & 16.956 & 15.905 & 13.358 & 12.232 & 11.498 & 10.402 & 9.915 & 9.491 & 8.707 & 5.422 \\
Mon-000341 & 17.927 & 17.147 & 15.744 & 14.951 & 13.101 & 12.279 & 11.800 & 10.971 & 10.507 & 9.965 & 8.772 & 5.248 \\
Mon-000406 & 17.151 & 16.193 & 15.229 & 14.613 & 13.188 & 12.407 & 12.084 & 11.302 & 10.952 & 10.457 & 9.586 & 6.503 \\
Mon-000412 & 17.872 & 16.624 & 15.461 & 14.699 & 12.770 & 12.002 & 11.694 & 11.063 & 10.720 & 10.333 & 9.726 & 6.692 \\
Mon-000469 & 18.291 & 18.154 & 16.692 & 15.353 & 13.121 & 12.248 & 11.452 & 10.673 & 10.207 & 9.739 & 9.005 & 5.877 \\
Mon-000474 &  \nodata  &  \nodata  &  \nodata  &  \nodata  & 10.434 & 9.533 & 8.858 & 8.160 & 7.621 & 7.210 & 6.496 & 3.49 \\
Mon-000510 & 16.568 & 15.699 & 14.952 & 14.614 & 12.719 & 11.934 & 11.458 & 10.737 & 10.276 & 9.830 & 9.016 & 5.579 \\
Mon-000567 & 19.064 & 17.601 & 15.995 & 15.254 & 12.775 & 11.786 & 11.015 & 10.087 & 9.616 & 9.259 & 8.579 & 4.771 \\
Mon-000808 & 18.098 & 16.360 & 15.213 & 14.758 & 12.993 & 12.203 & 11.786 & 11.274 & 11.081 & 10.854 & 10.297 & 6.295 \\
Mon-000860 & 19.434 & 18.609 & 16.978 & 16.350 & 14.328 & 13.640 & 13.405 & 13.293 & 13.190 & 13.103 & \nodata & 7.403 \\
Mon-000877 & 17.259 & 15.571 & 14.511 & 14.095 & 12.645 & 11.877 & 11.502 & 11.081 & 10.796 & 10.629 & 10.130 & 6.75 \\
Mon-000919 & 20.545 & 18.485 & 16.947 & 15.398 & 12.738 & 12.034 & 11.683 & 11.005 & 10.617 & 10.139 & 9.375 & 6.503 \\
Mon-000945 & 17.364 & 16.233 & 15.082 & 14.443 & 12.399 & 11.611 & 11.160 & 10.516 & 10.157 & 9.887 & 9.257 & 6.242 \\
Mon-000996 & 17.496 & 16.382 & 15.214 & 14.616 & 12.806 & 12.024 & 11.713 & 10.905 & 10.554 & 10.152 & 9.432 & 7.02 \\
Mon-001022 & 17.545 & 16.299 & 15.049 & 14.392 & 12.293 & 11.251 & 10.508 & 9.614 & 9.170 & 8.866 & 8.308 & 5.635 \\
Mon-001174 & 17.485 & 16.967 & 15.727 & 15.098 & 12.920 & 12.131 & 11.584 & 10.605 & 9.985 & 9.270 & 8.162 & 5.378 \\
Mon-001187 & 21.403 & 18.943 & 17.510 & 15.997 & 13.753 & 12.964 & 12.587 & 11.869 & 11.518 & 11.230 & 10.535 & 7.261 \\
Mon-001217 & 17.939 & 16.900 & 15.573 & 14.806 & 12.733 & 11.759 & 11.101 & 10.022 & 9.401 & 8.926 & 7.923 & 4.641 \\
Mon-001573 & 18.151 & 16.481 & 15.298 & 14.593 & 12.511 & 11.792 & 11.404 & 10.681 & 10.446 & 9.921 & 9.483 & 6.665 \\
\enddata 
\tablecomments{Broad-band photometry for the stars from
Table~\ref{tab:basicinformation}, all in units of magnitudes, though a mix of
Vega and AB magnitudes.  The $ugri$ data are from CFHT, are reported
in Venuti et al.\ (2014), and in AB magnitudes; the $JHK_s$ data are
Vega magnitudes from the on-line 2MASS all-sky point source catalog;
the IRAC and MIPS data are Vega magnitudes from {Sung et al. 2009}, based on Spitzer imaging
obtained in 2004.} 
\end{deluxetable*}

\subsection{Proof That These Flux Bursts are Real} 

Given the comparative novelty of the light curves in
Figure~\ref{fig:sixctts}d-f, it seems useful to begin by providing
arguments why we are certain that the flux bursts are real and not due
to some type of artifact in the {\em CoRoT} data.  As described in
Cody et al.\ (2014), not only do we have {\em CoRoT} light curves for
$\sim$550 NGC~2264 members (from both 2008 and 2011) but we also have
light curves for more than 500 field stars of similar magnitude
obtained simultaneously with the NGC~2264 members and reduced in the
same manner.  Events such as those shown in
Figure~\ref{fig:sixctts}d-f and in the Appendix
(Figure~\ref{fig:allcorotlcs}) are not present amongst the field
stars.   They are also not present, with one or two possible
exceptions, amongst the NGC~2264 members that do not have accretion
disks nor strong emission lines (i.e., among the WTTS). 

\begin{figure*}
\begin{center}
\epsscale{1.0}
\plottwo{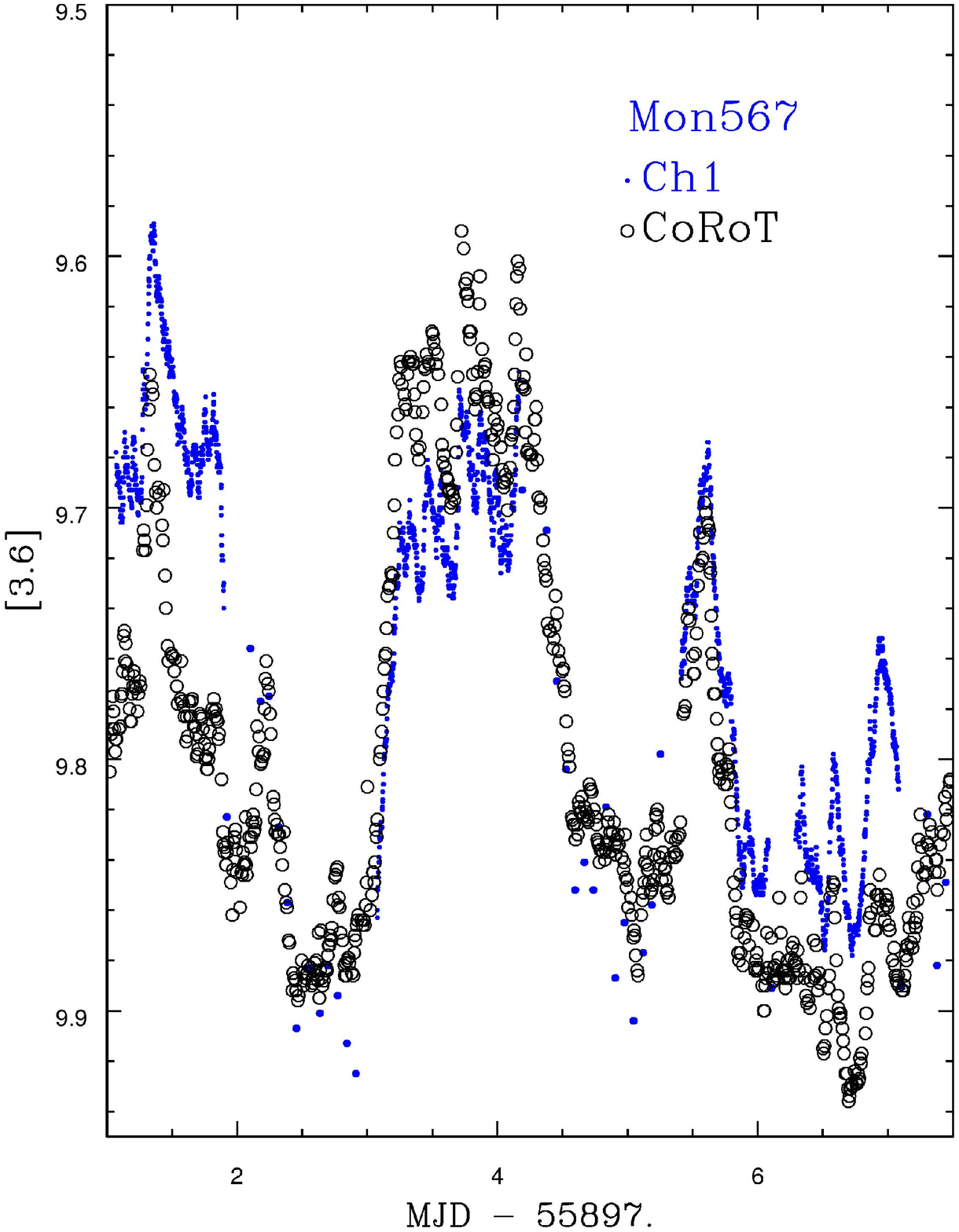}{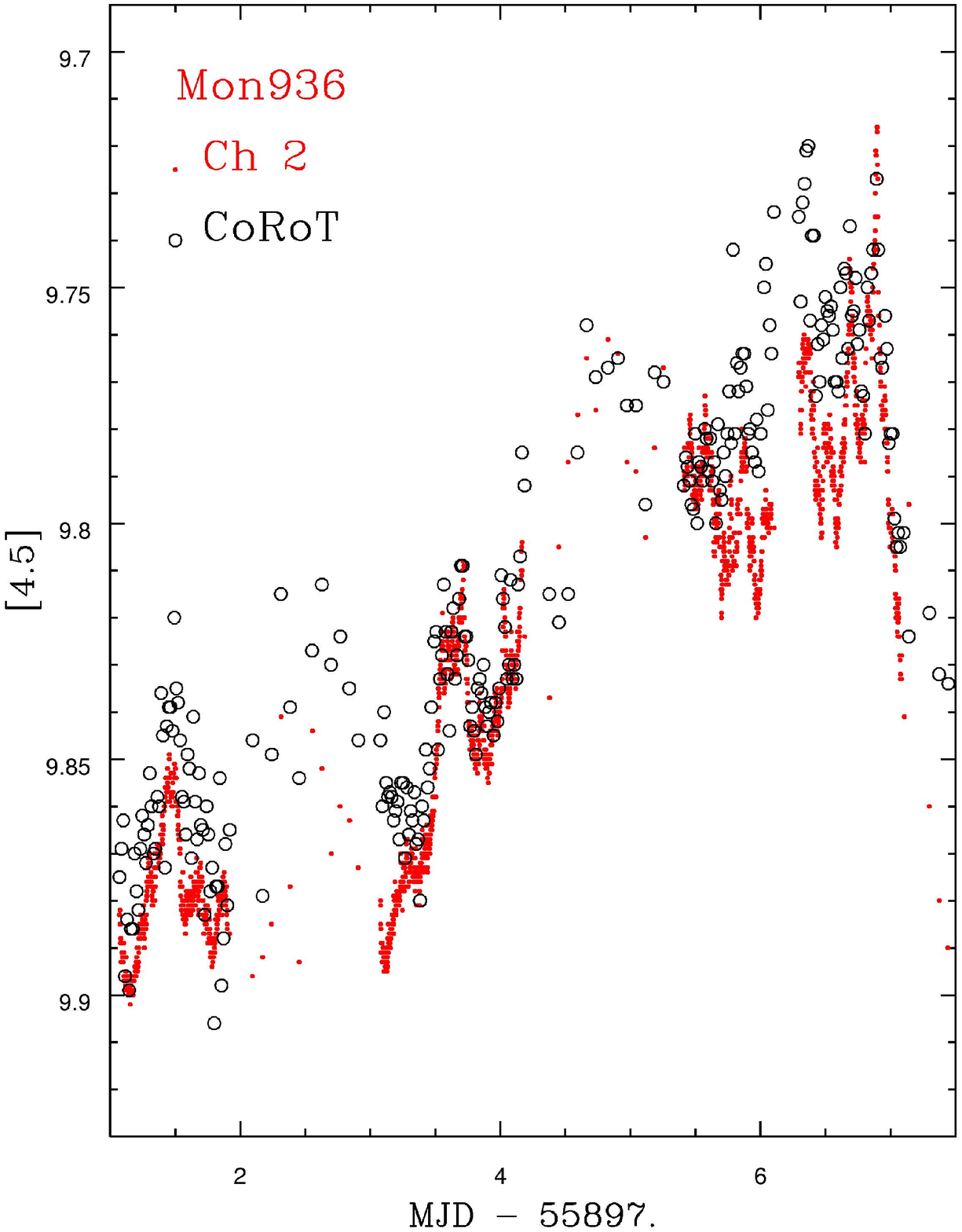}
\end{center}
\caption{{\em Spitzer} and {\em CoRoT} data for two stars from
Table~\ref{tab:basicinformation} (Mon-000567, left, and Mon-000936,
right) where we have IRAC staring-mode (high-cadence) data,
illustrating the often very good correlation between the optical and
IR light curve shapes. Ch1 and Ch2 refer respectively to IRAC's 3.6
and 4.5 $\mu$m channels.  For Mon 936, which is faint in the optical,
we have rebinned the {\em CoRoT} data to 20 minute sampling, centered
on the nearest IRAC data point.  \label{fig:spitzercorot}}
\end{figure*}

For a few of the stars in Table~\ref{tab:basicinformation}, we have
four days of  high-cadence IRAC data simultaneous with a portion of
the {\em CoRoT} light curves.  Figure~\ref{fig:spitzercorot} overplots
the IRAC and {\em CoRoT} light curves for two of these stars.  The
close correspondence between the two sets of data provides strong
confirmation of the reality of the flux bursts.

As we show in Section 4, based on their $ugri$ photometry, the stars
of Table~\ref{tab:basicinformation} all have UV excesses indicative of
active accretion.  More significantly, for the set of NGC~2264 YSOs
for which we have {\em CoRoT} light curves, more than half of the
stars with the largest UV excesses are included in
Table~\ref{tab:basicinformation}.

\subsection{Resolving Individual Flux Bursts}

\begin{figure*}
\begin{center}
\epsscale{1.0}
\plottwo{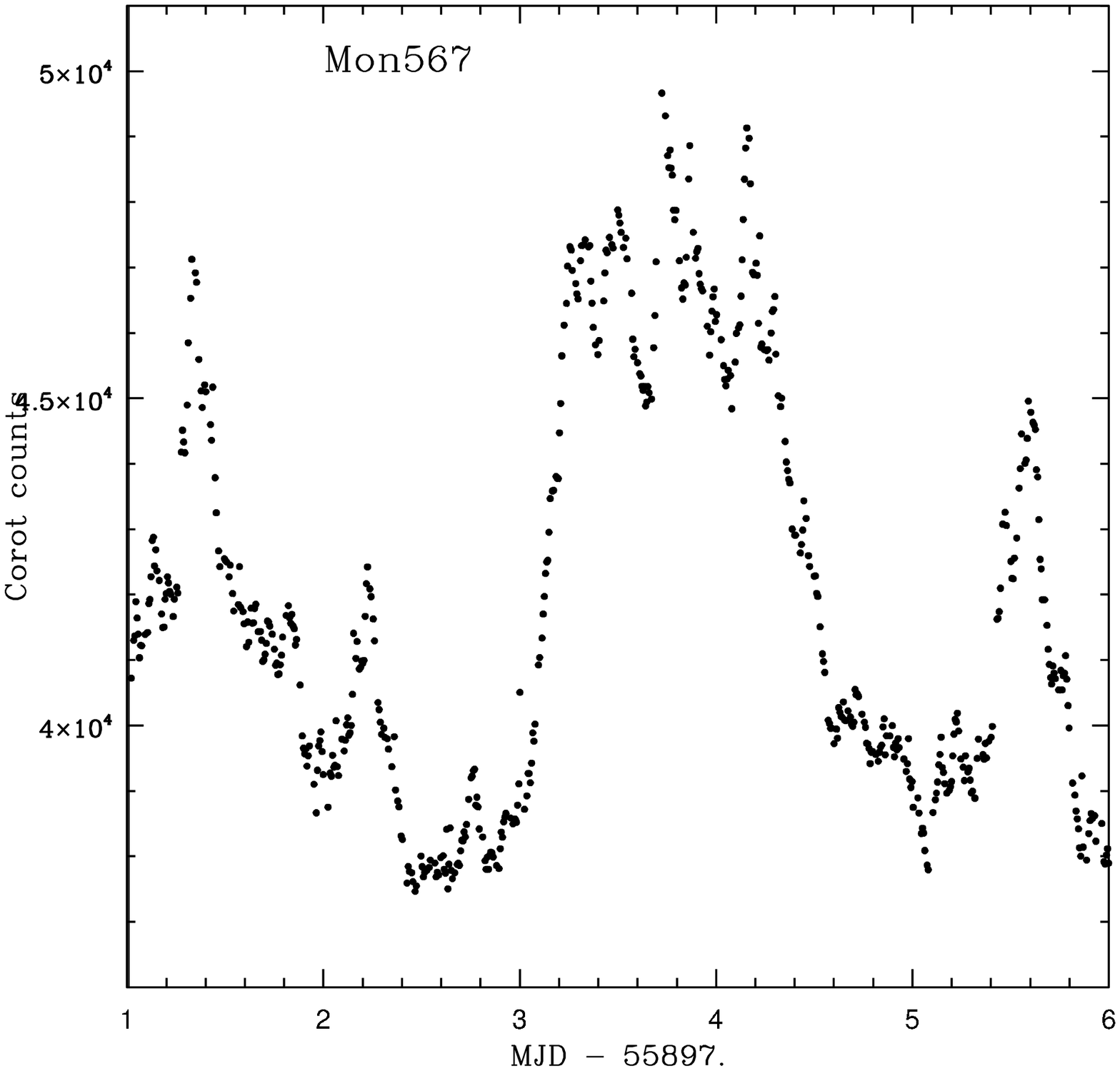}{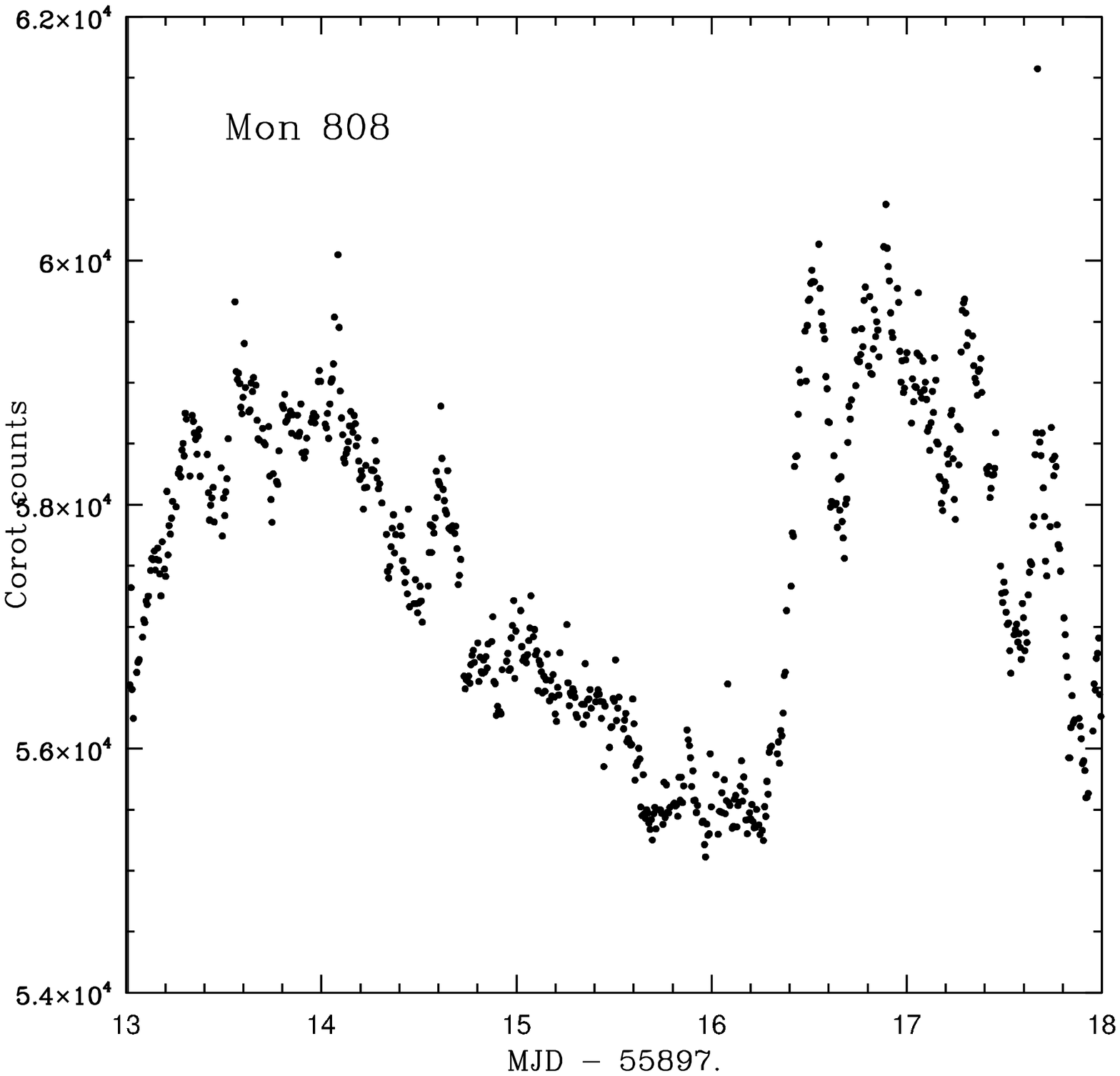}
\vspace{-1.0cm}
\end{center}
\caption{Expanded view of five day windows for light curves of two of
the accretion burst stars from Figure~\ref{fig:sixctts}, Mon-000567
(left), and Mon-000808 (right).   These plots illustrate that when
viewed at an expanded scale, the relatively broad (day or a few day)
flux excesses often break up into many much shorter duration events.
\label{fig:expandedlcs}}
\end{figure*}

The light curves shown in Figure~\ref{fig:sixctts}d-f are good at
illustrating the existence of the flux bursts. However, because of the
scale of those plots, they do not illustrate well the full structure
of those bursts.   Figure~\ref{fig:expandedlcs} shows expanded views
of portions of two of those light curves.  In each case, what appear
to be single bursts break up into many blended, shorter duration
bursts.   The shortest duration bursts visible in 
Figure~\ref{fig:expandedlcs} last a few hours.   Given the
standard {\em CoRoT} cadence of about 7.5 minutes and typical
signal-to-noise, such data could not resolve bursts of significantly
shorter duration.  See \S6 for a more quantitative discussion of the
burst timescales.  

We note for completeness the following. Many stellar  flares are
detected amongst the NGC~2264 {\em CoRoT} light curves, particularly
among the WTTS. These are usually distinguishable from the accretion
bursts because of their shorter timescales of generally $<$1 hour and
their characteristic very short rise time and relatively slow decay. 
Only about a half dozen likely flares are identifiable in the {\em
CoRoT} light curves of the stars in Table~\ref{tab:basicinformation}.

\subsection{Long-Term Stability of the Accretion Signature}

Ten of the stars in Table~\ref{tab:basicinformation} have {\em CoRoT}
light curves from both 2008 and 2011. Based on examination of the
light curves, for six of the stars (Mon-000007, 341, 945, 996, 1022,
and 1174)  we would classify them as members of the accretor class at
both epochs.   Mon-000185 appears to have a similar character at
both epochs, but none of the bursts in 2011 have an amplitude $\geq$\
5\%, and therefore we would not classify that light curve as
burst-dominated. Two of the stars (Mon-000011 and 510) were identified
as accretors based on their 2011 light curves, but have more
complicated light curves in 2008.    However, each of those stars have
a number of flux bursts in the 2008 data, thereby indicating that the
same mechanism dominating the 2011 light curves was also present in
2008. The only star with a much different light curve morphology at
the two epochs is Mon-001573, which had an approximately sinusoidal
light curve characteristic of a spotted star in 2008 and an accretion
burst dominated light curve in 2011.

\section{Characteristics of the Stochastic Accretor Class based on
Broad-Band Photometry and High Resolution Spectroscopy}

The stars in Table~\ref{tab:basicinformation} were identified as
having unusually active accretion solely on the basis of their light
curve morphology.   If that identification is correct, it seems likely
that their other observable characteristics such as IR and UV excesses
and Balmer emission line profiles should also be distinctive.   We now
show that is indeed the case.

For the remainder of this section, we will compare the YSOs in
Table~\ref{tab:basicinformation} to two other sets of YSOs in
NGC~2264.   The first of these are 81 WTTS, identified on the basis of
their having stable, sine-wave shaped, periodic light curves typical
of cold spots (Class I light curves of H94),  and small H$\alpha$
equivalent widths (we subsequently verified they also do not have IR
excesses). These WTTS are expected not to have actively accreting
circumstellar disks and hence little or no  UV excesses.  The second
group are 26 stars whose {\em CoRoT} light curves show periodic or
aperiodic flux dips which we believe are best ascribed to variable
extinction (Class III light curves of H94). The stars in
Figure~\ref{fig:sixctts}a-c are members of this variable extinction
group.   These stars must have disks if our interpretation of their
flux dips is valid.  The NGC 2264 stars that fall into these two 
light curve classes, particularly the variable extinction class, are
discussed further in Cody et al.\ (2014), and in Alencar et al.\
(2010).

\begin{figure*}
\begin{center}
\epsscale{1.0}
\plottwo{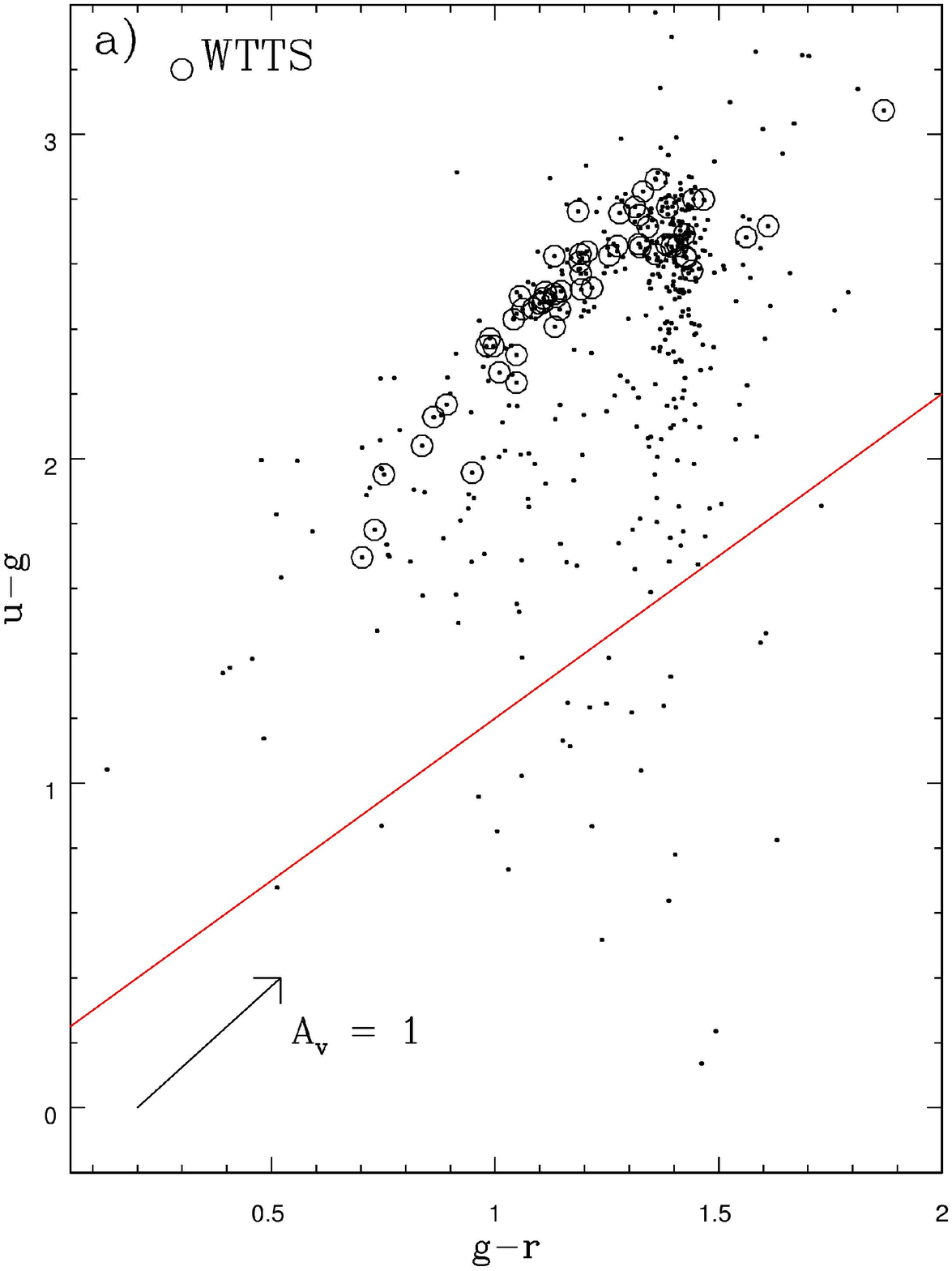}{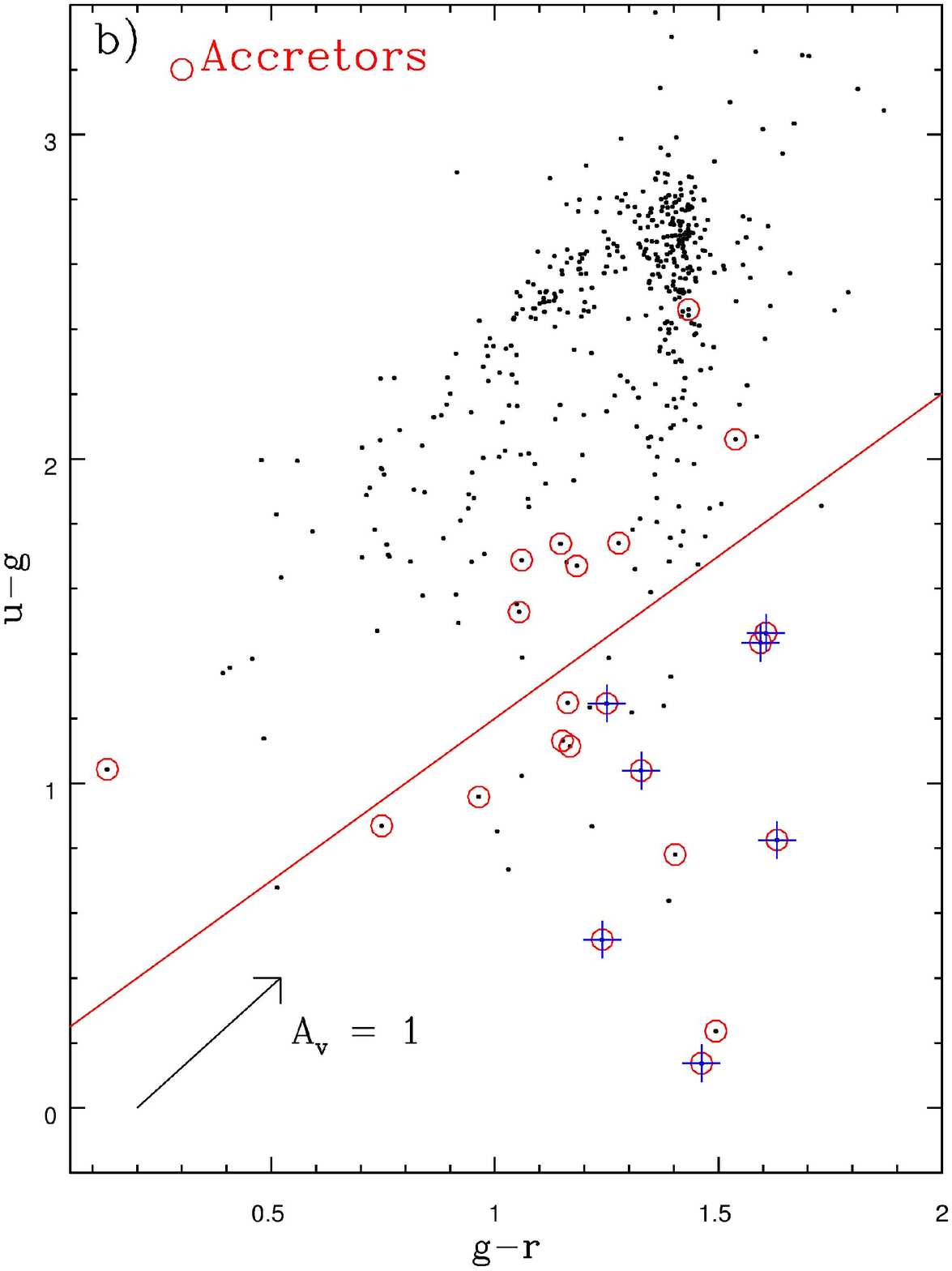}
\plottwo{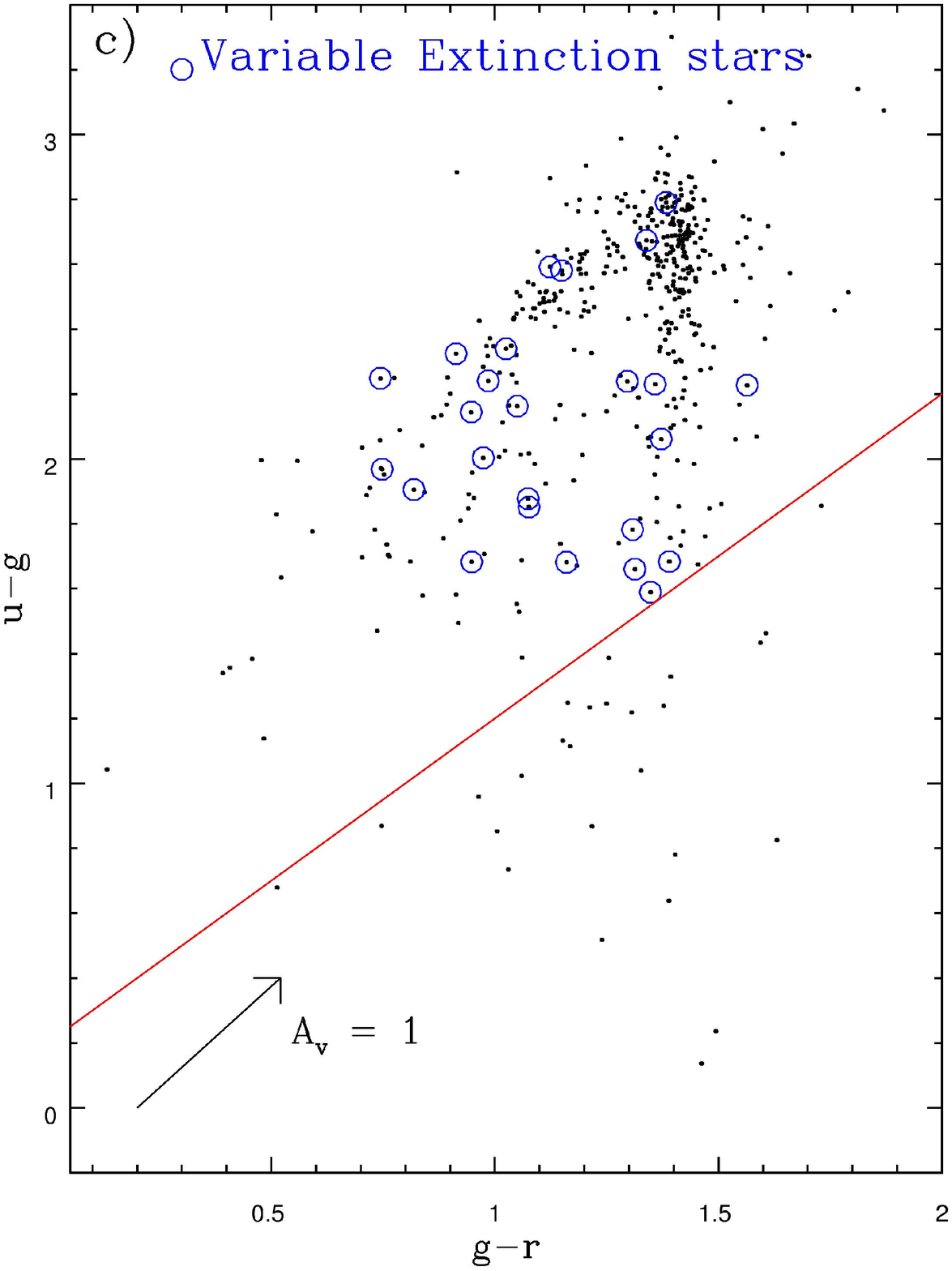}{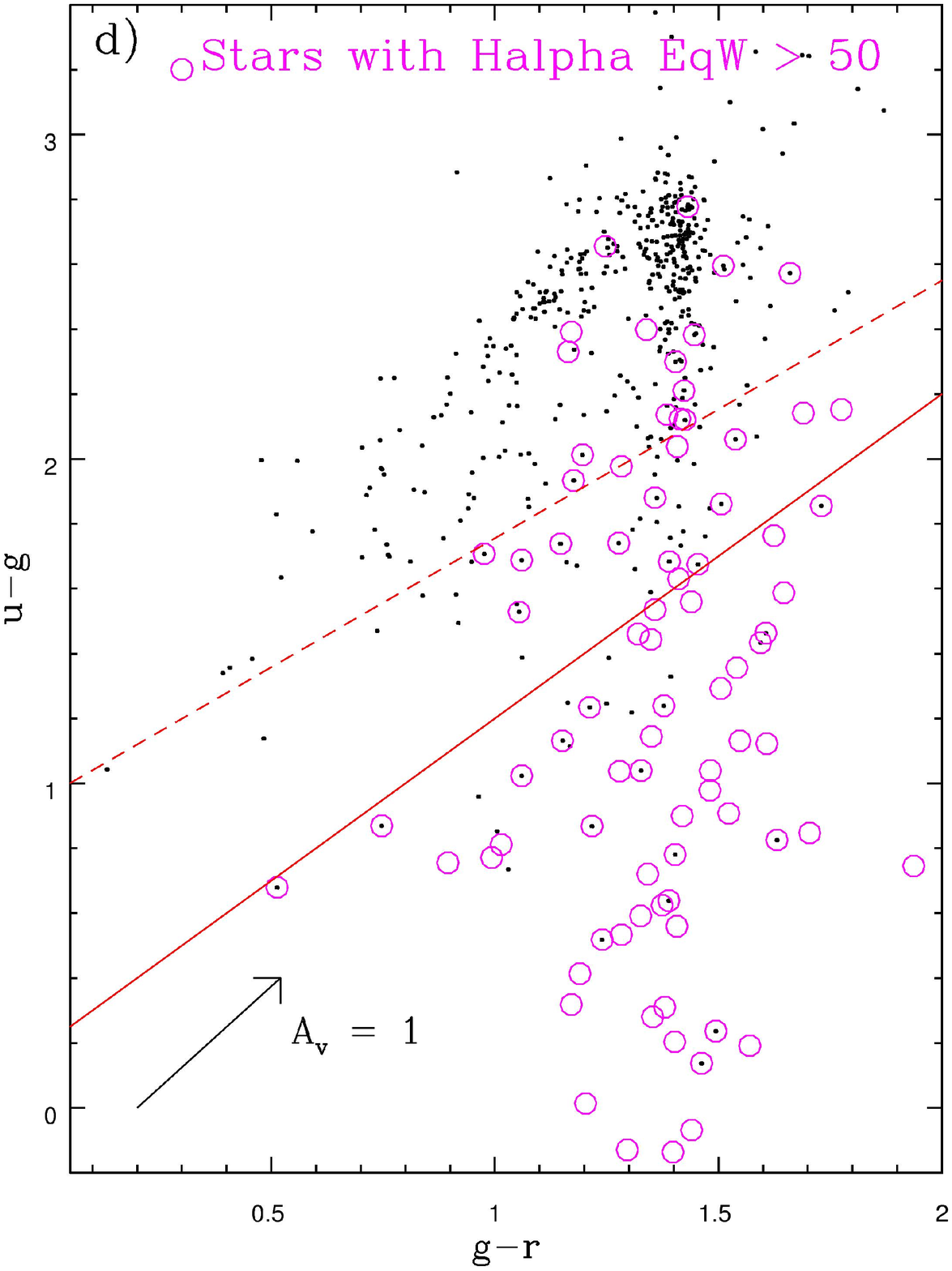}
\end{center}
\caption{Color-color diagrams derived from CFHT $ugri$ photometry of
NGC~2264 (Venuti et al.\ 2014). For panels a-c, only stars with {\em
CoRoT} light curves are plotted.  For panel d, the magenta circles correspond
to NGC~2264 members, regardless of whether there is a {\em CoRoT}
light curve or not. Small black dots correspond to all likely members
having CFHT photometry.  Panel (a) highlights stars that are probable
WTTS without warm circumstellar dust or active accretion; panel (b)
highlights stars with accretion burst dominated light curves
(accretors whose H$\alpha$\ profiles are of Reipurth type III-B
are further marked with a blue plus sign -- see Sec. 4.2); panel (c)
highlights the variable extinction stars -- stars with flux dips,
sometimes periodic, thought to be due to portions of the disk passing
through our line of sight, causing enhanced extinction; and panel (d)
highlights all NGC~2264 members with H$\alpha$ equivalent widths $>$
50 \AA.  The plots indicate that the accretion burst class members
have large UV excesses, while the variable extinction stars have a
range of UV colors, from nearly photospheric to mild UV excess.  The
red solid line provides a plausible boundary for stars with the
largest UV excesses -- stars below the red line should be the most
actively accreting members of NGC~2264.  The red dashed line is
discussed in \S7. \label{fig:colorcolorf4}}
\end{figure*}

The most certain property which the stars in
Table~\ref{tab:basicinformation} should have if they have unusually
active accretion is enhanced UV emission.   To determine if that is
the case, we use $ugri$ photometry obtained  at CFHT.  
Figure~\ref{fig:colorcolorf4} shows the $u-g$ vs.\ $g-r$ color-color
diagram for the NGC~2264 YSOs for which we have {\em CoRoT} light
curves. In panel (a), the WTTS define the locus of YSOs  without UV
excesses, with spectral types ranging from mid-G at $u-g \sim$ 1.7,
$g-r \sim$ 0.7, to early M at $u-g \sim$ 2.8 and $g-r \sim$ 1.4.  
Panel (b) shows the stars from Table~\ref{tab:basicinformation},
confirming that nearly all of them have UV excesses relative to the
WTT locus, with some of them having very blue $u-g$ colors.   Panel
(c) shows that the variable extinction stars have intermediate
properties -- the majority of them have UV excesses, but of
comparatively small degree, and a significant number of them have
essentially no UV excess.   Panel (d) indicates, not unexpectedly,
that stars with large  H$\alpha$\ equivalent widths also usually have
large UV excesses. Among the stars with large H$\alpha$ equivalent
widths and no {\em CoRoT}  data are likely many more stars with light
curves like those in Table~\ref{tab:basicinformation}.   Converting
the UV excesses to a mass accretion rate following the prescription of
Venuti et al.\ (2014), the mean accretion rate for the stars of
Table~\ref{tab:basicinformation} is $1.1\times 10^{-7}$ M$_{\sun}$
yr$^{-1}$ and that for the variable extinction  group of stars is
$2.5\times 10^{-8}$ M$_{\sun}$ yr$^{-1}$.

The red solid line in Figure~\ref{fig:colorcolorf4} serves as an
arbitrary but reasonable boundary below which are found the {\em
CoRoT} stars with the largest UV excesses in NGC~2264. Fourteen of the
twenty-six YSOs falling below that line are included in
Table~\ref{tab:basicinformation}. The light curves for the other
twelve stars are provided in Figure~\ref{fig:bigexcesses}.  One of
these stars (Mon-000273) is only slightly variable, and it is difficult
to characterize the features that are present given their low amplitude
relative to the noise fluctuations.  Another star (Mon-000423) has
several jumps or gaps in its light curve, also making it difficult
to separate real features from artifacts.  Of the
remaining ten, five have light curves that could be interpreted as
being dominated by short-duration flux bursts (Mon-000766, Mon-000893,
Mon-001061, Mon-000168, and Mon-001048); we did not include them in
Table~\ref{tab:basicinformation} because either their continuum level
is poorly defined or the noise level is too high. Depending on the
interpretation of these five stars,  stars with burst-dominated light
curves comprise at least 55\% and possibly up to $\sim$80\% of the
YSOs with large UV excesses and useful {\em CoRoT} light curves.  Only
one of the stars from Figure~\ref{fig:bigexcesses} (Mon-001132) has a
light curve shape which could correspond to a hot spot with lifetime
long compared to the rotation period, as predicted for the funnel-flow
accretion models of KR08.   That burst-dominated light curves would
dominate over stable hot spot light curves from funnel-flow accretion
at high accretion rates was one of the predictions of KR08; H94 had
also found this to be true in their sample of YSOs.

\subsection{Comparison of the Disk Properties of the NGC~2264 CTTS}

\begin{figure*}
\begin{center}
\epsscale{1.0}
\plottwo{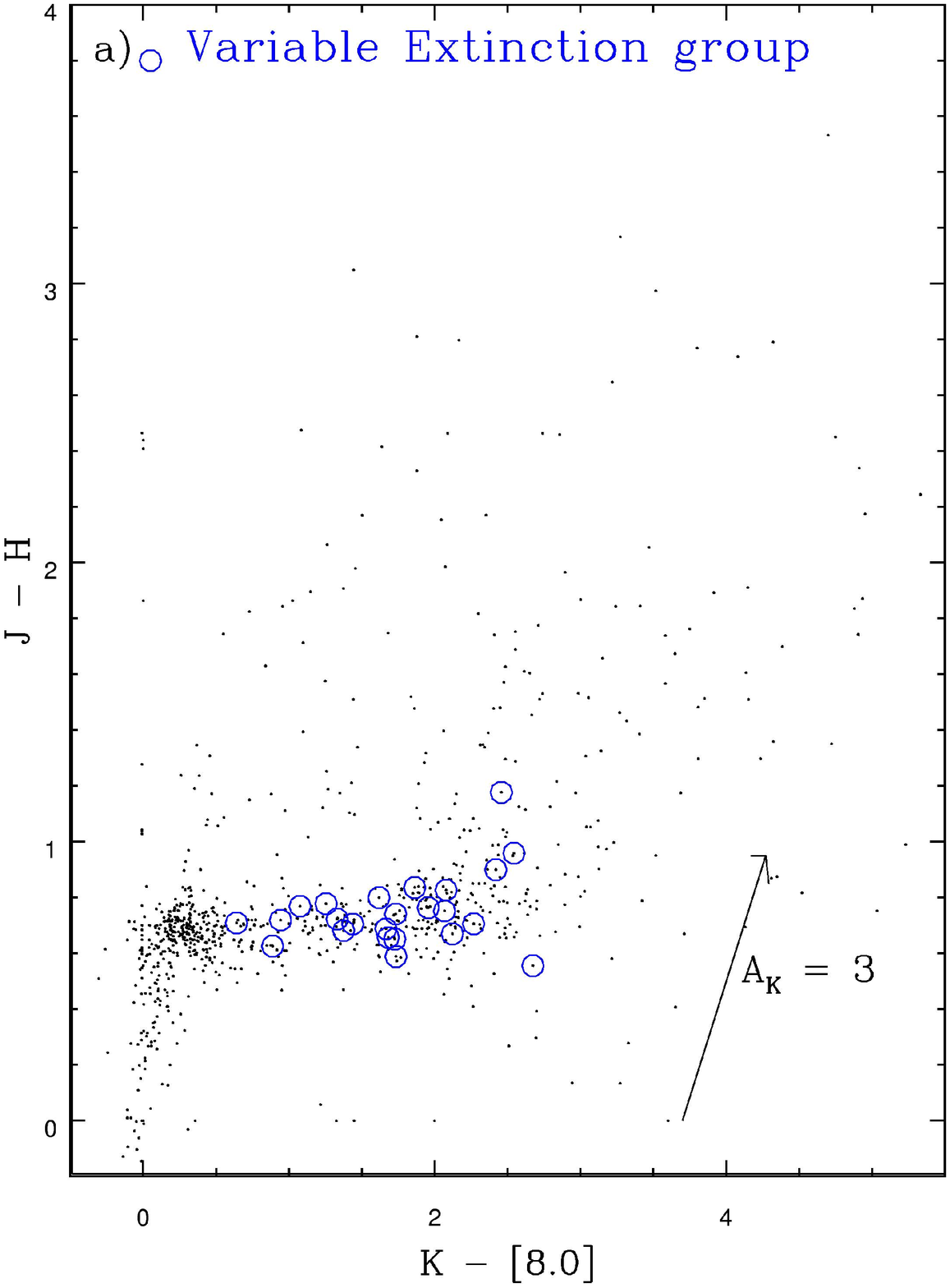}{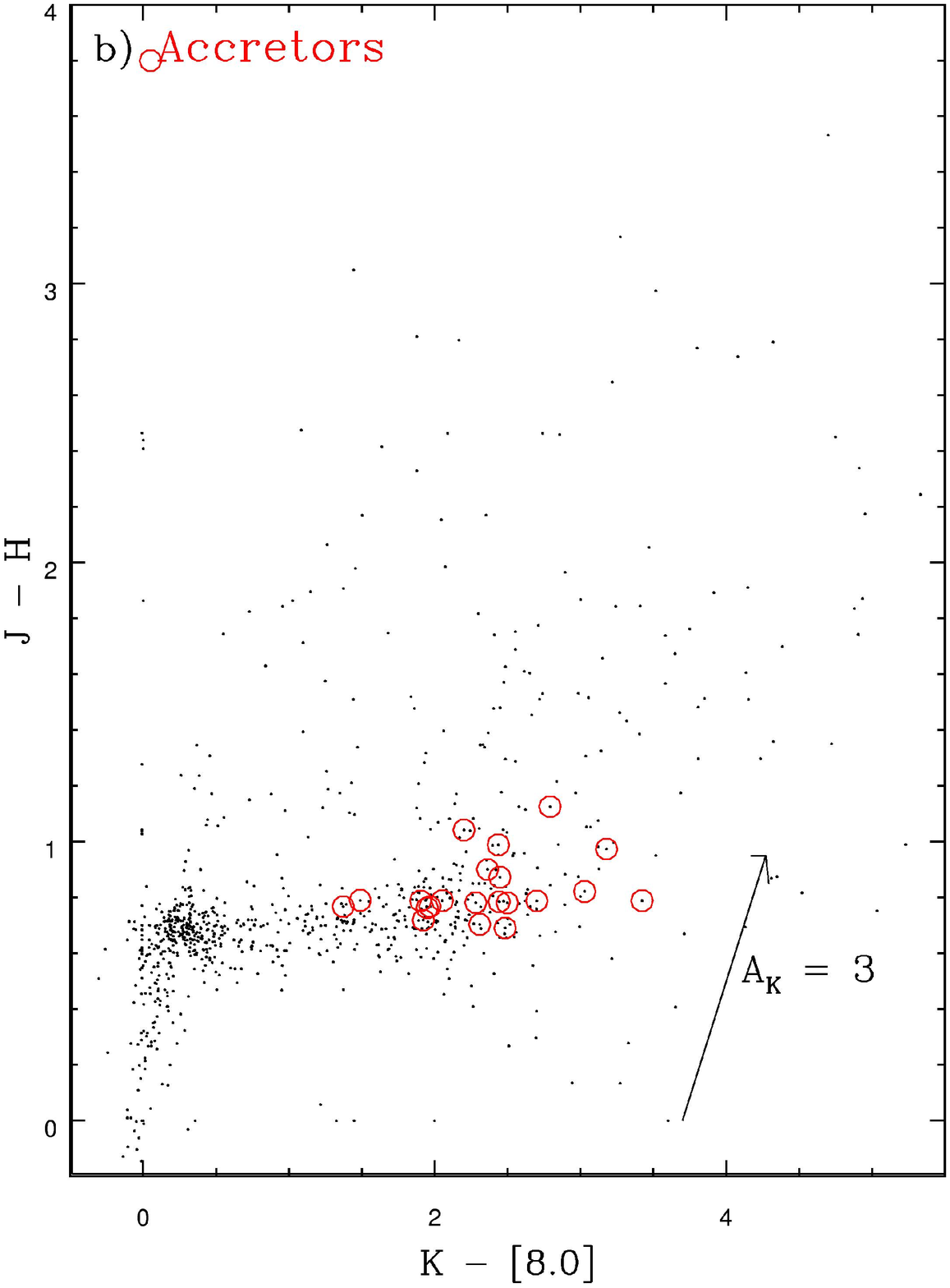}
\plottwo{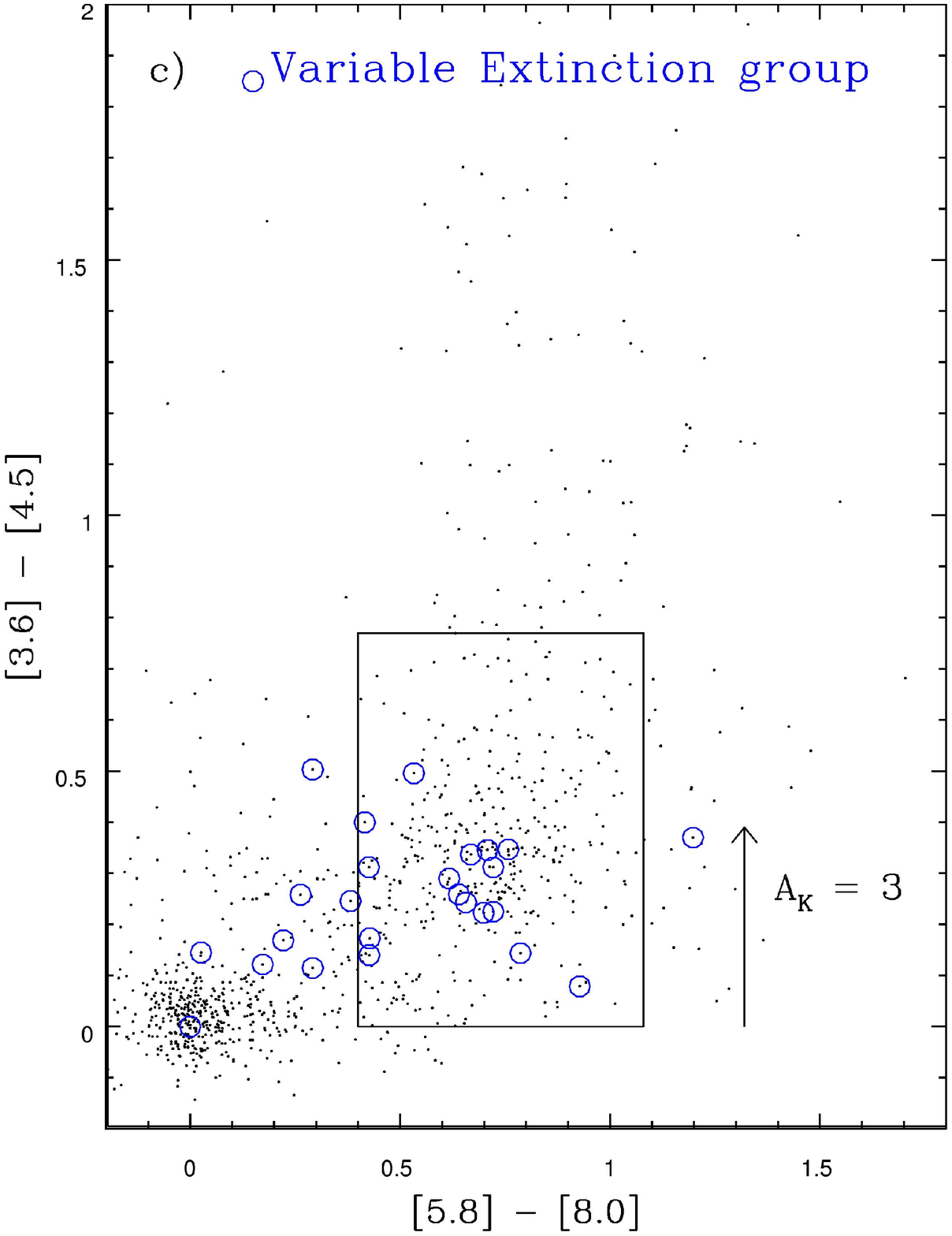}{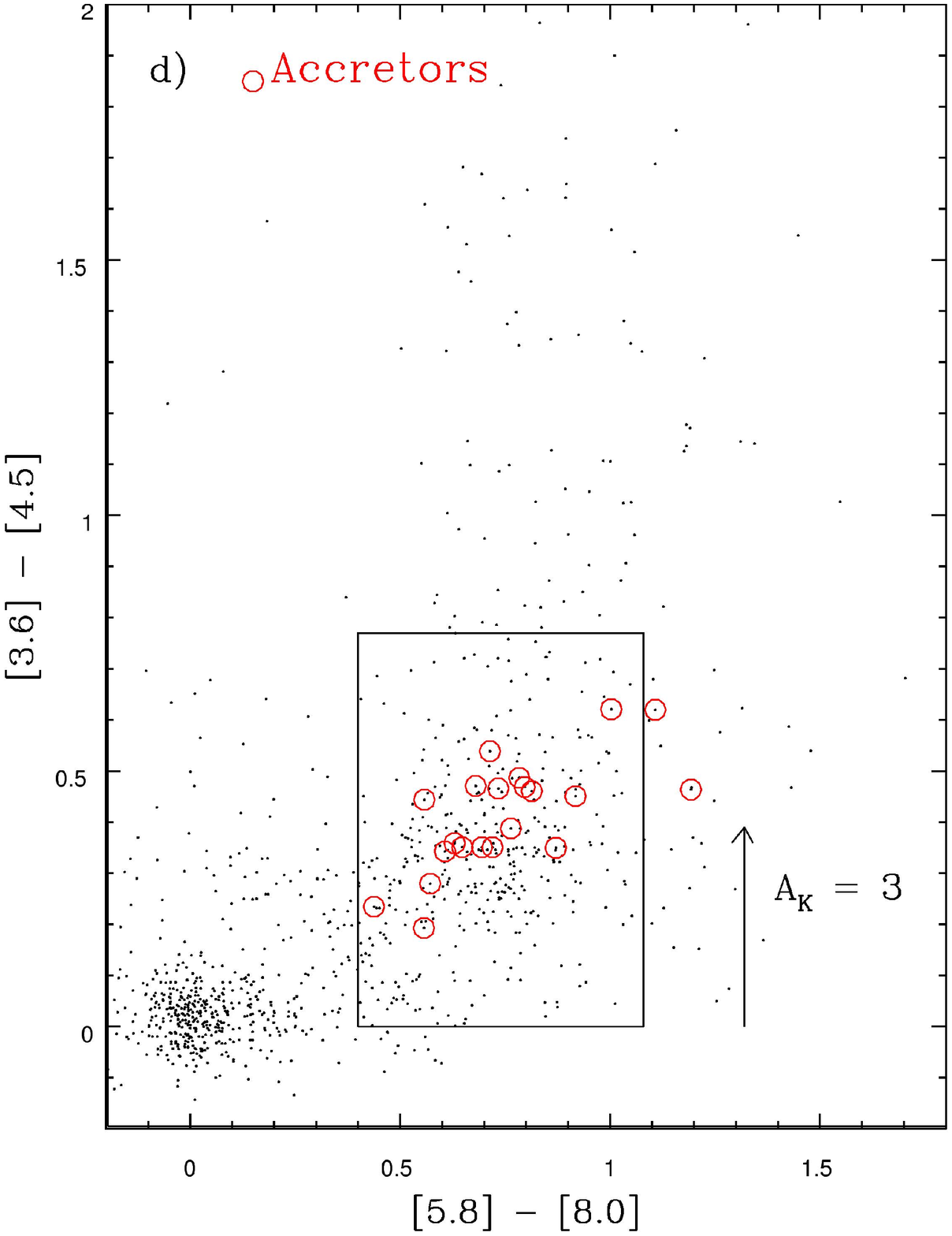}
\end{center}
\caption{Comparison of the distribution for the variable extinction 
stars (blue points) and 
members of the accretion burst class (red points) in two IR color-color
diagrams. Panels (a) and (b) are in $J-H$ vs.\ $K_s - [8.0]$; panels
(c) and (d) show the same set of stars in $[3.6] - [4.5]$ vs.\ $[5.8]
- [8.0]$ diagrams.  The box in panels c and d corresponds to the locus
for Class II YSOs as originally defined by Allen et al.\ (2004).  The
small black dots are all NGC 2264 members and candidate members, regardless
of whether we have {\em CoRoT} light curves or not.  The
diagrams show that the stars with accretion burst dominated light
curves have well-developed disks, whereas the variable extinction
stars often have weaker IR excesses, with nearly a third falling
blueward in [5.8]-[8.0] color than the Class II box of Allen et al.\
(2004). \label{fig:colorcolorf5}}
\end{figure*}

If the stars of Table~\ref{tab:basicinformation} have
accretion-dominated light curves, it must also be true that they
should have IR SEDs that indicate the presence of warm dust in their
inner circumstellar disks.  Figure~\ref{fig:colorcolorf5} examines
this point using two different IR color-color diagrams; the top two
panels show the variable extinction and accretion burst stars in a
$J-H$ vs.\ $K_s - [8.0]$ diagram, while the bottom two panels show the
same set of stars in a $[3.6] - [4.5]$ vs.\ $[5.8] - [8.0]$ diagram. 
Teixeira et al.\ (2012) have constructed theoretical models to predict
the IR colors of YSOs in the $J-H$ vs.\ $K_s - [8.0]$ diagram, showing
that the disk population is expected to form a narrow band of stars
stretching to the right (as observed).  They find that ``anemic" disks
are restricted to $K_s - [8] \sim$ 1.3 in the diagram, and Class I
sources to $K_s - [8] >$ 3. Bare photospheres have $K_s - [8] <$ 0.5;
embedded YSOs with large reddening or a disk viewed at large
inclination angle fan out to the upper right in this diagram.  The
IRAC color-color diagram was introduced in Allen et al.\ (2004) and
has been used  extensively to separate YSOs by disk class.   The box
shown in the lower two panels is from Allen et al., and is designed to
isolate Class II YSOs.  Bare photosheres are located near 0,0 in the
diagram, and Class I sources lie redward of the Class II box.  In both
diagrams, nearly all of the stars of Table~\ref{tab:basicinformation}
have colors corresponding to Class II YSOs.   The variable extinction
stars have IR colors that are significantly less red on average than
the Table~\ref{tab:basicinformation} stars, with nearly half of them
falling blueward of the Class II locus.   Several of the variable
extinction class members have disk colors consistent with
classification as transition or pre-transition disks (Espaillat et
al.\ 2007). Espaillat et al.\ (2012) showed that the accretion rates
of transition and pre-transition disks are lower than for normal Class
II YSOs, in accord with our finding that the mean accretion rate of
the variable extinction group as estimated from their UV excesses is
relatively low.

\subsection{Comparison of the H$\alpha$ Emission-line Profiles of the NGC~2264 CTTS}

One of the defining characteristics of CTTS is that they show strong
emission lines in their optical spectra.  At high spectral resolution,
the shapes of these emission lines are often quite complex, with
blue-shifted and/or red-shifted absorption components and asymmetries
(Reipurth et al.\ 1996 = R96; Alencar and Basri 2000 = AB00).  It is
expected that these profile shapes encode information on the
kinematics of the gas (infall of accreting gas; outflow of gas from
stellar and disk winds) and geometry of the system and our view angle
to it.  We have H$\alpha$ profiles for a  large fraction of the stars
in Table~\ref{tab:basicinformation}, though for most stars the spectra
are from 2004/2005.  We use these spectra here to help further compare
the stars of Table~\ref{tab:basicinformation} with other CTTS and
thereby to illuminate their physical nature.  We also discuss whether
the non-simultaneity of the  spectra and photometry significantly
reduces the utility of the MMT H$\alpha$\ data.  Throughout the
discussion, we primarily reference the theoretical models of CTTS
magnetospheres and disk winds of Kurosawa et al.\ (2006 = KHS06), but
we have also consulted other models (Lima et al.\ 2010; Kurosawa et
al.\ 2013).

\begin{figure*}
\begin{center}
\includegraphics[scale=0.8]{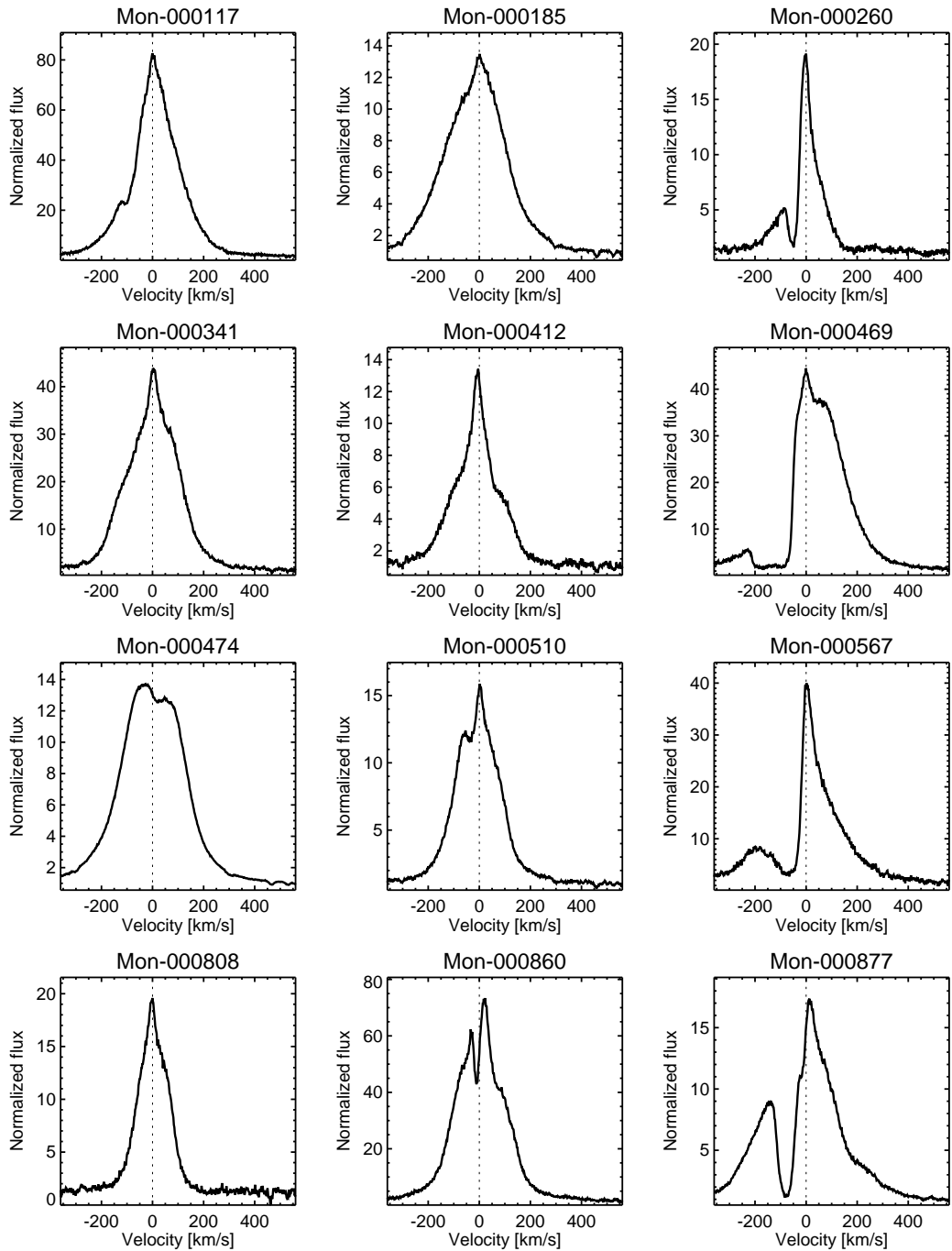}
\end{center}
\caption{H$\alpha$\ profiles of the stars whose {\em CoRoT} light
curves are categorized as being dominated by accretion bursts, for
those stars for which we have available high resolution spectra.  The
emission profiles have been shifted to the mean cluster velocity,
marked by the dashed line at $v = 0$. \label{fig:halphaprofiles}}
\end{figure*}

Figure~\ref{fig:halphaprofiles} shows the H$\alpha$ profiles for
twenty of the stars in Table~\ref{tab:basicinformation}. In all
but two cases, Mon-000945 and 996, the spectra are from the MMT in
2004/2005 (Furesz et al.\ 2006).  The spectral resolution is about
34,000 and the flux scale has been normalized so that the continuum
level near H$\alpha$ has been set to 1.0.  The MMT spectra have not
been sky subtracted, which can lead to spurious narrow emission
features at rest velocity from nebular emission.  This is only a
significant issue for Mon-000260, where the central peak in
Figure~\ref{fig:halphaprofiles} may appear much stronger than it
really is (see also Figure~\ref{fig:halpha2004-2012}).  
H$\alpha$\ equivalent widths from the literature for these
stars are reported in Table~\ref{tab:basicinformation}.

\addtocounter{figure}{-1}
\begin{figure*}
\begin{center}
\includegraphics[scale=0.8]{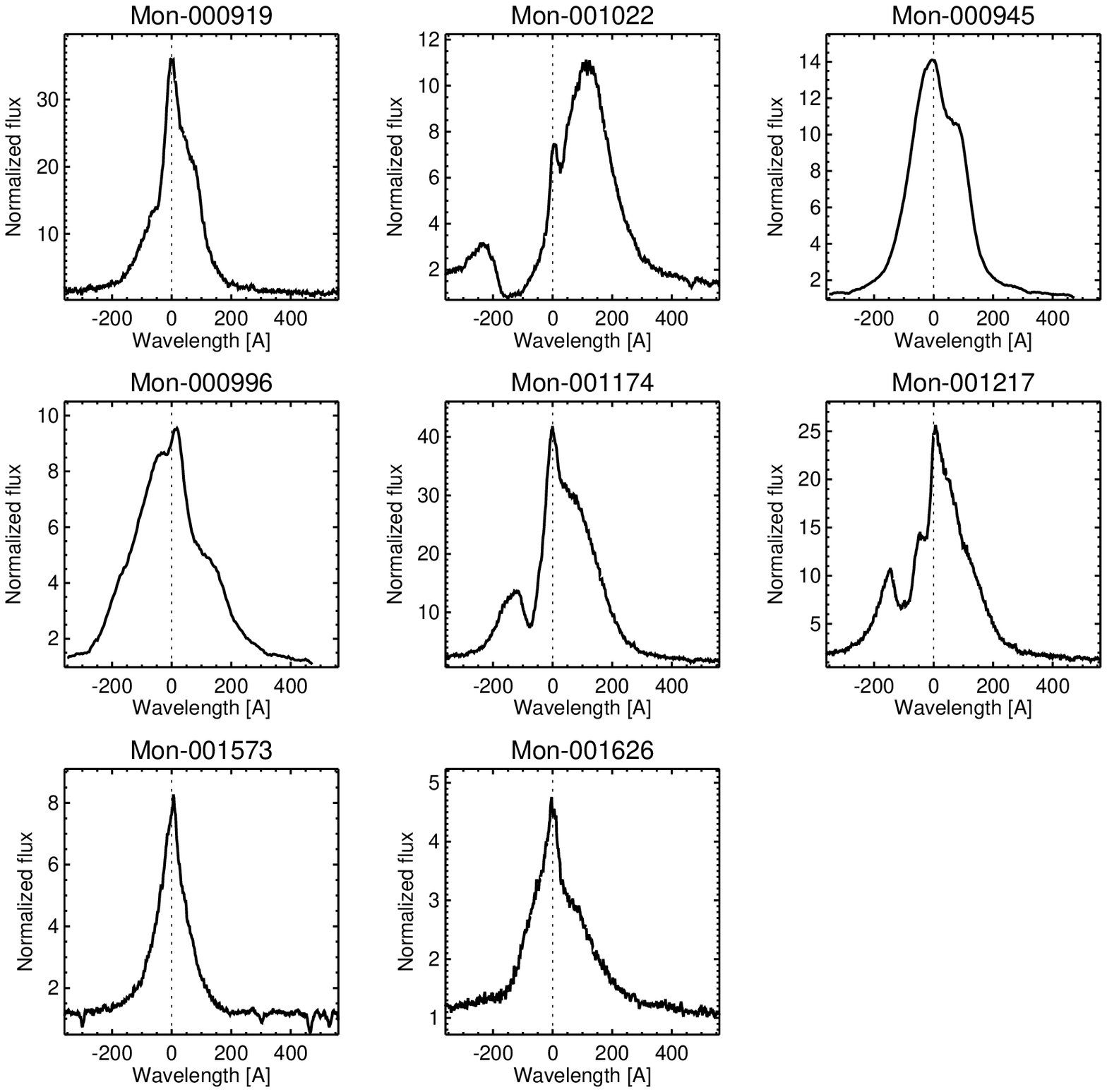}
\end{center}
\caption{H$\alpha$\ profiles for stars with accretion burst dominated
light curves, continued.}
\end{figure*}

We first address the issue of the non-simultaneity of the MMT spectra
and the {\em CoRoT} light curves.  In Figure~\ref{fig:halpha2004-2012} of
the Appendix, for the five stars where we have both MMT and VLT FLAMES
spectra, we compare the 2004/2005 H$\alpha$\ profiles to the ensemble
of profiles from the VLT (the dashed lines are individual nightly spectra;
the solid line is the average of all VLT epochs). For four of the five stars, the VLT
H$\alpha$\ profiles are relatively stable during the 2011 campaign,
and also are consistent with the shape of the profile as observed in
2004/2005.  Only for Mon-000260 is the MMT profile outside the range
of the profile shapes shown in 2011, but even then the difference is
relatively small; the ratio of the height of the blue peak to that of
the red peak is  significantly less than 0.5 for the MMT profile but
significantly more than 0.5 for the VLT profiles.  This may reflect a
real change, or as noted previously, it  could be due to nebular emission
affecting the MMT red peak.  For a qualitative discussion, we
therefore believe that use of the 2004/2005 MMT data should be
adequate for our purposes.

The H$\alpha$ profiles of Figure~\ref{fig:halphaprofiles} are typical
of CTTS, but only sample a relatively small fraction of the total
range of profile types shown by CTTS (eg. R96 or AB00).  Most of the
profiles can best be described as lumpy, broad and centrally peaked,
corresponding to Type I profiles of the R96 scheme.  Nearly all of the
other stars have  blue-shifted absorption features, with profile types
of either II-B (B/R peak ratio between 0.5 and 1) or III-B (B/R peak
ratio less than 0.5).   To illustrate the relatively small range of
profile types amongst the Table~\ref{tab:basicinformation} stars,
Figure~\ref{fig:halphaextinction} provides the H$\alpha$ profiles of
the  stars with transient extinction dips in their light curves   as
previously discussed in relation to Figures \ref{fig:colorcolorf4} and
\ref{fig:colorcolorf5}.  Many of these profiles are much more complex
than the Table~\ref{tab:basicinformation} stars, with a third having
red-shifted absorption features (profile type II-R and III-R)  and
none having the deep, highly blue-shifted absorption features normally
ascribed to disk or stellar winds (as seen particularly for
Mon-000469, 567, and 1022 for the accretors).

KHS06 explicitly compared their model profiles to the R96 H$\alpha$\ 
classification scheme.  They concluded that the type III-B profiles
must have quite high accretion rates ($\dot{M} \sim 10^{-7}$
M$_{\sun}$ yr$^{-1}$) and are most likely for systems seen at
relatively low inclination angle.   The stars with type III-B profiles
are marked with blue plus signs in Figure~\ref{fig:colorcolorf4}b, confirming that they
indeed all do have quite large UV excesses and hence presumably quite
large accretion rates.  The type I profiles require lower, but still
significant, accretion rates with little constraint on the inclination
angle.  The type III-R profiles found among some of the stars with
variable extinction light curves generally requires lower accretion
rates ($\dot{M} \sim 10^{-8}$ M$_{\sun}$ yr$^{-1}$ or less) and very
high inclination angles.   The high inclination angle requirement is
in good agreement with the physical model for producing the flux dips
in the AA Tau light curves, and the weak UV and IR excesses for these
stars also agrees with the prediction of a lower accretion rate
expected to match their H$\alpha$ profiles.

The KHS06 model profiles were calculated for stars with stable,
funnel-flow accretion, and so conclusions derived from those models do
not necessarily apply to our stars which appear to have more
stochastic accretion.  However, model Balmer-series emission profiles
for stochastic accretors have  recently been published by Kurosawa et
al.\ (2013).   Based on those models, the H$\alpha$\ profiles for
funnel-flow and instability driven accretion appear qualitatively the
same, with the primary difference being that the red-shifted
absorption dips, if present, should be periodic in the former case 
but aperiodic in the latter case.   Because we generally only have a
single epoch H$\alpha$\ profile, we cannot test this prediction for
the stars of Table~\ref{tab:basicinformation}.

\subsection{Veiling and Veiling Variability}

In addition to the H$\alpha$\ profiles, the high resolution
spectroscopy we have provides us with an additional means to
characterize the stars of Table~\ref{tab:basicinformation} and compare
them to other NGC~2264 CTTS.  In particular, we can measure
photospheric veiling and veiling variability (Bertout 1984; Hartmann
and Kenyon 1990).  The strong UV excesses in CTTS, which we have shown
to be especially true for the stars of
Table~\ref{tab:basicinformation}, is usually attributed to photons (a
combination of black body and  free-free/bound-free emission) from the
hot gas created when the accreting gas strikes the stellar surface. 
While the contrast of this emission relative to the photospheres of
CTTS is greatest in the UV, it is still present in the VLT/FLAMES
spectral bandpass.  The effect is both to add a slowly varying blue
continuum to the light from the photosphere, but also in some cases to
add emission cores to some of the absorption lines; both effects
result in weaker mean absorption line equivalent widths in CTTS
compared to stars with purely photospheric spectra (Petrov et al.\
2011).   This is normally quantified with a parameter called the
veiling factor, \begin{equation}
r = (EqW ({\rm photosphere})/EqW ({\rm CTTS})) - 1
\end{equation}
In order to provide a mean veiling index for the six stars from
Table~\ref{tab:basicinformation} for which we have VLT spectra, we
have coadded all available spectra from the 2011-2012 campaign,
corresponding to usually 15 to 22 epochs.  We have then measured the
equivalent widths for the seven strongest and least blended absorption
lines in each of those spectra (at 6448.9, 6461.6, 6470.7, 6624.7,
6707.8 and 6717.7 \AA - the latter two being LiI and CaI), 
and done the same for a set of WTTS of
the same spectral type for which we also have VLT spectra.  The
derived veiling factors at $\lambda\sim$6500 $\AA$\ for the
accretion burst stars are provided in
Table~\ref{tab:basicinformation}; based on the scatter in the results
for individual absorption features, the typical RMS uncertainty in
these values is about 0.1.  While significant, the derived veiling
factors are not huge -- a characteristic value of 0.65 means that the
veiling continuum at 6600 \AA\ is 65\% of the photospheric continuum,
or equivalently that 39\% of the total continuum at 6600 \AA\ is
non-photospheric.

\begin{figure}
\begin{center}
\epsfxsize=.99\columnwidth
\epsfbox{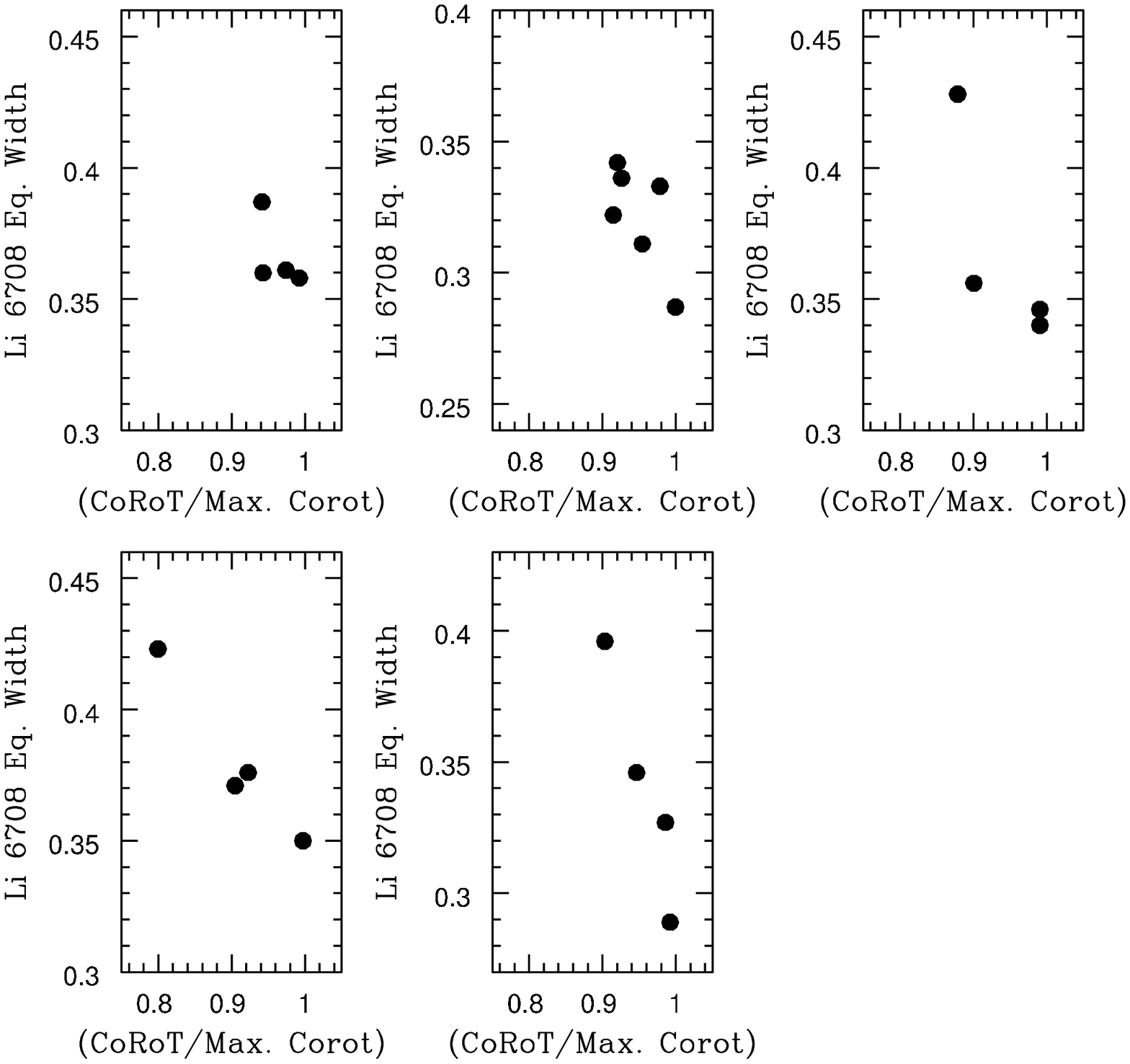}
\end{center}
\vskip-0.8truein
\caption{Correlation between the level of veiling (using Li
$\lambda$6708 equivalent as a proxy) and the observed broad-band
continuum level from the {\em CoRoT} light curves for the five stars with
accretion-burst dominated light curves for which we have VLT/FLAMES
spectra. From top-left to bottom-right, the stars plotted are
Mon-000341, Mon-000510, Mon-000945, Mon-000996 and Mon-001022.  Lithium
equivalent widths in Angstroms; the abscissa points are the CoRoT count
rate at the epoch of the VLT spectrum divided by the maximum CoRoT count
rate during the time window encompassed by the VLT observations.  Based
on similar spectra for three WTTS of similar brightness, the uncertainties
in the individual lithium equivalents are of order 0.011 \AA.
\label{fig:veilingaccretors}}
\end{figure}

\begin{figure}
\begin{center}
\epsfxsize=.99\columnwidth
\epsfbox{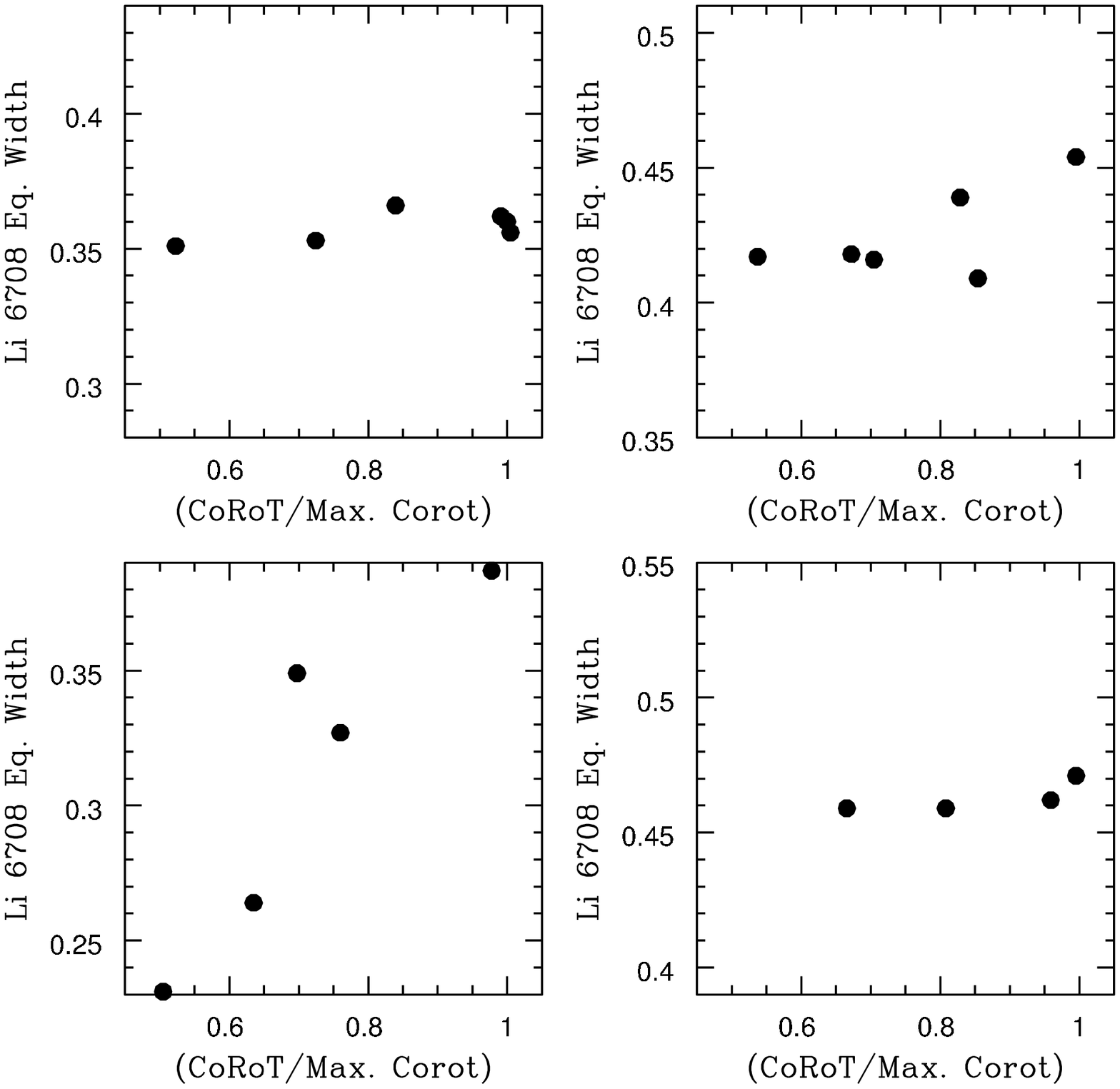}
\end{center}
\vskip-0.8truein
\caption{As for Figure~\ref{fig:veilingaccretors}, except plotting
data for the four stars from our variable extinction group where we
have sufficient data.
From top-left to bottom-right, the stars plotted are
Mon-000297, Mon-000660, Mon-001144 and
Mon-001199.\label{fig:veilingext}}
\end{figure}

We can use the individual spectra obtained during December 2011 when
{\em CoRoT} was observing the cluster to look for a correlation
between the spectral veiling and the broad-band optical flux.  There
are usually either four or six epochs where we have both VLT spectra
and {\em CoRoT} photometry.   Because the signal-to-noise ratio of the
individual spectra is low, only the \ion{Li}{1} 6708 \AA\ line can be
measured accurately, and we use its equivalent width as a proxy for
veiling (though see Stout-Batalha et al. 2000 for a cautionary note
re: Li $\lambda$6708 as a veiling proxy). 
Figure~\ref{fig:veilingaccretors} shows the correlation between our
veiling proxy and the {\em CoRoT} broad-band photometry for the five
stars from Table~\ref{tab:basicinformation} with such data;
Figure~\ref{fig:veilingext} shows a corresponding plot for the
variable extinction stars where we have sufficient data.   The two
sets of stars clearly behave differently -- the accretion burst group
show larger equivalent widths as the star becomes fainter, whereas the
variable extinction set shows either no correlation or larger
equivalent widths when brighter.  Having larger equivalent widths when
fainter is as expected if the driver of the variations is a changing
amount of veiling continuum.  To first order, variable extinction
should have no significant effect on equivalent widths, so the general
trends shown in Figure~\ref{fig:veilingext} are also as expected.  
The increase in lithium equivalent width as the star brightens is
perhaps unexpected, however a similar dependence has been seen
previously in AA Tau (Bouvier et al.\ 1999).   For AA Tau, the
suggested explanation was that the geometry of our line of sight and
the warped disk should lead to the enhanced extinction largely
affecting the low latitude portion of the star's photosphere.   Hot
spots at high latitudes dilute the lithium equivalent widths, and
therefore when the star is most extincted (the light is most dominated
by light from high latitudes), the lithium equivalent width should be
smallest.

\section{Age and View Angle Inferences for Members of the Accretion Burst Class}

NGC~2264 is normally thought of as slightly older than, for example,
the Orion Nebula Cluster (ONC) or Taurus.  A commonly quoted average
age for the region is about 3 Myr (e.g., Park et al.\ 2000).
However, it is quite clear that there are portions of the star-forming
region that are much younger and there is on-going star formation
along the molecular ridge extending northward from the Cone Nebula. 
Some of the YSOs in the youngest parts of NGC~2264 are likely to be
$\sim$1 Myr old or younger (Young et al.\ 2006; Teixeira et al.\
2012).   An age range from 0.1 Myr to 5 Myr was derived by Dahm \&
Simon (2005), with similar estimates by Rebull et al.\ (2002) and
Soderblom et al.\ (1999).   Given that the stars of
Table~\ref{tab:basicinformation} are heavily weighted towards high
current accretion rates, it seems plausible that they are on average
also comparatively young.  In this section, we examine to what extent
that speculation can be validated by data.

\subsection{Age Inferred from Color-Magnitude and HR Diagrams}

\begin{figure}
\begin{center}
\epsfxsize=.99\columnwidth
\epsfbox{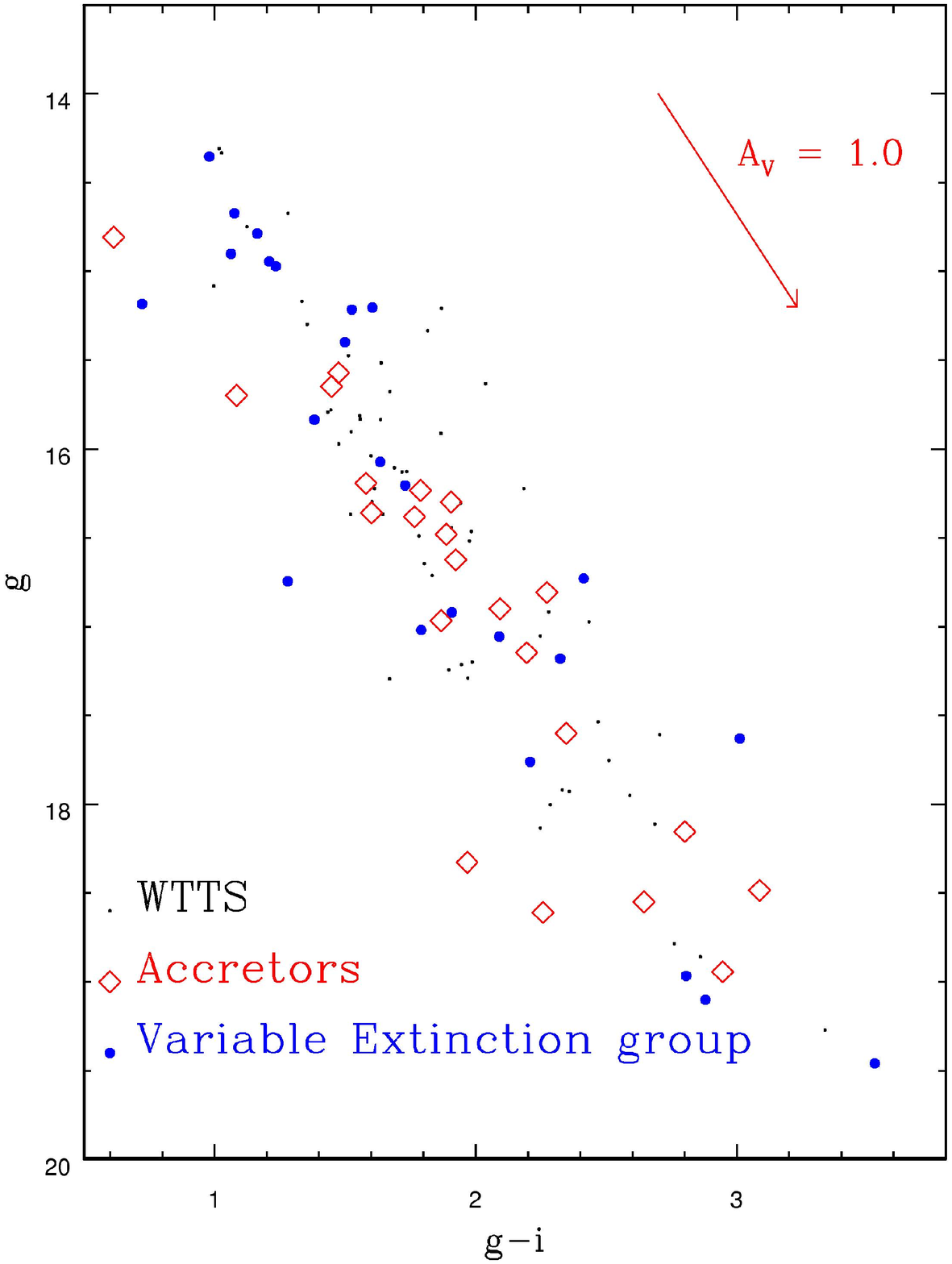}
\end{center}
\caption{$g$ vs. $g-i$ color-magnitude diagram, for the same set of
WTTS (small black dots), accretion burst stars (red diamonds), and
variable extinction group stars(blue dots) as
plotted in Figure~\ref{fig:colorcolorf4}. 
We see no systematic displacement between the three
groups shown, e.g., all three seem to be the same approximate age, or
at least that any systematic age difference is too small for us to
determine from this figure. \label{fig:colormagf9}}
\end{figure}

\begin{figure}
\begin{center}
\epsfxsize=.99\columnwidth
\epsfbox{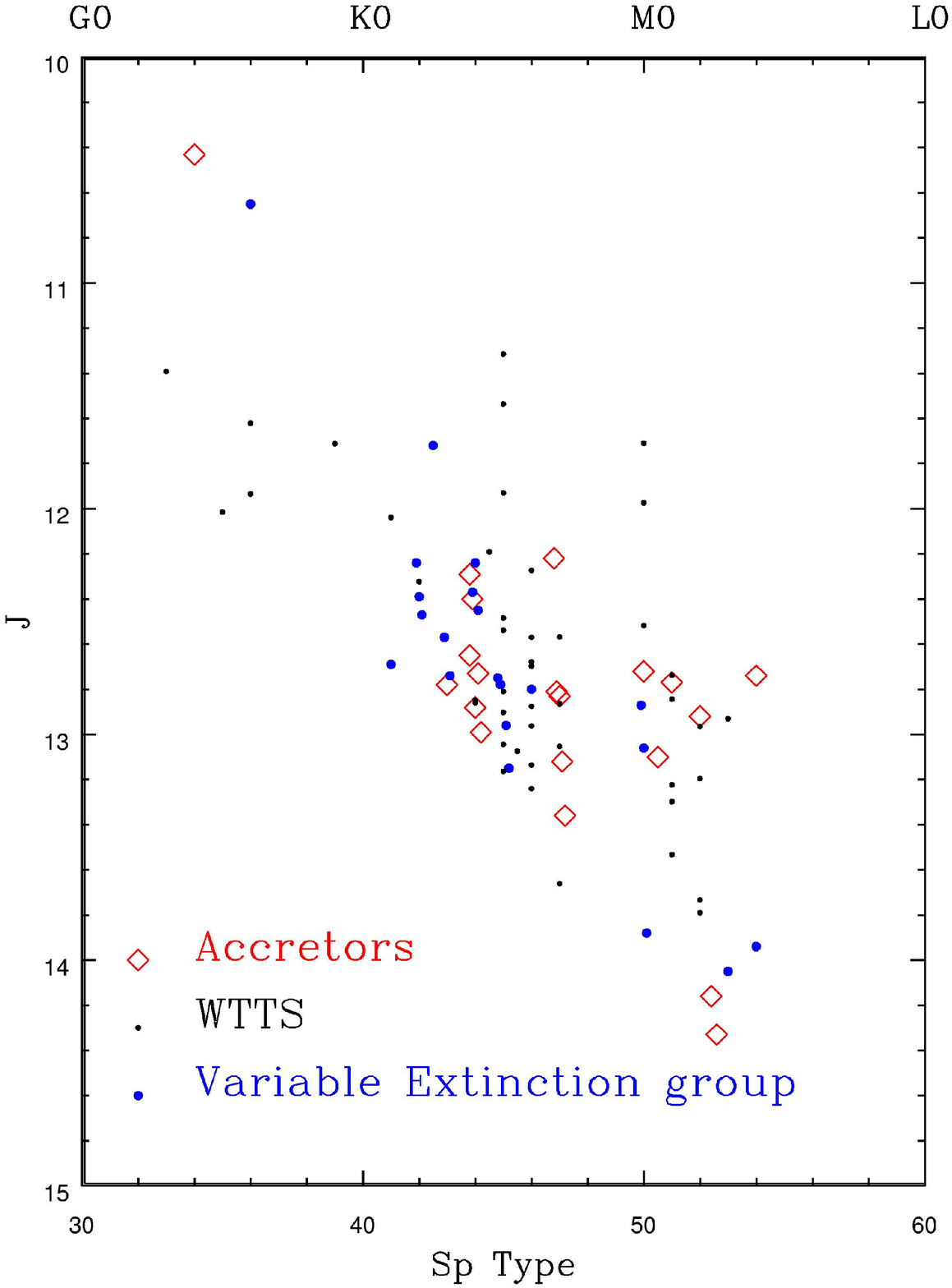}
\end{center}
\caption{$J$ vs.\ spectral type diagram, for the same set of stars
plotted in Figure~\ref{fig:colorcolorf4}.  As for
Figure~\ref{fig:colorcolorf4}, small, black dots are WTTS, red
diamonds are stars with accretion burst dominated light curves, 
and blue dots are from the variable extinction
group.  Spectral types have been converted to integer equivalents,
with G0 = 30, K0 = 40, M0 = 50, etc. We see no systematic displacement
between the three groups shown, e.g., all three seem to be the same
approximate age, or at least that any systematic age difference is too
small for us to determine from this figure. \label{fig:colormagf10}}
\end{figure}

Using our CFHT photometry, Figure~\ref{fig:colormagf9} compares the
location of the stars in Table~\ref{tab:basicinformation} to the stars
with extinction dips and to the WTTS of NGC~2264 in a $g$ vs.\ $g-i$
color-magnitude diagram (CMD).  If one of these sets of stars were
younger on average than the others, we would expect the stars of that
set to be displaced systematically above the other groups.  We see no
such systematic displacement between the three groups in the diagram;
all three follow essentially the same locus in the CMD.  The only
possible difference appears to be that the WTTS may have a smaller
dispersion about that locus than either of the two sets of CTTS
(similar results were found by L04).  One could argue that in both
axes of this plot, the location of the stars with accretion burst
dominated light curves in particular would be affected by their hot
spots, plausibly causing the added dispersion and possibly causing a
systematic displacement relative to the other stars.  A diagram with
perhaps less sensitivity to accretion luminosity is $J$ vs.\ spectral
type, which we provide as Figure~\ref{fig:colormagf10}.   However, the
basic result is the same -- the three sets of stars appear to follow
the same locus, suggesting no significant age difference or at least
that any systematic age difference is too small for us to determine
with these proxies for stellar luminosity and effective temperature.

\subsection{Age Inferred from Spatial Location}

\begin{figure*}
\begin{center}
\epsscale{1.0}
\plottwo{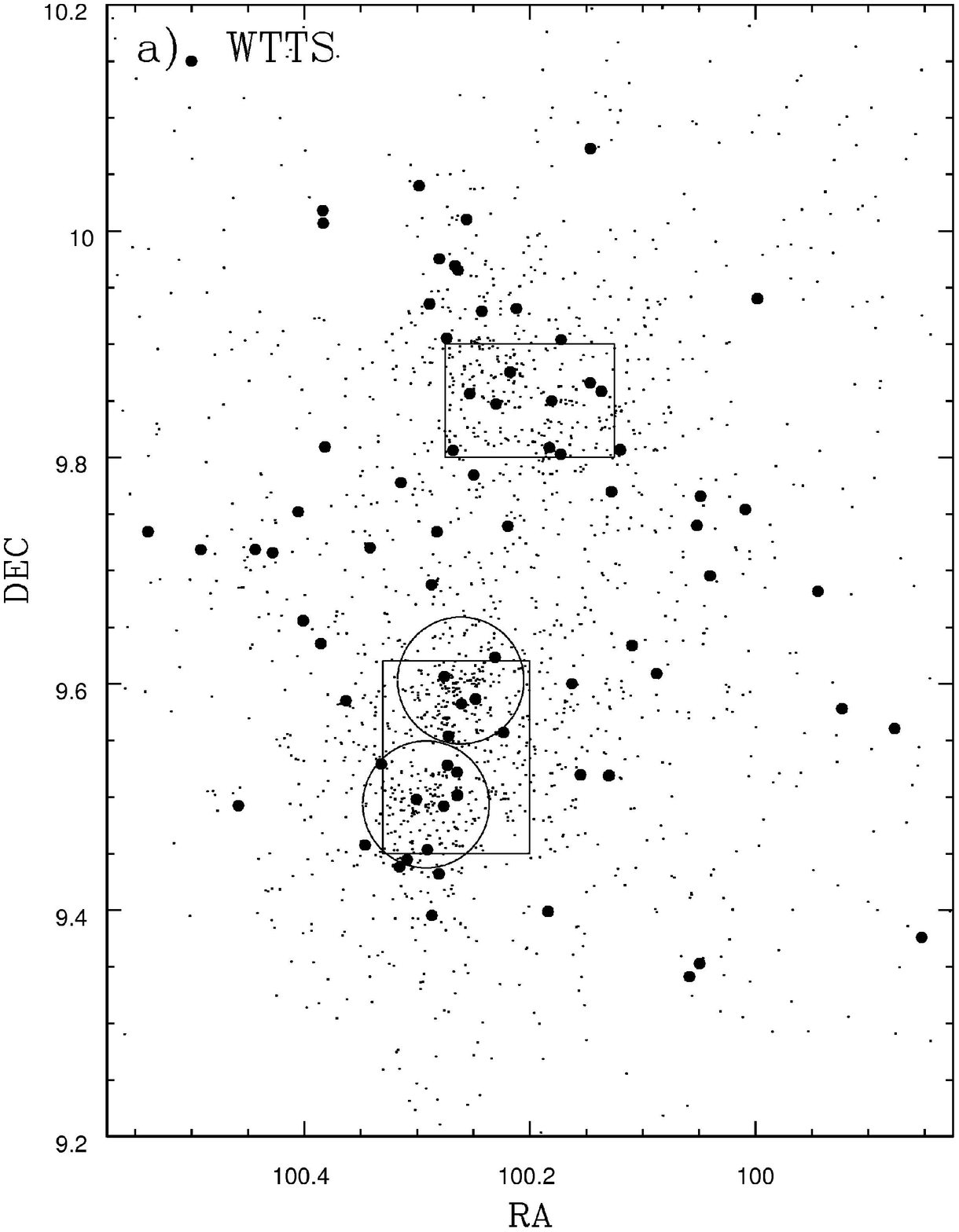}{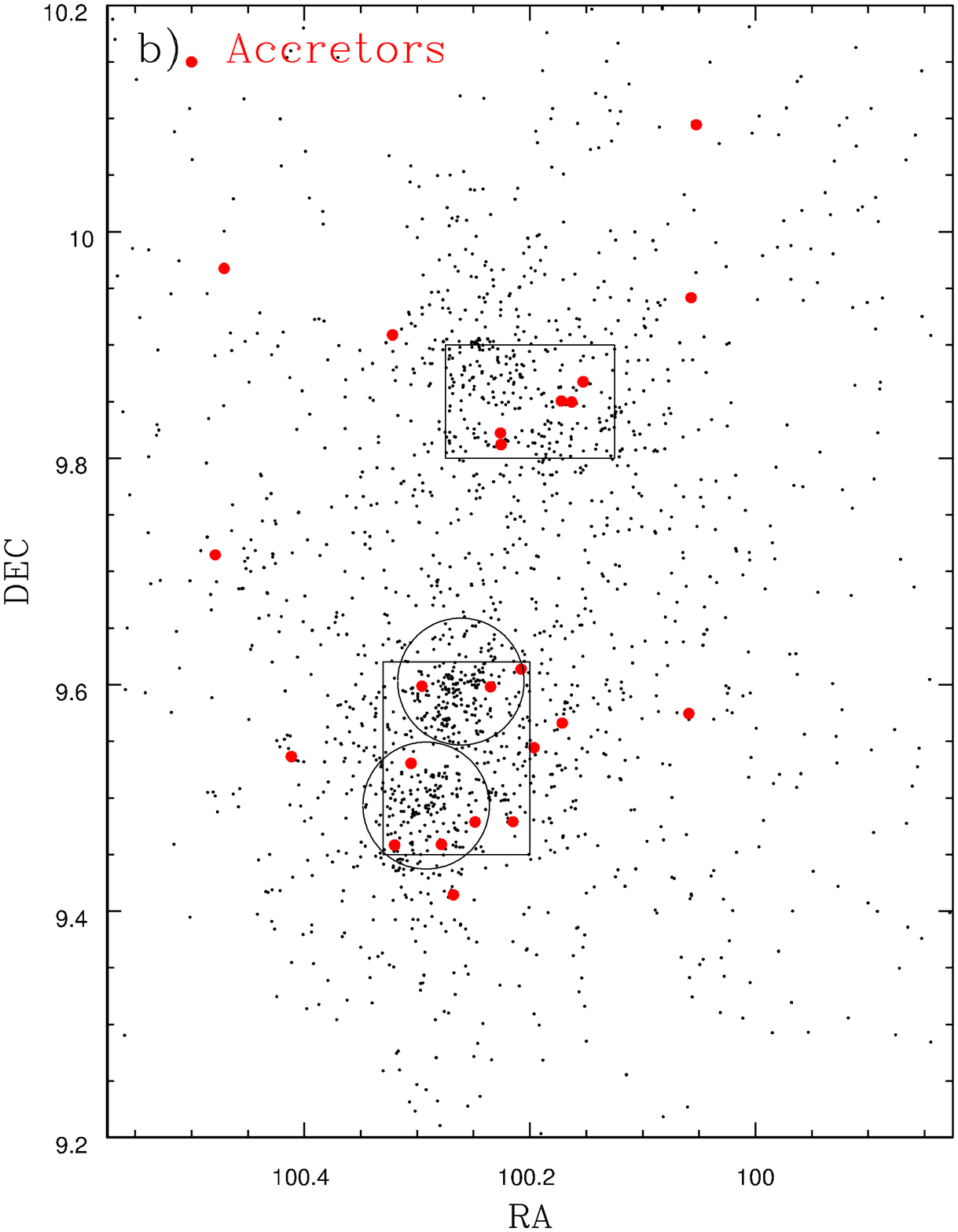}
\plottwo{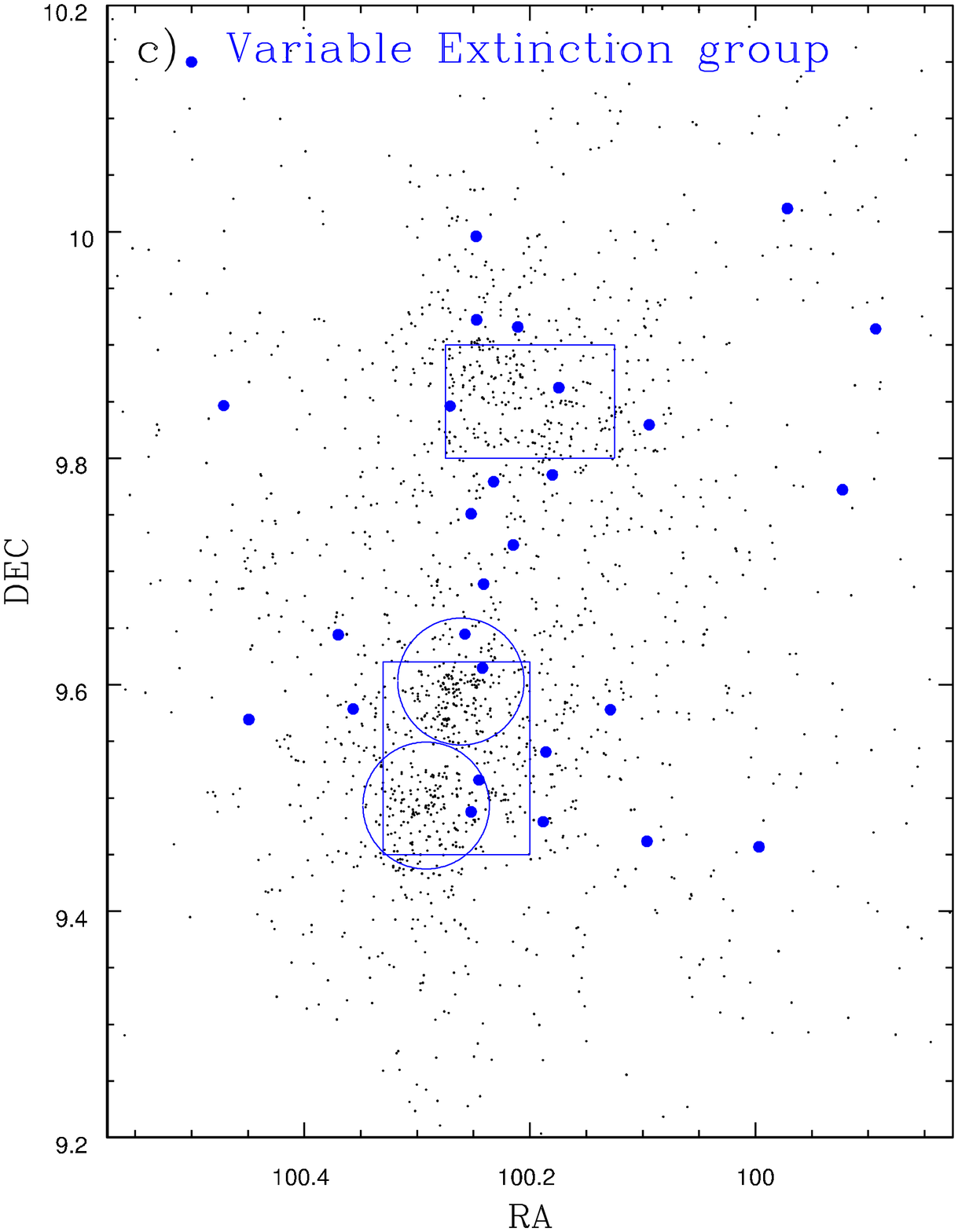}{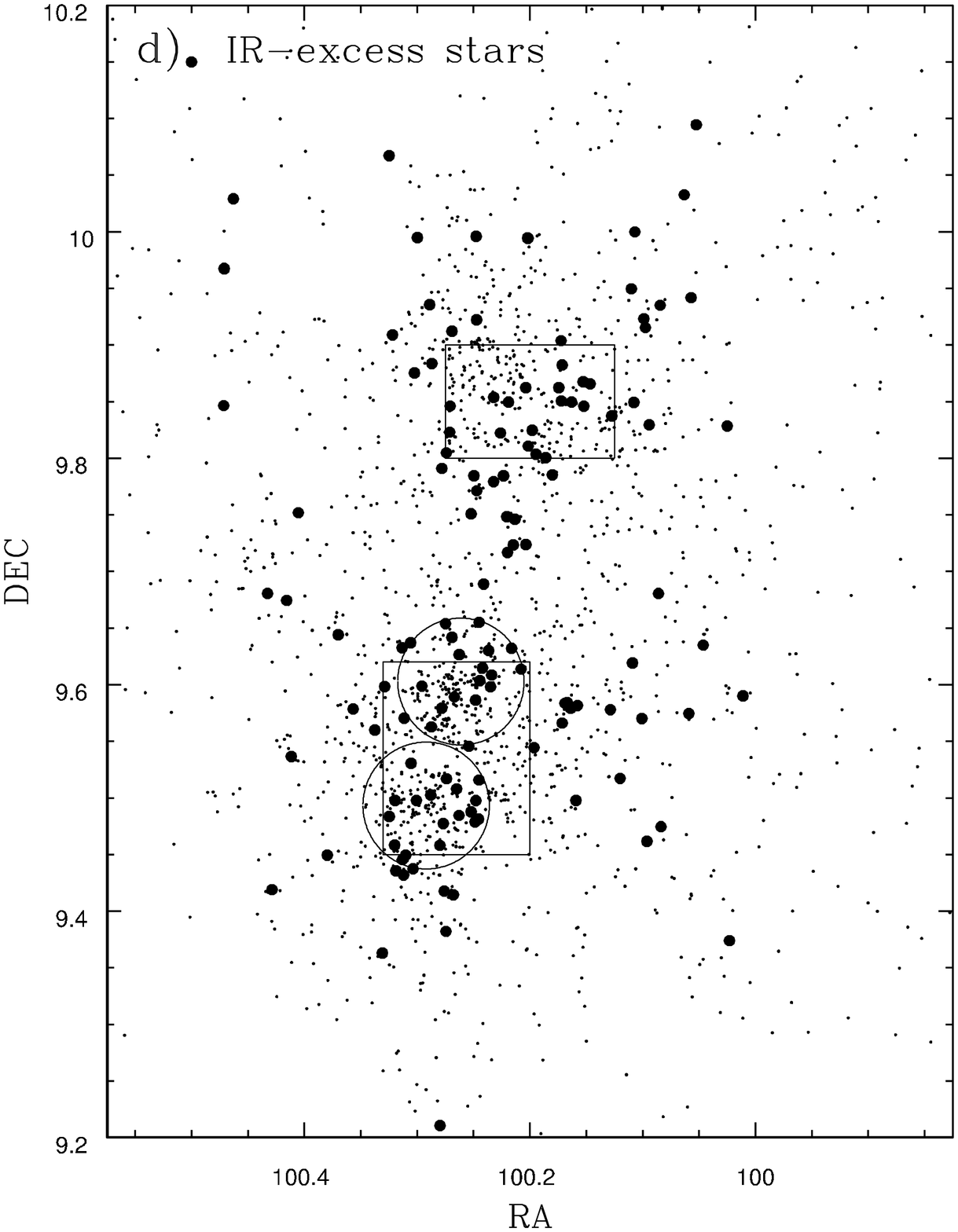}
\end{center}
\caption{Spatial distribution of YSOs in NGC~2264.  The four panels
correspond to the same sets of stars as plotted in
Figure~\ref{fig:colorcolorf4} for the $u-g$, $g-r$ plots. Panel (a)
highlights stars that are probable WTTS; (b) highlights members of the
accretion burst class; (c) highlights stars with extinction dips in
their light curves; and (d) highlights all NGC~2264 members classified
as Class I or II.  The circles and squares enclose the regions of most
active current star formation (Sung et al.\ 2008; L04) - see text
in \S 5.2.  The plots indicate that the accretion burst class
members are more concentrated in the sites of active star formation in
NGC~2264 than WTTS or other types of CTTS. \label{fig:spatial}}
\end{figure*}

As emphasized by Sung et al.\ (2009) and Teixeira et al.\ (2012), the
regions of current active star formation occupy a relatively small
fraction of the total area on the sky for the entire population of
cluster members.  If, for example, light curves of the type shown in
Figure~\ref{fig:sixctts}d-f were to occur primarily amongst the
youngest YSOs, then one should find a larger fraction of such stars in
the currently star-forming cores.  Figure~\ref{fig:spatial} shows the
spatial locations of the accretion burst stars, variable extinction
stars and WTTS from Figure~\ref{fig:colorcolorf4}, plus (in
panel d) the set of stars whose SEDs indicate they are Class I or II. 
The circled regions are the most active star-forming regions from Sung
et al.\ (2008), while the boxed regions are the most active
star-forming regions from L04.   Table~\ref{tab:spatial} provides
the fraction of stars in each group that fall within the active
star-forming cores, and provides circumstantial evidence that
the stars with accretion burst dominated light curves are indeed
younger, on average, than the other three groups.

\begin{deluxetable*}{cccccc}
\tabletypesize{\scriptsize}
\tablecolumns{8}
\tablewidth{0pt}
\tablecaption{Spatial Distribution of YSOs by Type \label{tab:spatial}}
\tablehead{
\colhead{Group}  & 
\colhead{Total Number} &  
\colhead{N(Sung)} & 
\colhead{Fraction(Sung)} &
\colhead{N(Lamm)} &
\colhead{Fraction(Lamm)
} }
\startdata
Accretion burst stars & 23  &  7  & 0.30 & 13  &  0.56 \\
Extinction dip stars  & 27  &  4  & 0.15 &  5  &  0.19 \\
WTTS                  & 81  &  13 & 0.16 & 20  &  0.25 \\
Class II              & 140 &  33 & 0.24 & 48  &  0.34 \\
\enddata
\tablecomments{``Sung" and ``Lamm" refer to the circled and boxed
regions of the most active current star formation as shown in  
Figure~\ref{fig:spatial}, as originally defined in Sung et
al.\ 2008 and L04.  N(Sung) is the number of stars of a given group
that fall within the circular regions defined by Sung; Fraction(Sung) is
the ratio of N(Sung) to the total number of members in that group. }
\end{deluxetable*}

\subsection{Rotational Velocity Distribution}

\begin{figure}
\begin{center}
\epsfxsize=.99\columnwidth
\epsfbox{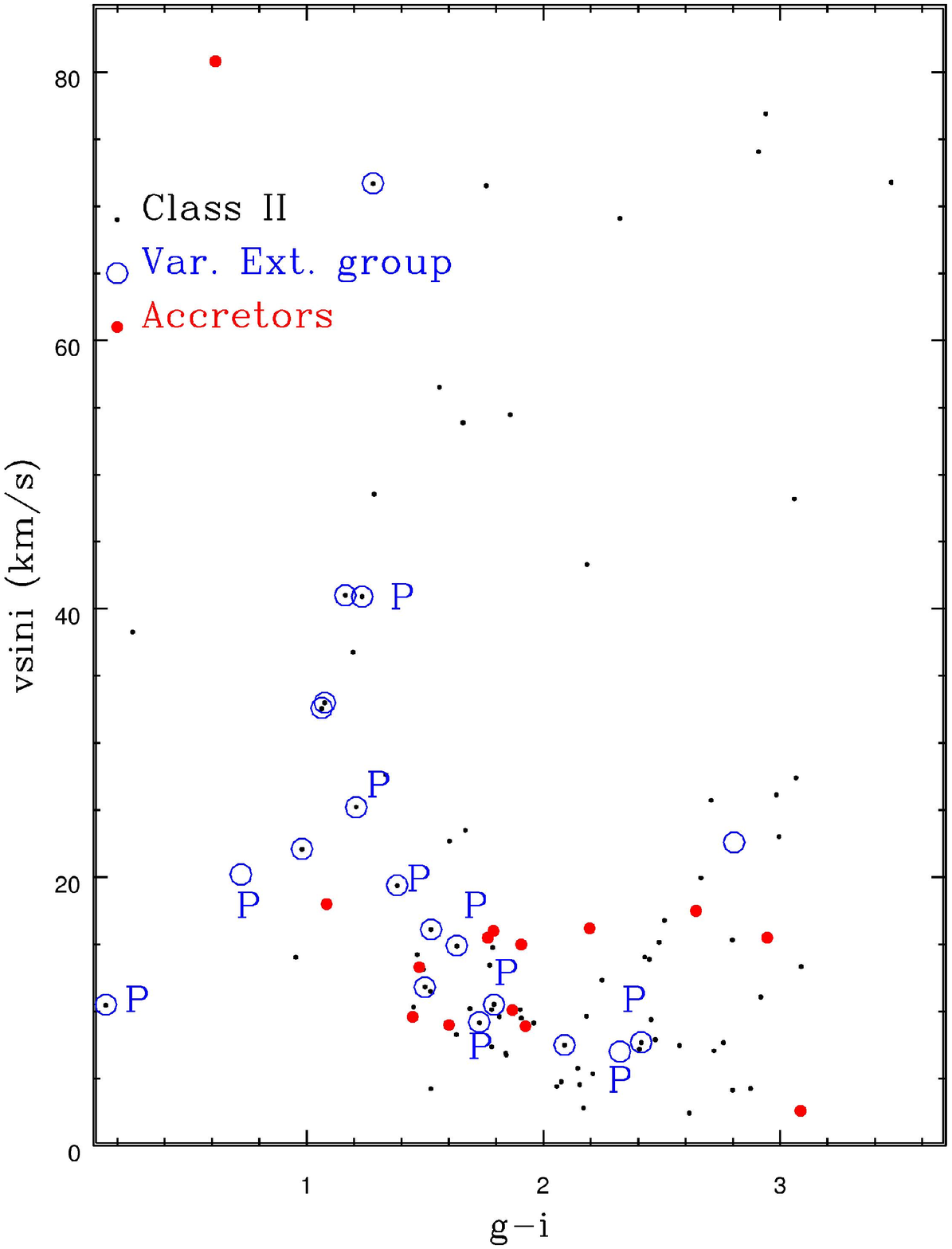}
\end{center}
\caption{Rotational velocity distribution ($v \sin i$) as a function
of $g-i$ for the stars with accretion burst dominated light curves
from Table~\ref{tab:basicinformation}, compared
to that for the stars of the variable extinction group, and a set of
all the CTTS in NGC~2264 for which we have $v \sin i$\ estimates.
Black dots are Class II obejcts, open blue circles are objects from
the variable extinction group, and filled red dots are objects from
the accretor group.  Members of the variable extinction group with
periodic flux dips are labelled with ``P". 
The variable extinction dip group includes a much
larger fraction of relatively early spectral type stars with
relatively large rotational velocities. \label{fig:vsini}}
\end{figure}

In principle, the spectroscopic rotational velocities ($v \sin i$) for
a group of young stars should depend on the ages and initial angular
momenta of the stars, the degree to which their surface rotation is
coupled to their circumstellar disks, and the orientation of their
rotation axes to our line of sight.  Based on our previous discussion,
it is likely that the stars of Table~\ref{tab:basicinformation} are
comparatively young and that our line of sight to their rotational
axes may be weighted towards small inclinations -- both of which would
result in relatively small projected rotational velocities.   We have
$v\sin i$ data for more than half of the stars with {\em CoRoT} light
curves, so we may be able to check this prediction.  
Figure~\ref{fig:vsini} compares the rotational velocity distributions
for the same groups of stars considered in Figure~\ref{fig:spatial},
for the subset of stars for which we have high resolution spectra.  

The primary difference in Figure~\ref{fig:vsini} between the stars
with accretion  burst dominated light curves and the stars with
extinction dips, to our eyes, is that the variable extinction dip
group includes a larger fraction of comparatively blue
stars with relatively large rotational velocities.  Only one of
the Table~\ref{tab:basicinformation} stars seems to be in this
extension to high rotational velocities at higher (inferred) masses, and
that is Mon-000007 at $g-i$ = 0.6, $v \sin i$ = 80 km s$^{-1}$.  This
difference in colors is not, for example, due to a difference in
extinctions.   The variable extinction group includes a larger 
fraction of stars with relatively early spectral types -- half of the
extinction dip group members with spectral types and $v \sin i$ values
have spectral types earlier than K4, compared to only one of fifteen
such stars from Table~\ref{tab:basicinformation}.  We believe this
dependence of the projected rotational velocity on spectral type
reflects the initial conditions which lead to the Kraft main-sequence
dependence of mean rotational velocity with mass.   At 2 Myr, the
inferred mass for spectral type K1 (from the Siess isochrone, Siess,
Dufour \& Forestini 2000) is 2.5 M$_{\sun}$, and 2.1 M$_{\sun}$ for
K2.5.  These stars will arrive on the main sequence as A stars, with
mean rotational velocities of about 150 km s$^{-1}$.

Ignoring the high mass stars, if one compares only the stars with $g-i
>$ 1.3, there is no obvious difference between the $v \sin i$
distributions for the accretor group and the variable extinction
group.  This seems surprising, since the variable extinction stars are expected to be
viewed at high inclination angles and to probably be older on average
-- both of which should lead to larger spectroscopic rotational
velocities.  Do accretion models offer an explanation?  KR08 discuss
the factors that favor burst-dominated light curves, which include
high accretion rates, relatively slow rotation, and relatively weak
dipole B fields.  Favoring slow rotation for the burst-dominated
sources only exacerbates our problem.  For disk-locked CTTS, Long et
al.\ (2013) predict rotation periods which scale as M$_{\odot}^{3/7}$\ -
predicting relatively slow rotation for stars with high accretion
rates.  Our expectation remains, therefore, that the variable
extinction stars should have larger $v \sin i$'s at a given mass
compared to the burst-dominated sources, and we have no explanation
for why this is not reflected by the data shown in
Figure~\ref{fig:vsini}. We note, however, that a similar problem was
also present and discussed  amongst the Class III (or UX Ori)
variables of H94, whose $v \sin i$'s also did not appear to be larger
on average than non-variable YSOs of the same spectral type.

\subsection{Inferences on System Geometry from Optical/IR Light Curve Correlations}

We have concentrated so far on the {\em CoRoT} light curves, because
they have the highest cadence and directly detect photons from the
accretion powered hot spots.  Our {\em Spitzer} light curves also have
relatively high cadence and accuracy, and based on model fits to the
optical-IR SEDs, for the stars with accretion burst-dominated light
curves of Table~\ref{tab:basicinformation}, the 4.5 $\mu$m flux is
expected to come primarily  from warm dust in the inner disk.   

Our model for the accretion bursts is that they arise from individual,
short-lived mass infall events at small spots on the stellar
photosphere.  Given that model, it follows that the correlation
between the optical and IR light curves will depend on the longitude
and latitude of the hot spot, our view to the system, the alignment of
the stellar and disk rotation angles, and the detailed structure of
the inner disk.  Empirically, we see considerable variation in the
degree of correlation between the {\em CoRoT} and {\em Spitzer} light
curves.   This is illustrated in Figure~\ref{fig:spitzercorotf13} for
four stars of Table~\ref{tab:basicinformation};
Figure~\ref{fig:corot2011} in the appendix shows the entirety of the
data for the stars of Table~\ref{tab:basicinformation}.

We will not attempt here to enter into a detailed analysis of the {\em
Spitzer} light curves.  Instead, we will only attempt to show that the
primary features shown in Figure~\ref{fig:spitzercorotf13} are
compatible with our basic model for the bursts.  In order to motivate
that discussion, we first describe the expected 4.5 $\mu$m\ appearance
of a standard flared disk with an inner disk wall as a function of our
view angle to that disk, based on disk models by co-I B.\ Whitney
(Whitney et al.\ 2013).    
Figure~\ref{fig:corotmodel} shows images of a typical disk system for
a 1 solar
luminosity star  with $T_{\rm eff}$=4000 K.   The radius of the inner
disk is set by the dust  destruction temperature of 1600 K, and is
about  7.9 stellar radii or  0.075 AU.   The wall height is set by the
hydrostatic equilibrium scale  height at the same temperature.   The
disk accretion rate is 8$\times$10$^{-8}$\  solar masses yr$^{-1}$. 
The ratio of disk to photospheric flux at 4.5 $\mu$m\ is about 3.0 for
these models, nearly independent of inclination angle.
Figure~\ref{fig:corotmodel} shows this disk as seen from three
different view angles -- 0$\arcdeg$, 30$\arcdeg$, and 70$\arcdeg$. As
the view angle increases, both the total surface area of the 4.5
$\mu$m\ emitting dust increases and the fraction of that flux emitted
by the back wall increases. Table~\ref{tab:barbmodel} quantifies the
dependence of the 4.5 $\mu$m\ emission on view angle for this specific
disk model.  At 70$\arcdeg$\ view angle,  80\% of the 4.5 $\mu$m\ flux
we receive from the circumstellar dust comes from the back side of the
disk.  For inclinations greater than about 75$\arcdeg$, the  4.5
$\mu$m\ emission from the system drops sharply due to extinction from
cold dust in the flared outer disk.   

\begin{deluxetable*}{ccc}
\tabletypesize{\scriptsize}
\tablecolumns{3}
\tablewidth{0pt}
\tablecaption{Predicted Disk Properties at 4.5 $\mu$m\ as a Function of View Angle \label{tab:barbmodel}}
\tablehead{
\colhead{Inclination}  & \colhead{Back to Front Flux Ratio} & \colhead{Relative Total 4.5 $\mu$m Flux} 
 }
\startdata
0    &  1.00  &  1.00   \\
30   &  1.52  &  1.27    \\
60   &  2.55  &  1.35   \\
70   &  4.03  &  1.34    \\
\enddata
\end{deluxetable*}

Based on this model, our expectation for the optical and IR light curve
correlations are:

* For a star with no disk, the {\em CoRoT} and {\em Spitzer} light curves must be
    almost perfectly correlated, with the only difference being the
    wavelength dependence of the different components at or near the
    star's surface (spots, flares, etc.)

* For a star with a disk, an important factor is the fraction of the
    star's flux at 3.6/4.5 microns that comes from warm dust in the
    disk.  Stars with an inner disk hole (no warm dust) should have a
    different degree of correlation between their {\em CoRoT} and {\em Spitzer}
    light curves compared to those with significant warm dust.  For
    this reason, the last column of Table~\ref{tab:basicinformation}
    shows our estimate of the ratio of the 4.5 micron flux from the
    disk to that from the star, based on fits to their SEDs for stars
    where we have spectral types.  Not surprisingly given their other
    characteristics, with only one exception (Mon 860 - for which we
    have no synoptic IRAC data),
    the disk dominates the flux at 4.5 microns.  However, many of the
    other stars have disk/star flux ratios of only 2-3, so even though
    the disk dominates, large features in the {\em CoRoT} light curves
    should still have something present in the {\em Spitzer} light curve.

* At most view angles to the
    disk, it is quite possible to have a transient hot spot on the
    back side of the star illuminate the back-side inner disk, thereby yielding
    a flux burst in the IRAC light curve but because the hot spot is outside
    our direct view there is no imprint on the
    {\em CoRoT} light curve (e.g. the features in the Mon 117 light curve
    near day 25 and day 30).

 * For disks seen at large inclination angles, it is possible to have
     a transient hot spot on the side of the star facing us
     cause a {\em CoRoT} flux burst, but because the 4.5 $\mu$m\ emission
     is heavily dominated by the back-side of the disk where we have a better
     view to the inner disk wall, there would be little or no imprint of
     the burst on the IRAC light curve 
     (e.g. the feature in the Mon 1022 light curve near day
     24).

\begin{figure*}
\begin{center}
\epsscale{1.0}
\plottwo{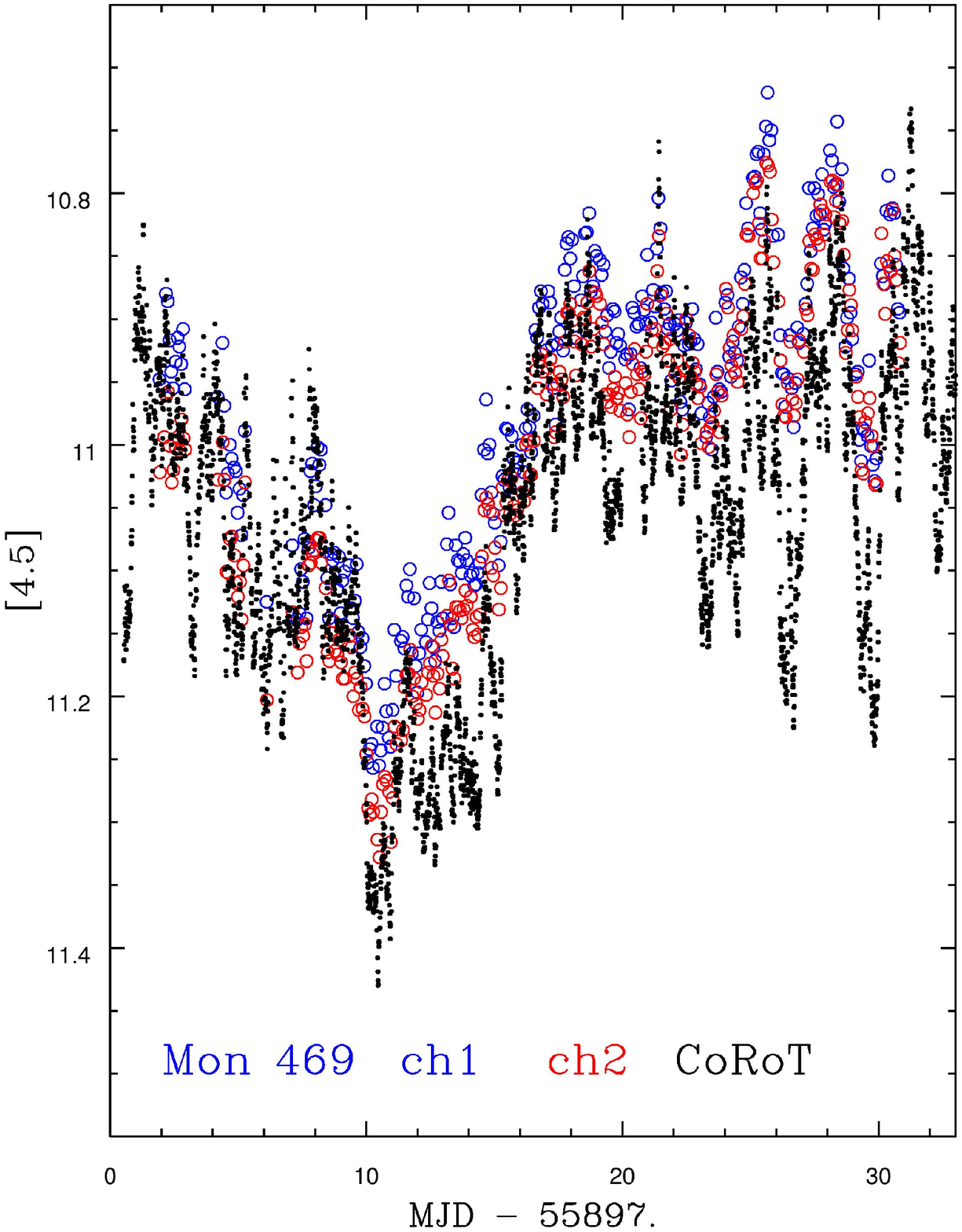}{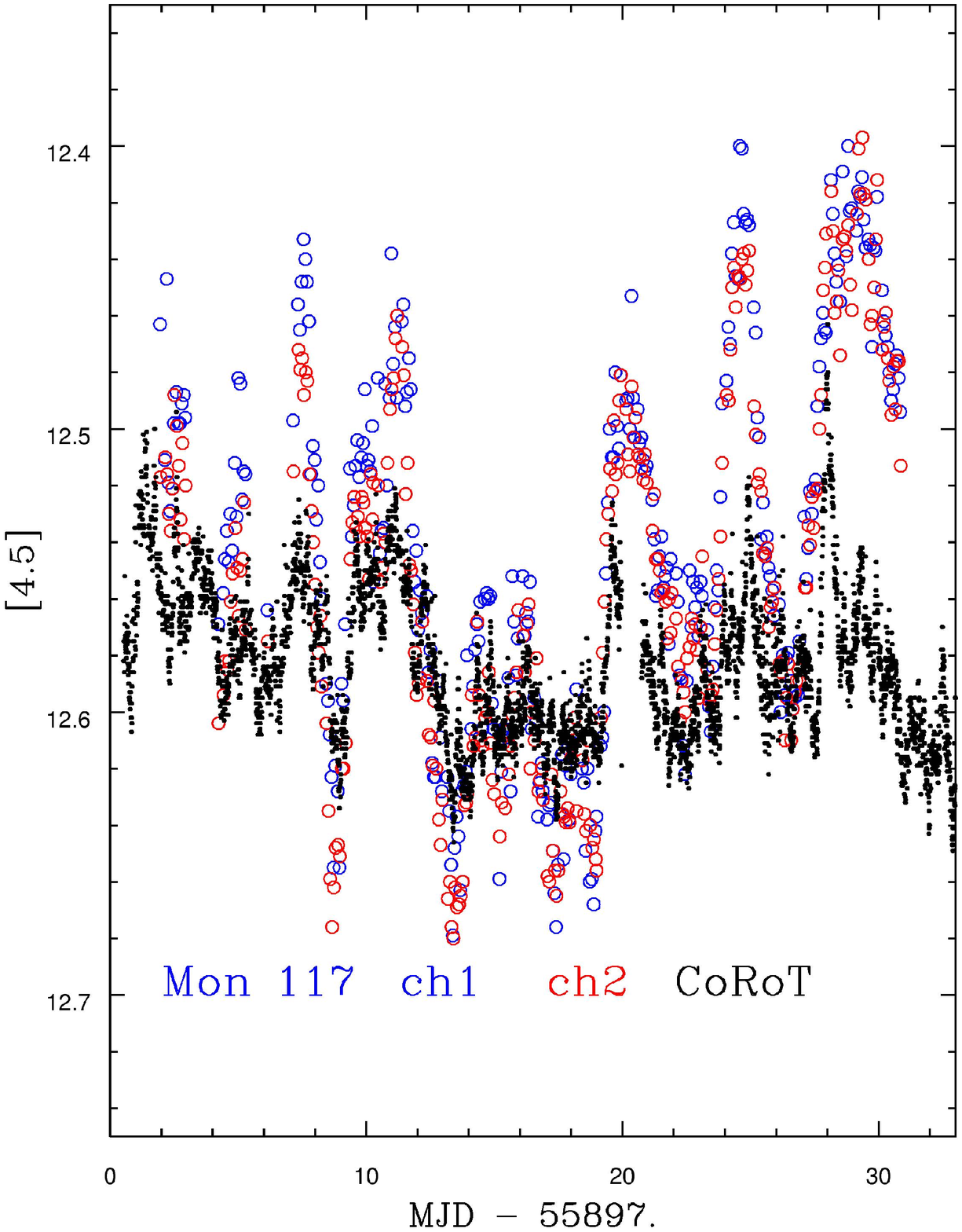}
\plottwo{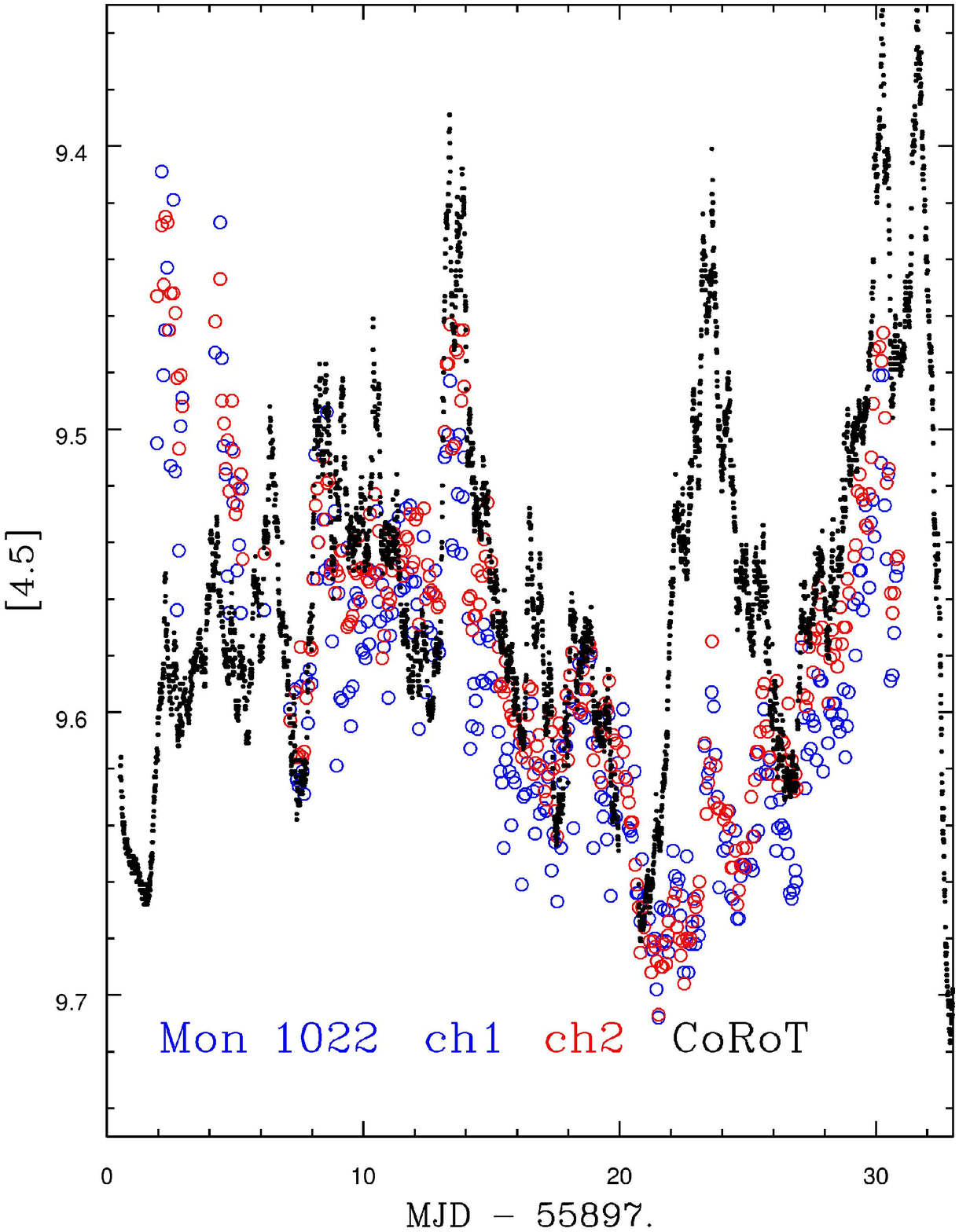}{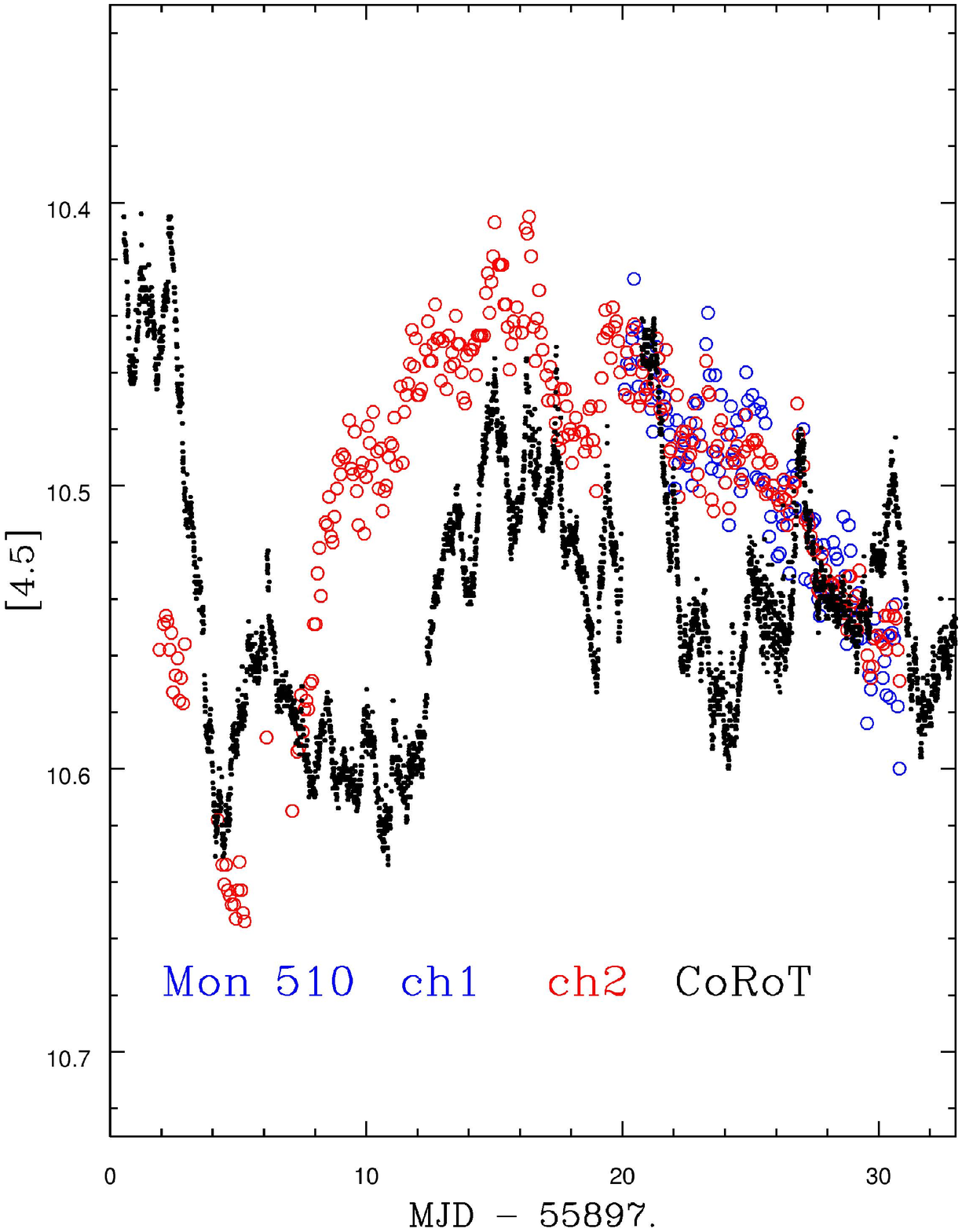}
\end{center}
\caption{Illustration of the correlation between {\em Spitzer} and {\em CoRoT}
light curves for four of the stars with accretion burst dominated
light curves.  The stars are arranged more or less from most
correlated (upper left) to least correlated (lower right).  
\label{fig:spitzercorotf13}  }
\end{figure*}

\begin{figure*}
\minipage{0.32\textwidth}
  \includegraphics[width=\linewidth]{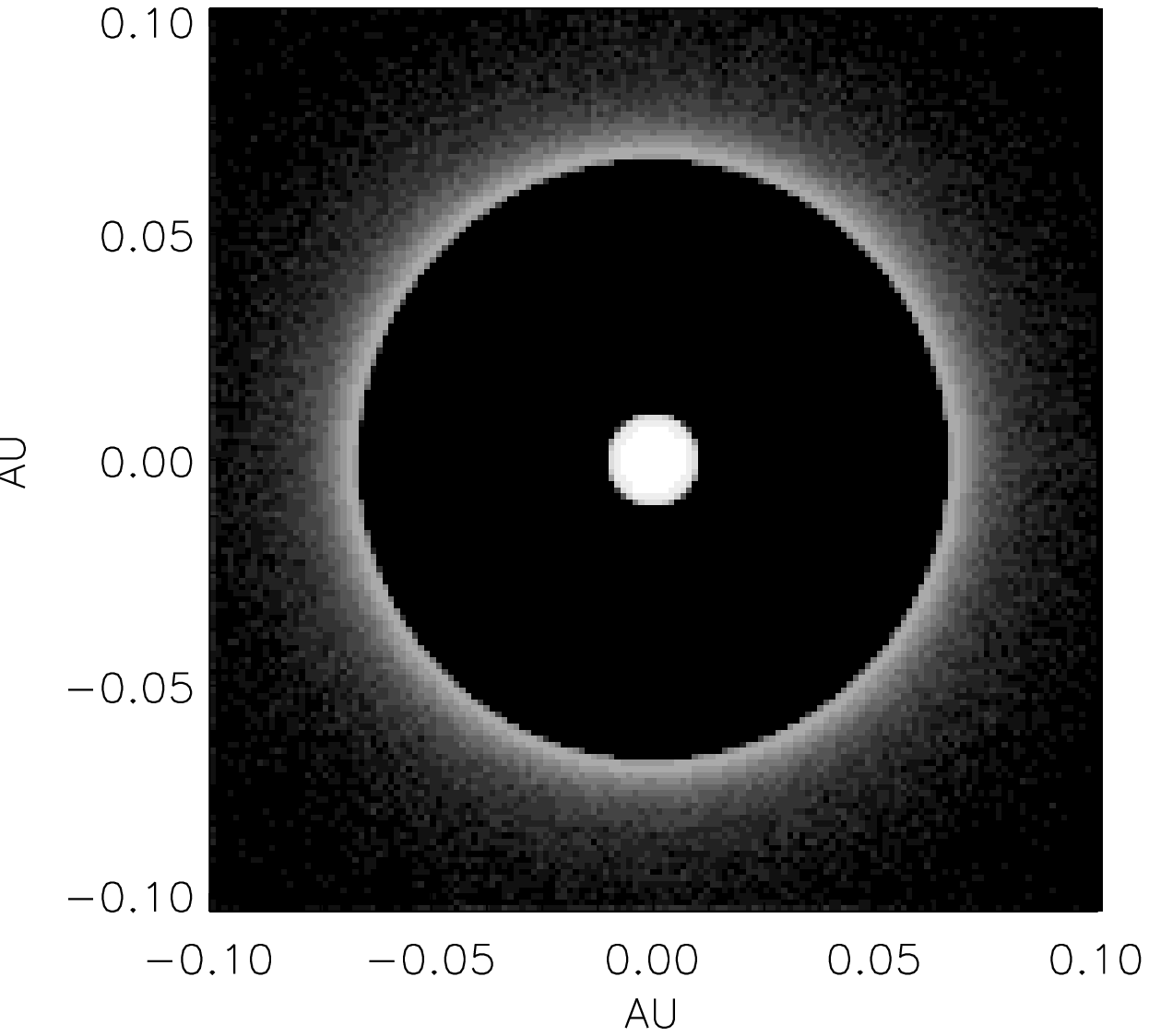}
\endminipage\hfill
\minipage{0.32\textwidth}
  \includegraphics[width=\linewidth]{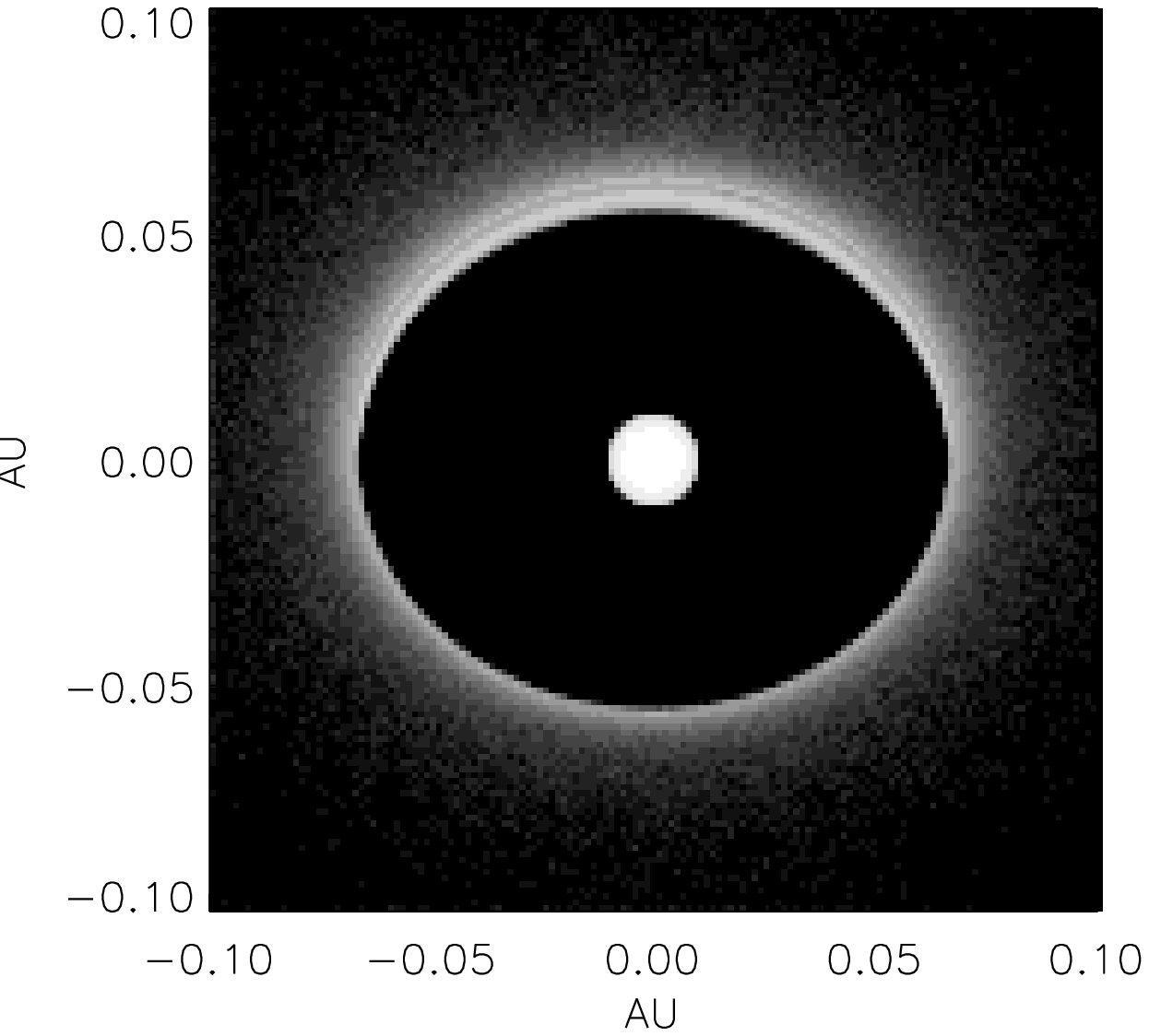}
\endminipage
\minipage{0.32\textwidth}
  \includegraphics[width=\linewidth]{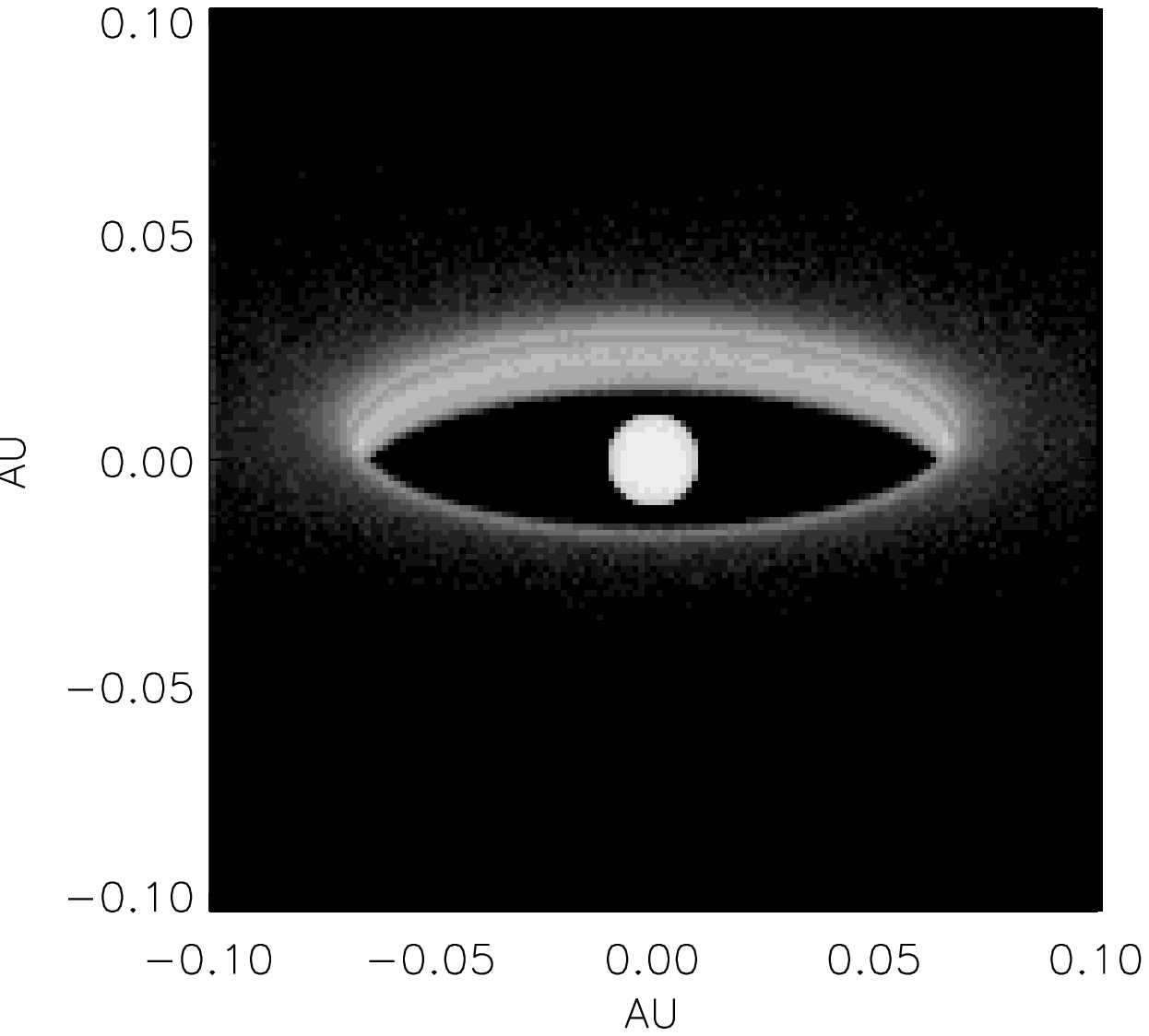}
\endminipage\hfill
\caption{Illustration of the dependence of 4.5 $\mu$m\ disk emission
on view angle, as predicted from a standard flared-disk model (Whitney
et al.\ 2013). The YSO pictured here has L(photosphere) = 1
L$_{\odot}$, T$_{\rm eff}$ = 4000 K, and a mass accretion rate of
8$\times$10$^{-8}$ M$_{\odot}$ yr$^{-1}$.  The three panels correspond
to disk inclinations of 0, 30 and 70 degrees.  See \S 5.4 for model
details. \label{fig:corotmodel}  }
\end{figure*}

\section{Accretion Properties Derived from the Light Curves}

The {\em CoRoT} light curves contain a considerable amount of quantitative
information related to the accretion events.   We discuss a number of
physical constraints on the accretion process derived from the light
curves in the next several sections.

\subsection{Burst durations}

The duration and amplitude of the individual flux bursts in the light
curves for the stars from Table~\ref{tab:basicinformation} encode
information related to the mass and size scale of the individual gas
streams accreting onto the photosphere. While it is often the case
that the light curve structures are too complicated to identify
individual events, in some cases what appear to be individual bursts
are present.  We have carefully examined all of the {\em CoRoT} light curves
of the stars in Table~\ref{tab:basicinformation}, and report those
results here.

\begin{figure}
\begin{center}
\epsfxsize=.99\columnwidth
\epsfbox{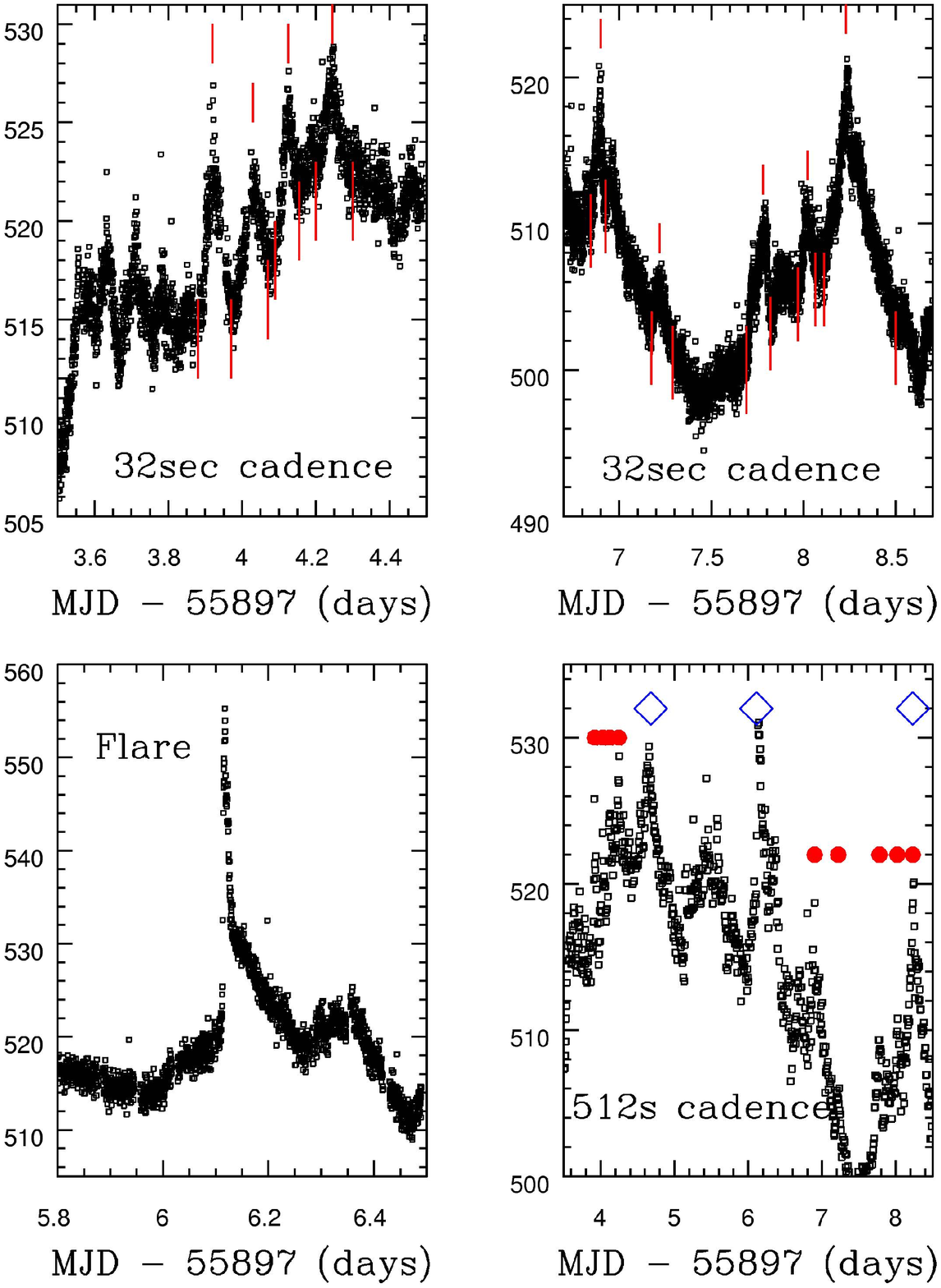}
\end{center}
\caption{(a, upper left) Expanded portion of the high cadence
Mon-000474 light curve.  The four flux bursts we have measured in this
segment of the light curve are marked with red bars indicating the
beginning, center, and end of the burst; (b, upper right) Same as (a)
for a different portion of the light curve; (c, lower left) One of two
stellar flares identified in the Mon 474 light curve; (d, lower right)
A resampled light curve, now at 512 second cadence, for the time
period covered in (a) and (b).  Red dots mark the location of bursts
identified in panels (a) and (b); blue diamonds mark bursts that would be
identified at this cadence. \label{fig:mon474}}
\end{figure}

The {\em CoRoT} high cadence mode (sample rate of once per 32 seconds,
or sixteen times faster than the normal cadence) was utilized for only
one star in Table~\ref{tab:basicinformation}; that star was
Mon-000474.  Mon-000474 is also (not coincidentally) by far the
brightest of the Table~\ref{tab:basicinformation} stars, with an $R$
magnitude more than three magnitudes brighter than the average for the
other stars of Table~\ref{tab:basicinformation}\footnotemark[3]\footnotetext[3]
{Mon-000474 is also the earliest spectral type star in the accretion burst group.  
It is possible that burst duration is a function of spectral type, in
which case the inferences we draw from Mon-000474 may not be
applicable to our entire sample.  With only one star observed in
high-cadence mode, we ignore this possibility for now.}.  This
combination makes the Mon-000474 light curve much more sensitive to
shorter duration, lower luminosity light curve structures.  The
complete {\em CoRoT} light curve for Mon-000474 is shown in the
Appendix in  Figure~\ref{fig:allcorotlcs}. Figure~\ref{fig:mon474}
(top row) shows two very small segments extracted from that light
curve, with nine short duration flux bursts which we ascribe to
accretion variability marked with red bars. Also present in this time
window in the {\em CoRoT} light curve is another event of a different
character, shown in Figure~\ref{fig:mon474}c.  This event is a
well-observed stellar flare, with a very fast rise time (the
``impulsive" phase), followed by an initially fast fall and then a
much slower decay (the ``gradual" phase).  At {\em CoRoT}'s high
cadence, stellar flares are easily distinguished from even
short-duration accretion bursts. At the more normal low cadence (512
second sampling), our ability to identify short duration bursts in the
Mon-000474 light curve and to discriminate them from flares would  be
much reduced.  Figure~\ref{fig:mon474}d provides a direct comparison
by showing a portion of the Mon-000474 light curve, but only plotting
every sixteenth point, thereby better matching both the cadence and 
signal-to-noise of the other burst-dominated light curves..   The red
dots mark the accretion bursts we had identified in
Figure~\ref{fig:mon474}ab; only one of those bursts (the last one) is
clearly identifiable at this cadence. The blue dots mark features we
would identify as isolated accretion bursts at this cadence.
Unfortunately, one of them (the point at day 6.1) is the stellar flare
illustrated in Figure~\ref{fig:mon474}c. At the lower, normal cadence,
a stellar flare can be mistaken for a short-duration accretion burst.

\begin{figure*}
\begin{center}
\epsscale{1.0}
\plotone{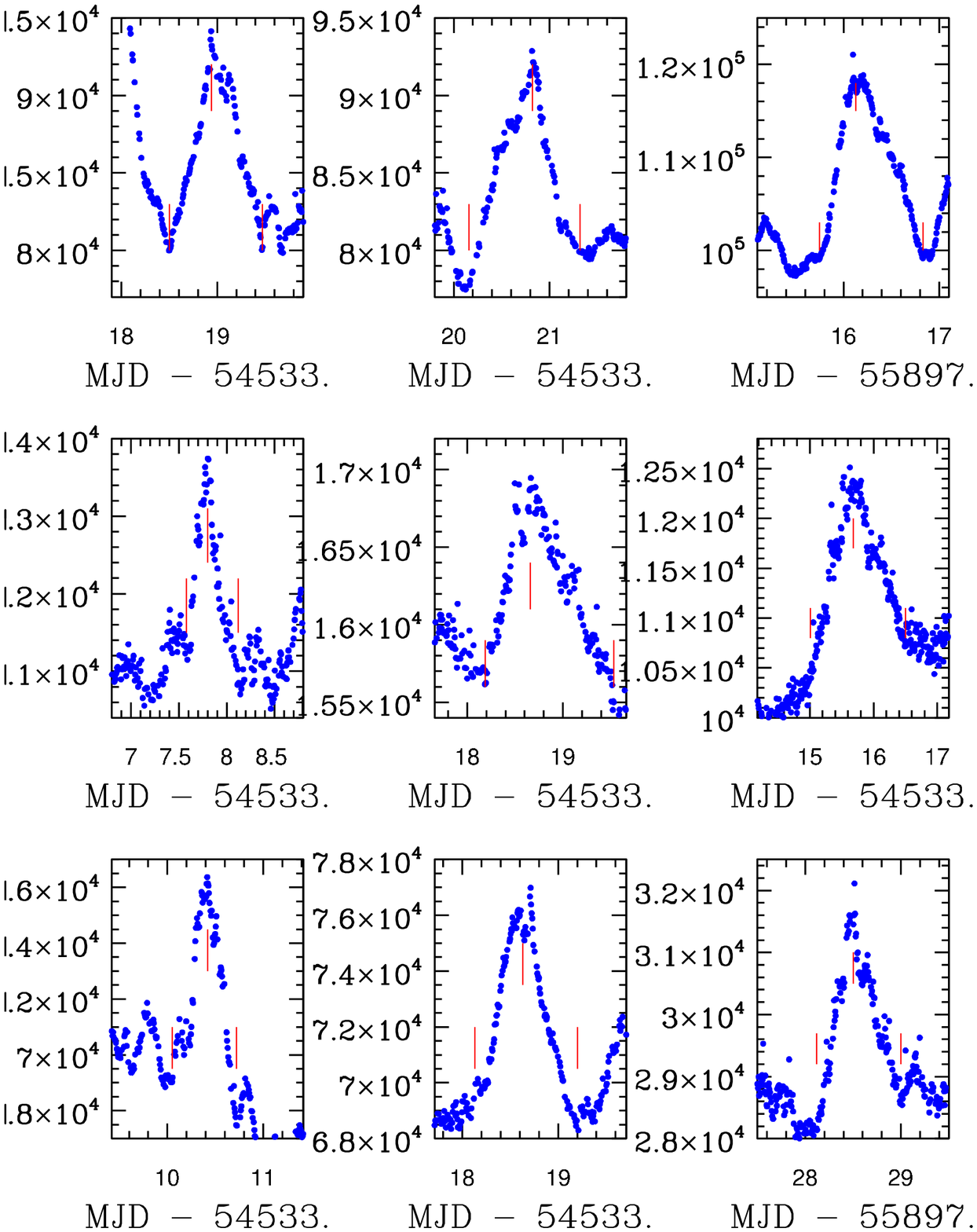}
\end{center}
\caption{Nine examples of relatively isolated flux bursts in the {\em CoRoT}
light curves obtained at the normal (512 second sampling) cadence. 
The red vertical lines mark our estimates of the begin, mid-point, and
end of each burst.\label{fig:accretionbursts}}
\end{figure*}

With the experience gathered from the Mon-000474 light curve, we have
made a careful examination of the lower cadence {\em CoRoT} light
curves for all of the other stars from
Table~\ref{tab:basicinformation}.  In order to avoid possible
contamination from flares, we have not included any feature with width
less than 0.2 days. Figure~\ref{fig:accretionbursts} shows nine of the
events classified as accretion bursts in the low cadence data.

\begin{figure*}
\begin{center}
\epsscale{1.0}
\plottwo{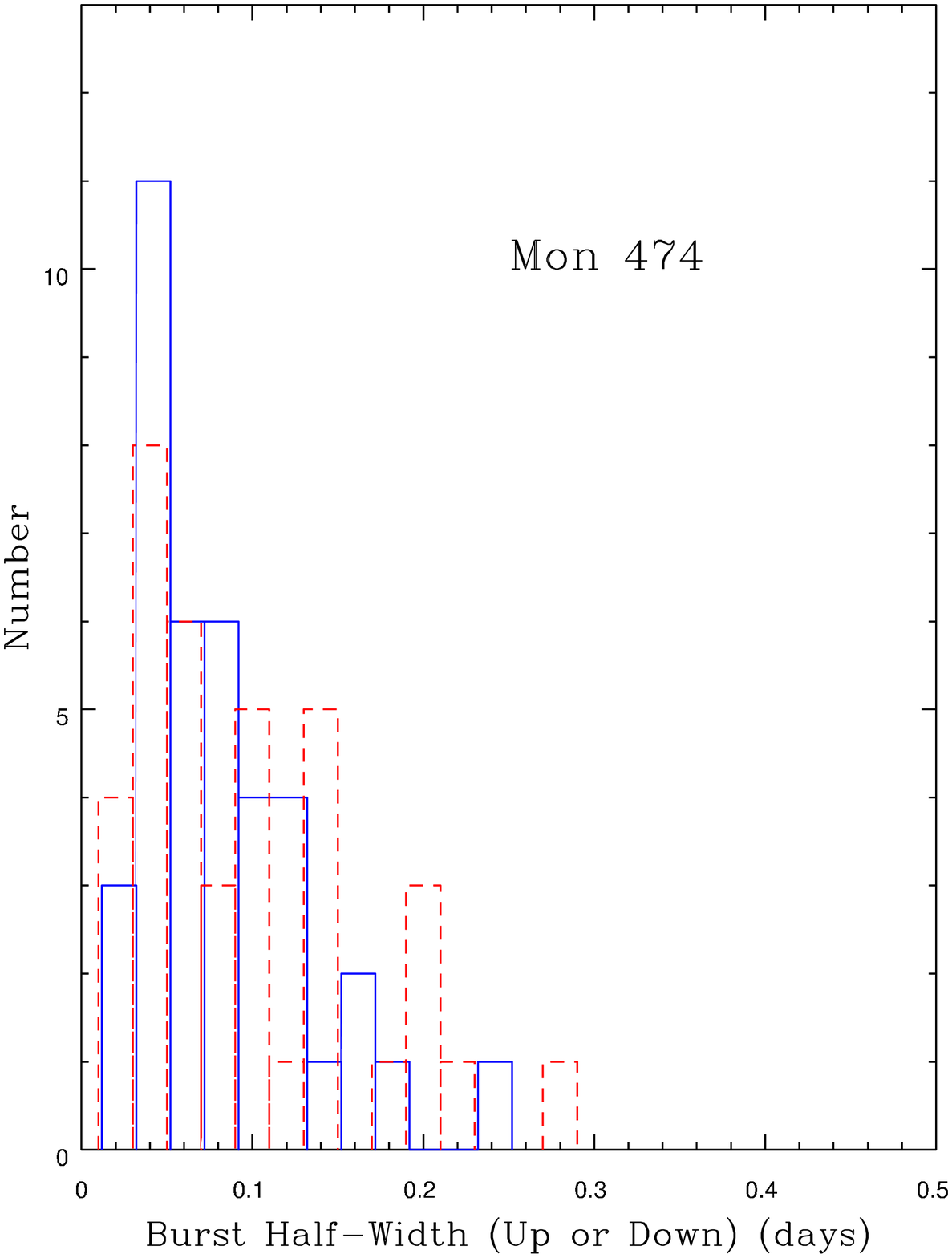}{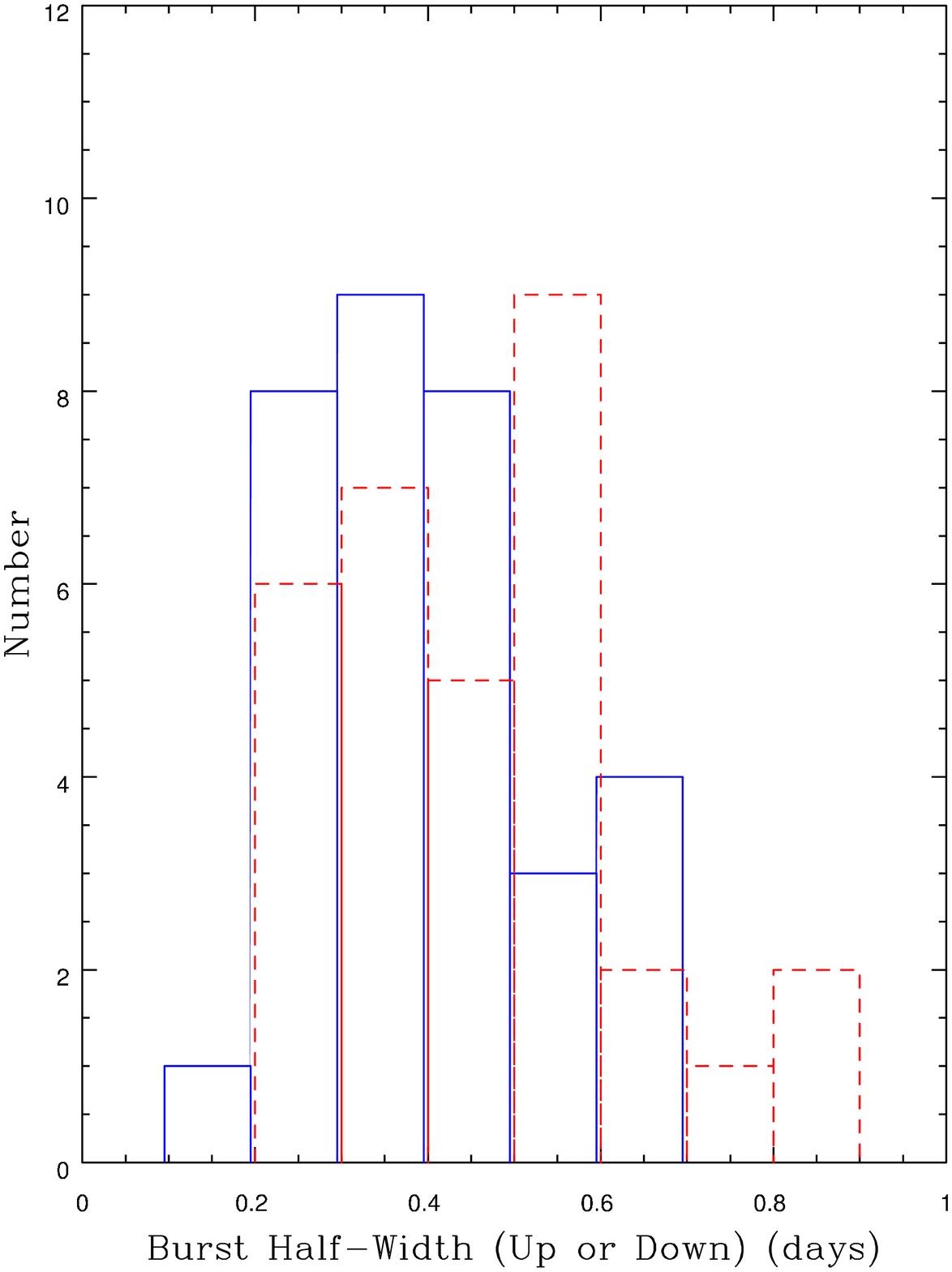}
\end{center}
\caption{ Distribution of rise and fall times for the comparatively
isolated flux bursts identified in the light curves shown in 
Figure~\ref{fig:allcorotlcs}.
Blue, solid lines correspond to the rise times; 
red, dashed lines to the fall times. Left
plot shows data from the high cadence observation of Mon-000474; right
plot shows burst widths from the remaining stars of
Table~\ref{tab:basicinformation}.  We believe the difference between the two
histograms can largely be ascribed to the higher cadence and signal-to-noise 
of the Mon-000474 data; however, Mon-000474 is also much earlier in spectral type
than the other stars with burst-dominated light curves, so there could 
be a dependence on some mass-related characteristic (e.g. outer convective
envelope depth).  \label{fig:mon474widths}}
\end{figure*}

Figure~\ref{fig:mon474widths} shows histograms of the rise times (time
from our estimate of  when the burst begins to the mid-point) and fall
times (time from the mid-point to our estimate when the burst
completes) for the events measured for Mon-000474
(Fig.~\ref{fig:mon474widths}a) and for the remaining 23 accretion
burst stars (Fig.~\ref{fig:mon474widths}b).  For Mon-000474, the
median rise and fall times are both 0.08 days; for the other stars,
those medians are 0.4 and 0.5 days.  The amplitudes of the peaks
identifiable in Mon-000474 and the other stars are also systematically
different.  Using the ratio of the height of the peak relative to the
local continuum as the measure, the mean for the Mon-000474 bursts is
2.3\%, whereas for the other 23 stars observed at low cadence, the mean
is 10.6\%.   The burst durations and amplitudes measured in the standard
cadence data match well to the models of KR08 and R12; the much shorter
duration and lower amplitude bursts visible in the high-cadence, high
signal-to-noise Mon-000474 data were not expected based on those models.

\subsection{Contribution to the Stellar Flux from Accretion Bursts}

\begin{figure*}
\begin{center}
\epsscale{1.0}
\plottwo{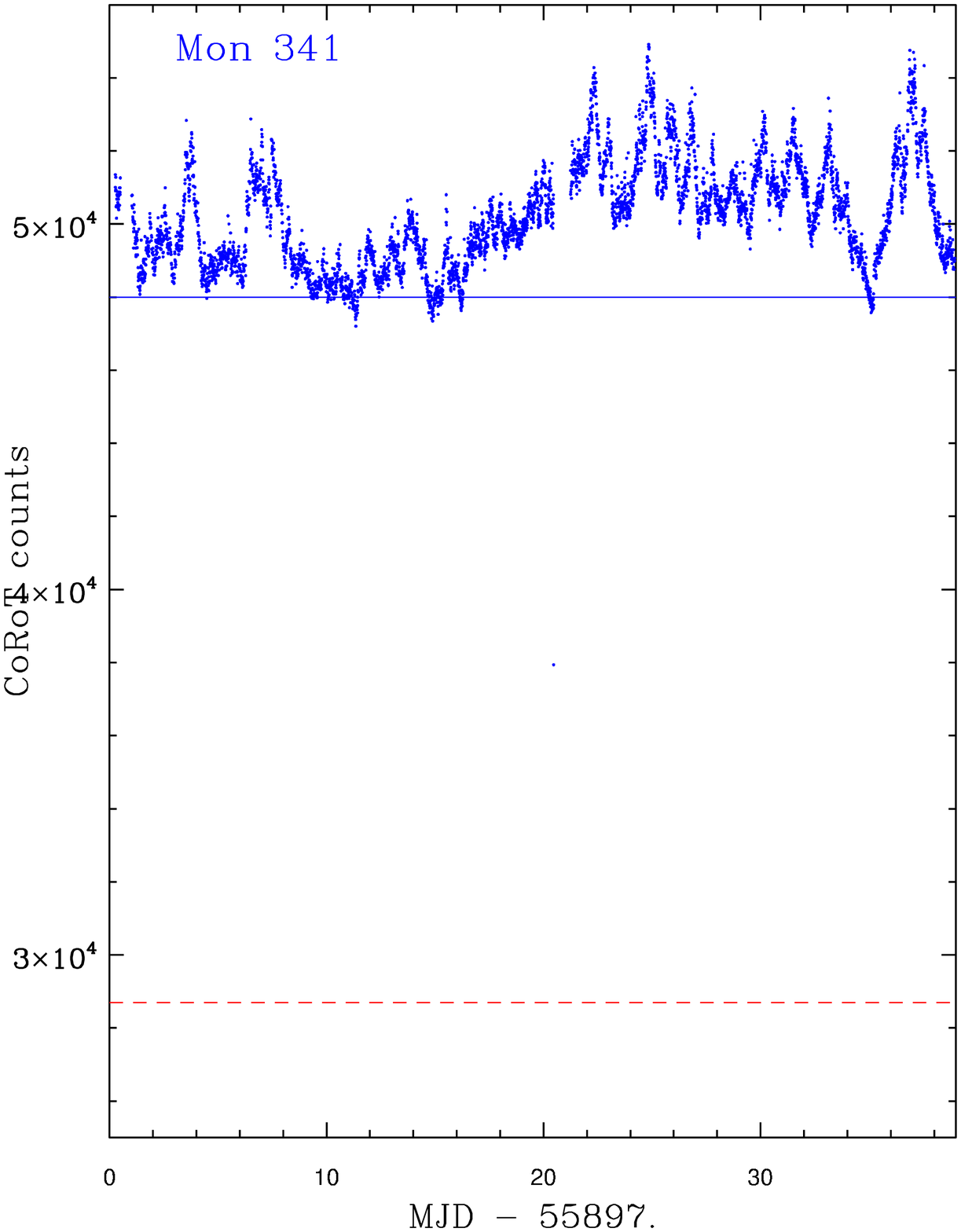}{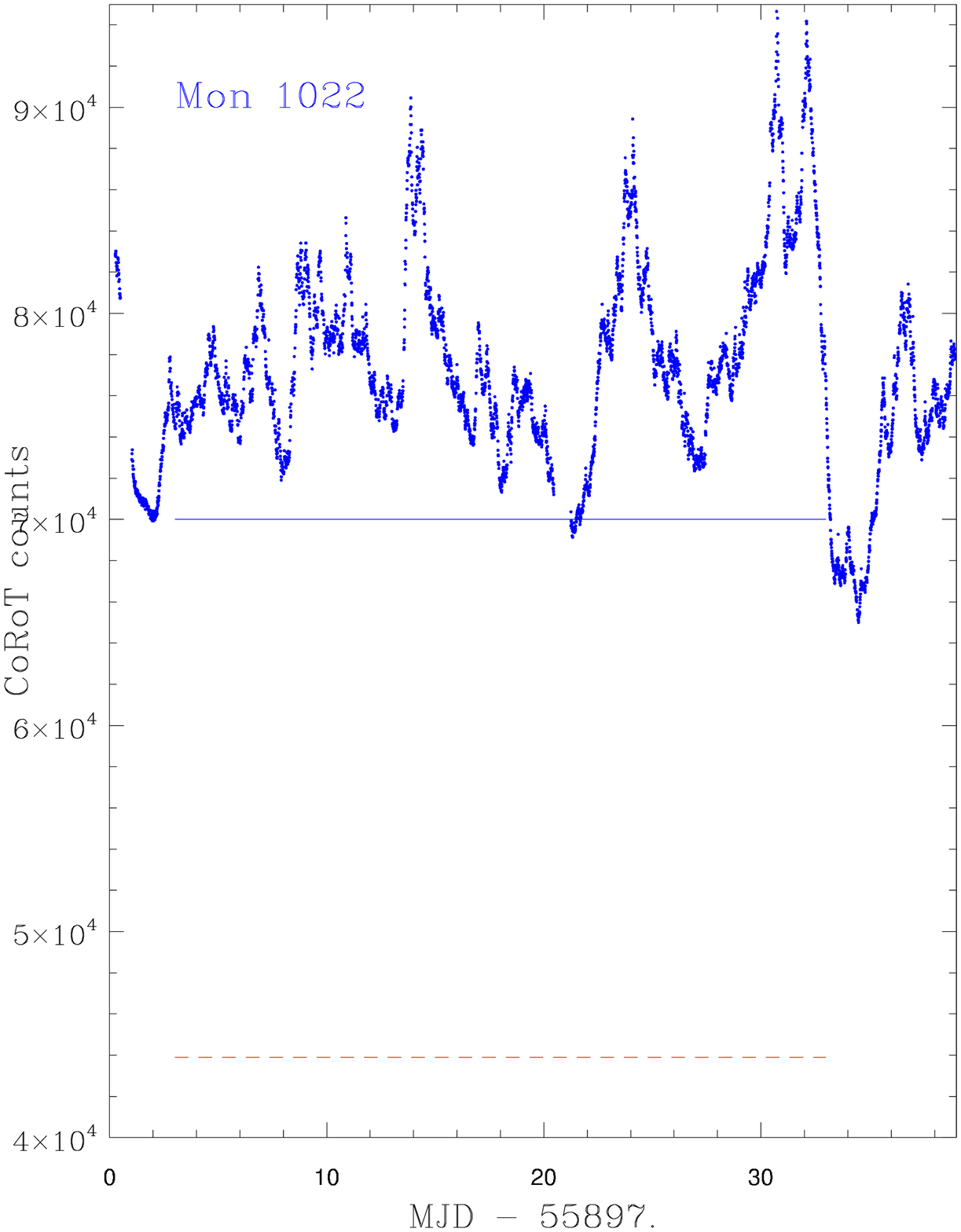}
\end{center}
\caption{Illustration of the method used to estimate the fraction of
the star's broad-band optical light possibly coming from accretion
bursts, for two of the stars with VLT/FLAMES spectra.   The solid
horizontal line is our estimate of the photospheric continuum level.  
The red dashed line is the photospheric continuum level derived from
the estimated veiling factor. \label{fig:accfrac}}
\end{figure*}

What fraction of the broad-band optical light output of the stars from
Table~\ref{tab:basicinformation} can be ascribed to the short duration
accretion bursts that are visible in the CoRoT light curves?  How does
the integrated flux from the discrete bursts that are identifiable in
the  light curves compare to the veiling flux inferred from the VLT
spectra? We can make an estimate of this fraction by adopting a
pseudo-continuum level as a function of time for the {\em CoRoT} light
curves, and then integrating the ``flux" above that level.   Setting
that continuum level is somewhat arbitrary, particularly for stars
where there is considerable structure not necessarily due to accretion
bursts.  For our purposes, however, we argue that it is sufficient to
do this estimate in a perhaps ``generous" and simplified manner.  
Specifically, we adopt a constant flux level for a portion (or all) of
the light curve where that constant level seems at least plausible as
the continuum above which the identifiable bursts are superposed.  We
have made these estimates for all six stars of
Table~\ref{tab:basicinformation} for which we have both {\em CoRoT}
light curves in 2011 and VLT/FLAMES spectroscopy. We illustrate our
method in Figure~\ref{fig:accfrac}.   The solid line is our estimated
pseudo-continuum level.   For these six stars, the estimated fraction
of the broad-band optical light from accretion bursts ranges from
4.9\% to 14\%.  In other words, the individually identifiable
accretion bursts are responsible for only a small fraction of the
optical broad-band light, at least by this metric.

As we have noted previously, the characteristic veiling factor for
these stars is $r$ = 0.65, corresponding to a prediction that 39\% of
the optical flux derives from hot gas.   The red dashed lines in
Figure~\ref{fig:accfrac} correspond to the continuum level which would
be needed to match this estimate for the veiling continuum
contribution to the {\em CoRoT} light curves.  We conclude that either
there are a large number of small accretion bursts which blend
together to contribute to the veiling continuum, or there is some
other, more continuous accretion mode which is contributing the
majority of the veiling continuum.   The large number of small bursts
present in the Mon-000474 light curve suggests that the former
explanation may be correct.

\subsection{Timescales and Periodicities for the Accretion Bursts}

Autocorrelation function analysis for time series data provide a
sensitive means to search for periodic signals. We have derived
autocorrelation functions for all of the stars in
Table~\ref{tab:basicinformation}.  Prior to doing that, as is
standard, we have resampled the data to a regular cadence 
and we have subtracted the median (see Cody et al.\ 2014).  

\begin{figure}
\begin{center}
\epsfxsize=.99\columnwidth
\epsfbox{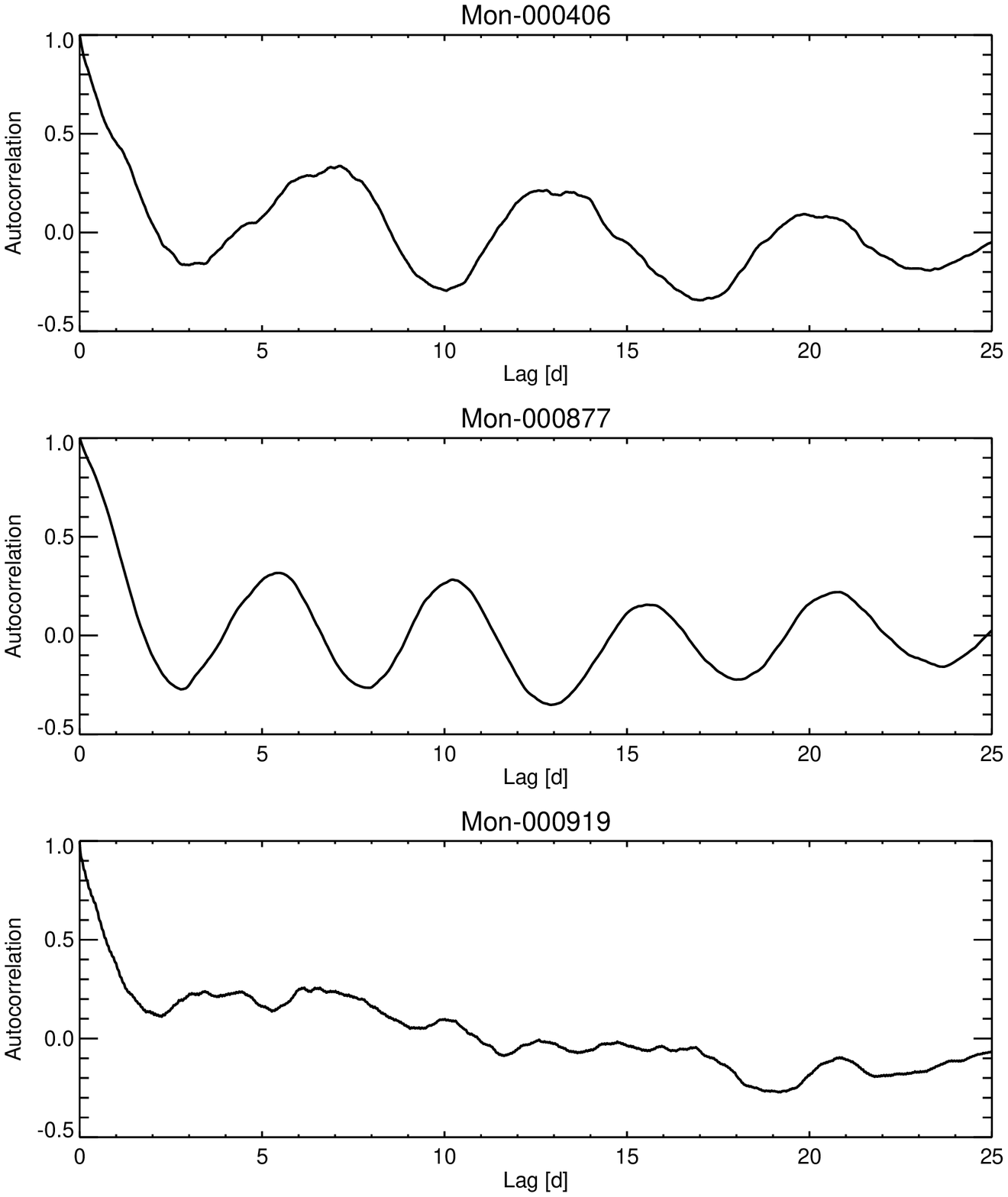}
\end{center}
\caption{Auto-correlation functions (ACF) for two of the accretion burst
stars, Mon-000406 (top) and Mon-000877 (middle). This illustrates
that the bursts for these stars are preferentially clumped together
with a periodic spacing.  The ACF for most of the burst-dominated
light curves do not show any significant evidence of periodic structures,
as illustrated by the ACF for Mon-000919 (bottom). \label{fig:acf}}
\end{figure}

Most of the burst-dominated light curves do not show a significant
periodic signal, i.e., the bursts occur stochastically.   However,
four of the stars do have strong periodicities: Mon-000406 (P=6.2d), 
412 (P=6.6d), 877 (P=4.6d) and 1217 (P=7.0d).  The autocorrelation
functions for two of these stars are shown in Figure~\ref{fig:acf}.
Because of their periodicity, one might associate these light curves
with funnel-flow accretion channeled by the stellar dipole magnetic
field. However, these light curves are quite unlike what is commonly
envisioned for stable funnel-flow accretion (see Figure 7a in KR08)
where the variations are more nearly sinusoidal, but they could
represent an intermediate state where there is a dominant matter flow
controlled by the magnetic dipole, but disk instabilities lead to an
unsteady flow of gas towards the stellar surface (quasi-periodic
variability of this type was also predicted by KR08; D'Angelo \&
Spruit 2010 propose another accretion disk instability which can
produce periodic accretion bursts).    Whatever is driving the
periodicity in these stars is apparently long-lived, because three of
the four (Mon-000406, 877 and 1217) have very similar periods
published in the literature (L04; Makidon et al.\ 2004).   We
assume the derived periods correspond to the photospheric rotation
period for these stars.

An interesting, but difficult to quantify, statistic to determine for
the burst-dominated light curves is the average frequency for the
bursts.  Based on the analysis of the high signal-to-noise, high
cadence light curve for Mon 474, the number of bursts one can identify
is a strong function of the light curve quality.  Therefore, it is
important to utilize a quantitative algorithm to provide a consistent,
though somewhat arbitrary, burst frequency.  For this purpose, we use
the PeakFind routine defined and described in Cody et al.\ (2014).  
This algorithm counts the number of peaks in a light curve that differ
in  magnitude by more than a particular amplitude. The number of peaks
is then divided by the total time baseline of the light curve to
produce a timescale. Repeating for a grid of amplitudes results in a
timescale distribution for peaks of different minimum size. For the
burst-dominated light curves, PeakFind does a reasonably good job of
identifying the same relatively strong, relatively isolated peaks that
were measured to create Figure~\ref{fig:mon474widths}. For the three
burst-dominated light curves shown in  Figure~\ref{fig:sixctts},
PeakFind identified five (Mon-000567), eight (Mon-000808) and ten
(Mon-001187) bursts.   For the entire set of stars in
Table~\ref{tab:basicinformation}, the median burst frequency
determined by PeakFind was 0.2 peaks per day, corresponding to $\sim$8
peaks in the CoRoT observing window.  This is roughly comparable to
the burst frequency displayed in the model stochastic light curves of
R12 and KR08.

\section{Population Statistics - the Frequency of CTTS Light Curve Types}

In Section 4, we noted that amongst the stars with the largest UV
excesses, the stars with accretion burst light curves were by far the
dominant light curve class (at least 15 out of 27 stars), and that only 
one of the extreme UV excess stars could be interpreted as being
dominated by a stable high-latitude hot spot as is expected from
funnel-flow accretion.   Here we conduct a more complete comparison of
the frequency of light curve types as it  relates to the accretion
process.

\begin{figure*}
\begin{center}
\includegraphics[scale=1.2]{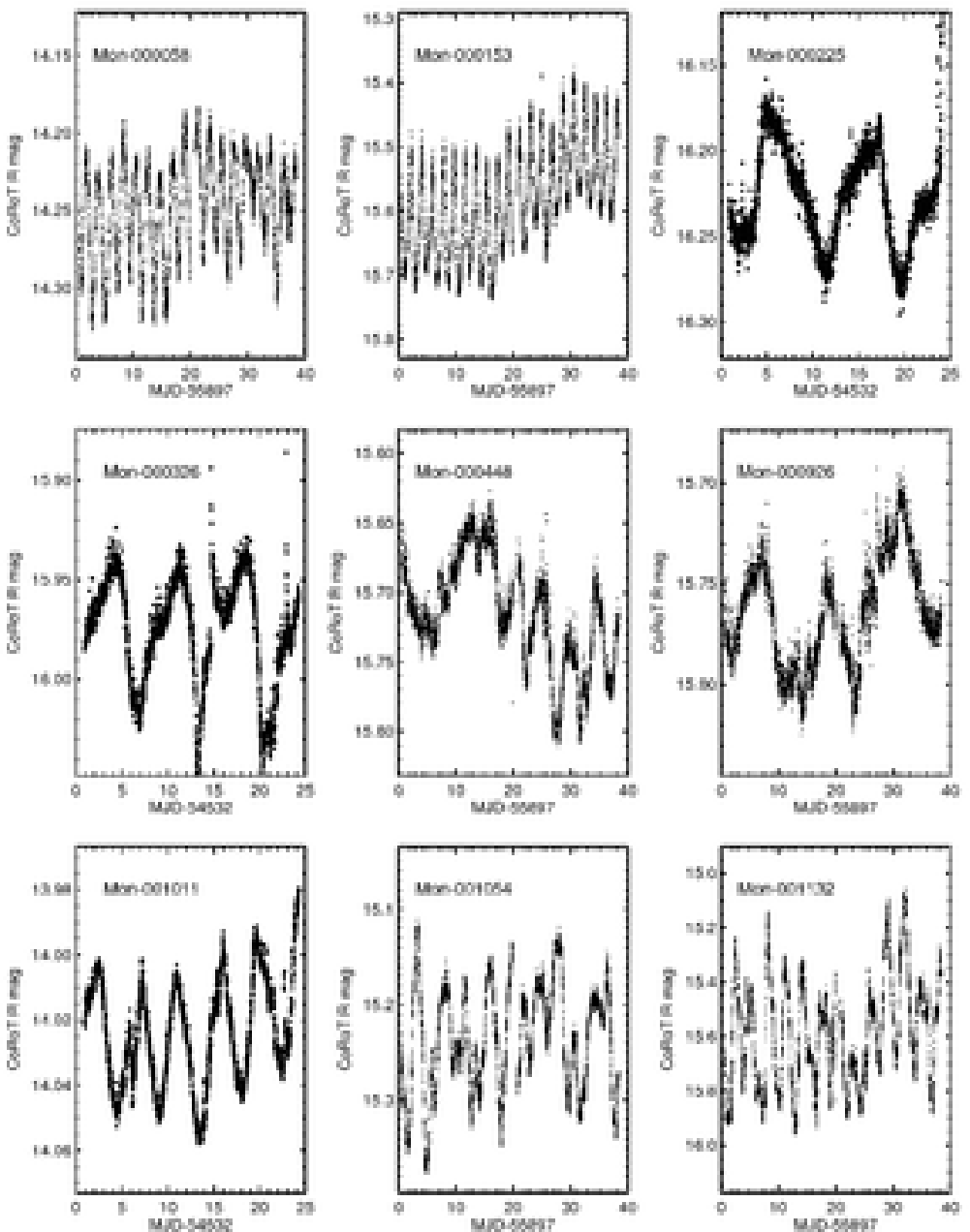}
\end{center}
\caption{Hot spot candidates selected from the $u-g$, $g-r$ diagram.  Based
on their UV excesses and other data we have, these are the most likely
NGC 2264 YSOs to have relatively stable hot spots due to funnel-flow
accretion. \label{fig:hotspot1}}
\end{figure*}

To create a larger sample of CTTS from which to determine the relative
frequency of various light curve types, we have drawn a second
boundary in the $u-g$, $g-r$ plane, meant to separate stars with
certain UV excesses from those which have either no UV excess or only
a slight UV excess. This line, approximately parallel with the locus
of WTTS in the $u-g$, $g-r$ plane, is shown as a dotted line in
Figure~\ref{fig:colorcolorf4}d.   There are 60 stars which we believe
are NGC~2264 members and for which we have {\em CoRoT} light curves whose
$ugr$ colors place them below the line.   We have made a visual
characterization of all 60 light curves, which we summarize in
Table~\ref{tab:uvx}.  Only nine of these stars have light curves whose
shapes are possibly consistent with that expected for a relatively
stable high-latitude hot spot; these light curves are shown in
Figure~\ref{fig:hotspot1}.   To fall into this category, we only
require that the light curve shows apparently periodic structure, and
that it is not better categorized as an AA Tau-analog or an accretion
burst star. 

\begin{deluxetable*}{ccc}
\tabletypesize{\scriptsize}
\tablecolumns{3}
\tablewidth{0pt}
\tablecaption{Light Curve Types for Stars with Definite UV Excess
\label{tab:uvx}}
\tablehead{
\colhead{Light Curve Type}  & \colhead{N} & \colhead{Description} 
}
\startdata
Constant &   2  &  Essentially constant \\
Low amplitude variable  &   5  & definitely variable, but noisy and ampl. small  \\
Stochastic & 9   & highly variable (see Cody et al.\ 2014) \\
Burst-dominated &  24  & 22 stars from Table~\ref{tab:basicinformation} + 2 additional low-ampl bursts   \\
Extinction dip stars   &  10  &    \\
Hot Spot Candidates   &  9 & see Figure~\ref{fig:hotspot1}   \\
\enddata
\tablecomments{Stochastic = adopted from
Cody et al.\ 2014,  stars whose light curves show often large
amplitude variability but without preference for positive or negative
flux changes.   These stars probably have several mechanisms
contributing to the observed variability.}
\end{deluxetable*}

Can we be sure that the light curve shapes in
Figure~\ref{fig:hotspot1} are, in fact, the result of hot (and not
cold) high-latitude spots? For WTTS, cold spots are generally stable
over many rotation periods.  Herbst et al.\ (1994) and others have
reported that spotted light curves on CTTS are much less stable,
possibly because they are the result of hot spots that result from
intrinsically variable accretion. There is, therefore, a prejudice to
suspect that the more consistent the light curve shape from one period
to the next, the more likely that cold spots are responsible. 
However, cold spots may behave differently on CTTS, so this is not a
certain conclusion.  Independent of light curve shape, well-sampled
synoptic $U$-band photometry or high resolution spectroscopy can
discriminate between cold and hot spots, but our access to such data
for these stars is limited.   What we can do is simply ask that these
hot spot candidates have properties that are consistent with having
accretion contribute significantly to their light curve shape, using
the accretion burst stars as a guide. 

First, we require that the star
have evidence of warm dust (as do all of the stars in
Table~\ref{tab:basicinformation}) -- that is, it does not lie in the
Class III locus in the IRAC color-color diagram, or if not all four
IRAC bands have data, then there is evidence in the SED for an IR
excess beginning somewhere shortward of $\sim$10 microns.  If we have
VLT spectra (true for five of these stars), then the star must be
detectably veiled -- $r >$ 0.2; we also expect these  stars to show
\ion{He}{1} $\lambda$6678 in emission (as do all of the stars from
Table~\ref{tab:basicinformation} with VLT spectra).  If we have
H$\alpha$ profiles, then the profile must be detectably structured or
broad. All nine of these candidates pass these tests, where we have
the appropriate data.  Table~\ref{tab:hotspotinfo}
provides the coordinates and other information for these
stars.  These nine stars are the best candidate
funnel-flow accretors among the 60 stars with certain UV excesses,
though it is possible that for some stars the light curve shape could
be due to evolving cold spots or other mechanisms.   We conclude that
for the NGC~2264 members with certain UV excesses, light curve
morphologies suggestive of funnel-flow accretion occur only about 1/3
as often as light curves dominated by short-duration accretion
bursts.   Depending on what fraction of these stable hot spot
candidates are confirmed, funnel-flow accretion {\bf may} dominate at
relatively low accretion rates, matching the prediction of KR08.

Cody et al.\ (2014) used an autocorrelation function technique to
identify NGC~2264 CTTs whose CoRoT light curves are periodic or
quasi-periodic.  The latter are light curves where there is a
periodic signal, but either the shape or amplitude of the waveform
changes significantly over the 40 day CoRoT short-run (see Cody
et al.\ 2014 for the quantitative details).  Only five CTTs were identified
as having periodic CoRoT light curves.  All five of these stars fall
within the WTT locus in the $u - g$ vs. $g - i$ color-color diagram,
indicating very low (or no) accretion and making it likely that their
light curve shapes are probably due to cold spots.  All but two or
three of the 27 quasi-periodic CTT's identified by Cody et al.\ (2014) fall in the
$u - g$ vs.  $g - i$ diagram above the dashed red line of Fig. 4d,
but in most cases below the WTT locus.  Therefore, they are probably accreting
but at a low rate; their light curve shape could be due either to cold
or hot spots.   In any event, they reinforce the conclusion
that stable, funnel-flow accretion - when present - is likely only to
occur for stars with low accretion rates.

\section{Conclusions and Speculation} 

Previous synoptic photometric surveys of star-forming regions have
provided many useful results constraining the physical properties of
YSOs.   However, ours is the first such survey to have the necessary
cadence, sensitivity, duration, wide-field and multi-band nature to
unmask the dominant modes by which YSOs accrete gas from their
circumstellar disk.   Our study has also benefited from having
high-quality, multi-wavelength photometry from $u$ to 24 $\mu$m for
most of the NGC~2264 members, and high-resolution spectroscopy at
least at H$\alpha$, also for a large fraction of the stars of
interest.

Using the $u-g$, $g-r$ two-color diagram to identify  the most
actively accreting YSOs in NGC~2264, 
we have shown that amongst the YSOs in NGC 2264 with the largest UV 
excesses, the dominant lightcurve signature  is short duration
accretion bursts as exemplified by the stars of
Table~\ref{tab:basicinformation}.   Usually these accretion bursts
occur in a seemingly stochastic pattern at least over 40 day
timescales; however, for a small fraction of the stars, the accretion
bursts are clumped  together, with the clumps recurring at a period
which is plausibly the rotation period of the star.  Based on
theoretical models, at lower accretion rates we expected to find other
accreting stars with long-lived high latitude hot spots (from
funnel-flow accretion) whose light curves would more closely resemble
the spotted light curves of WTTS.   We do find a set of candidate
stable hot-spot light curves which may correspond to this prediction,
but we do not have sufficient data to confirm that hot rather than
cold spots are responsible for their variability. 

The two most frequent light curve types among the CTTS with
significant UV excesses are the accretion burst group and the variable
extinction group. The latter consists of stars with periodic flux dips
(AA Tau analogs) and stars with similar flux dips but no obvious
periodicity; in both cases, the flux dips are likely associated with
transient structures in their inner circumstellar disk.   By their
nature, the variable extinction stars must have their disks oriented
at relatively large inclination angles to our line of sight.  However,
we have also discovered several other physical properties which
distinguish the two groups.  Based on their spatial locations, the
variable extinction stars appear to be, on average, older than the
accretion burst stars.   The variable extinction stars also have
smaller UV and IR excesses, perhaps reflective of their greater age. 
The H$\alpha$\ profiles of the accretion burst group are, for the most
part, centrally peaked and modestly structured, with the most common
additional feature being blue-shifted absorption troughs.  The
H$\alpha$\ profiles of the variable extinction group are significantly
more structured, with about a quarter showing red-shifted absorption
troughs. When we have the appropriate data, the accretion burst stars
all have HeI 6678 in emission, whereas the stars with variable
extinction generally  show an absorption feature, presumably due to
FeI, at that wavelength.

Using the normal cadence {\em CoRoT} data, the typical total durations
for isolated accretion bursts amongst the stars of
Table~\ref{tab:basicinformation} are about one day, as predicted by
the models of R12.   However, for the one star from
Table~\ref{tab:basicinformation} where we have much higher cadence
{\em CoRoT} data, we identify many shorter duration, low amplitude 
bursts that we believe are also best explained as accretion bursts,
with total durations as small as a few hours.  For at least this one
star, which has a much earlier spectral type than the other stars with
burst-dominated light curves, the burst distribution function is
dominated by these small bursts.   For the six stars from
Table~\ref{tab:basicinformation} for which we have VLT spectra and
therefore can estimate veiling, we find that only a small fraction of
the $R$-band excess light that arises from hot gas can be attributed
to the directly identifiable bursts.  This  could indicate that most
of the accretion flux arises from a large number of short duration,
low amplitude bursts (a hypothesis reminiscent of the debate over
whether nano and microflares power the coronae of low mass stars
-- Hudson 1991; Tajfirouze \& Safari 2012.

\section*{Appendix}

\subsection*{CoRoT and Spitzer Light Curves for CTTS in NGC~2264}

While we display light curves for small subsets of NGC~2264 CTTS in
the main body of the text, we believe it is important to provide light
curves for the full set of stars we have discussed.   The next three
figures provide the full light curves for all of the stars identified
as having burst-dominated light curves, plus light curves for the
stars with the most extreme UV excesses whose light curves were not
identified as being burst-dominated (see \S 4).

\begin{figure*}
\begin{center}
\epsscale{1.0}
\includegraphics[scale=1.2]{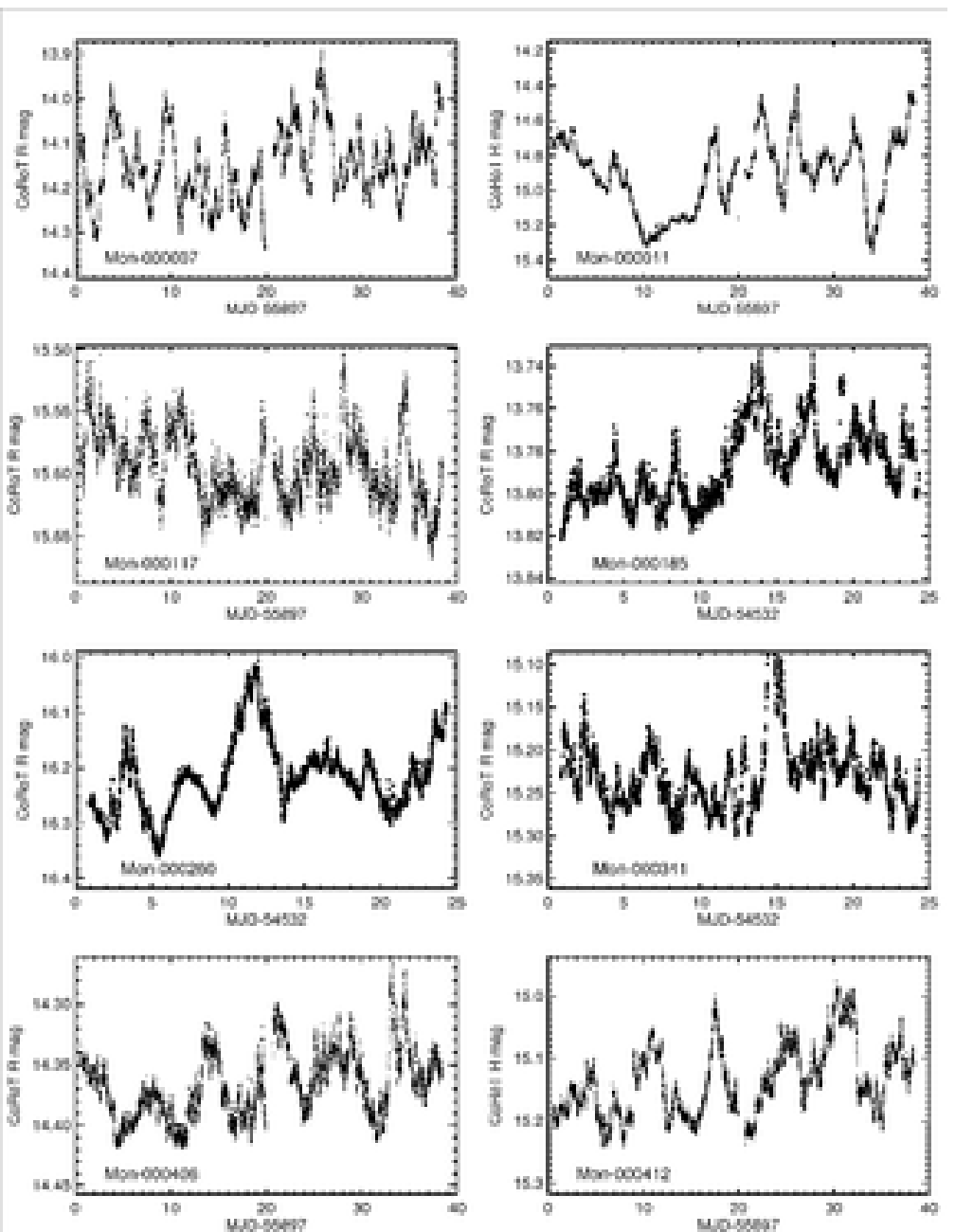}
\end{center}
\caption{{\em CoRoT} light curves for all the stars of
Table~\ref{tab:basicinformation}.  We have converted the {\em CoRoT}
``counts" into an estimated $R$ magnitude using a zero point of 26.74.
\label{fig:allcorotlcs}}
\end{figure*}

Figure~\ref{fig:allcorotlcs} provides the {\em CoRoT} light curves for
all stars from Table~\ref{tab:basicinformation}.  Where we have data
from both epochs, we show the light curve that best illustrates the
accretion burst class.  

\addtocounter{figure}{-1}
\begin{figure*}
\begin{center}
\includegraphics[scale=1.2]{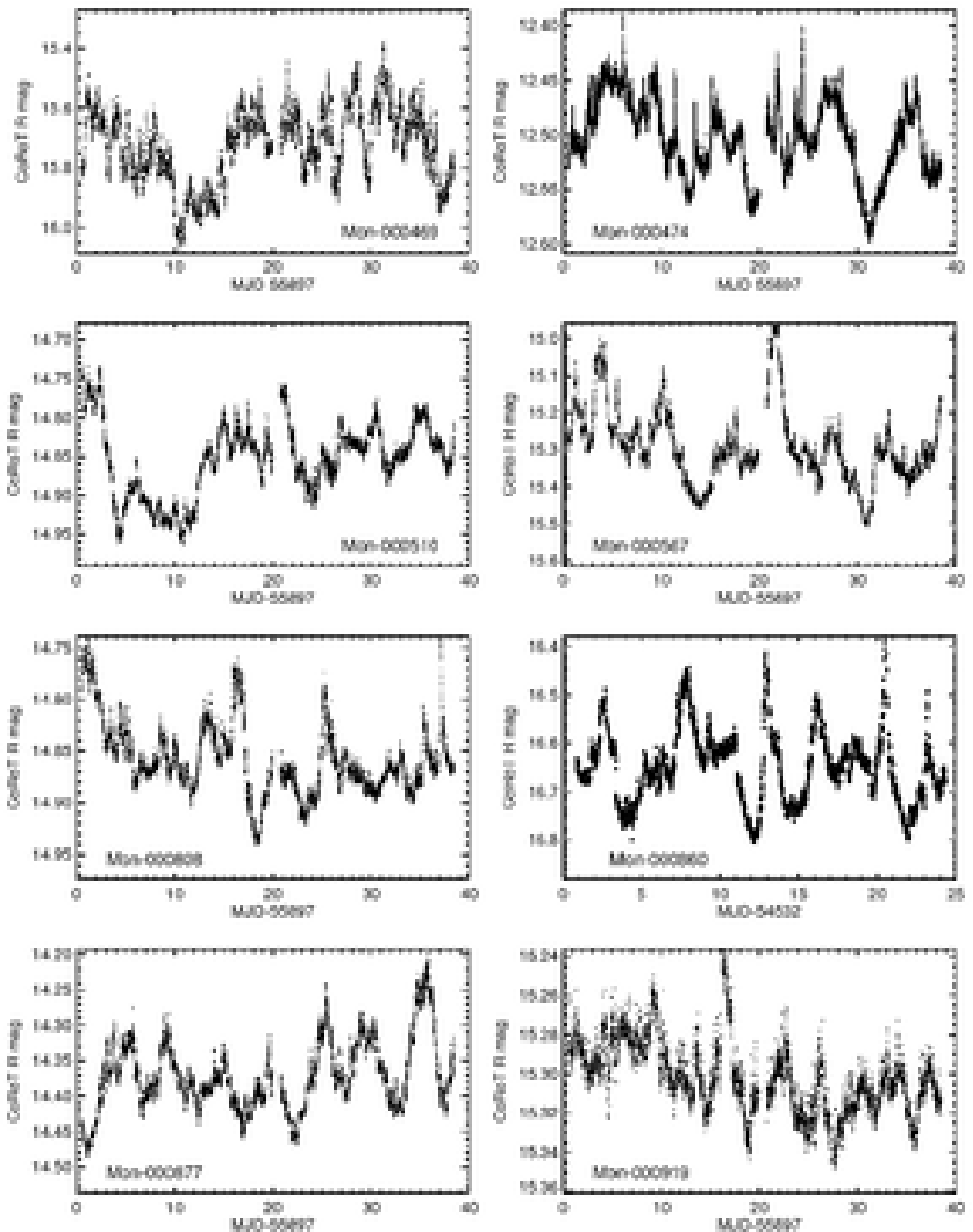}
\end{center}
\caption{{\em CoRoT} light curves, cont'd.
\label{fig:allcorotlcs2}}
\end{figure*}

\addtocounter{figure}{-1}
\begin{figure*}
\begin{center}
\includegraphics[scale=1.2]{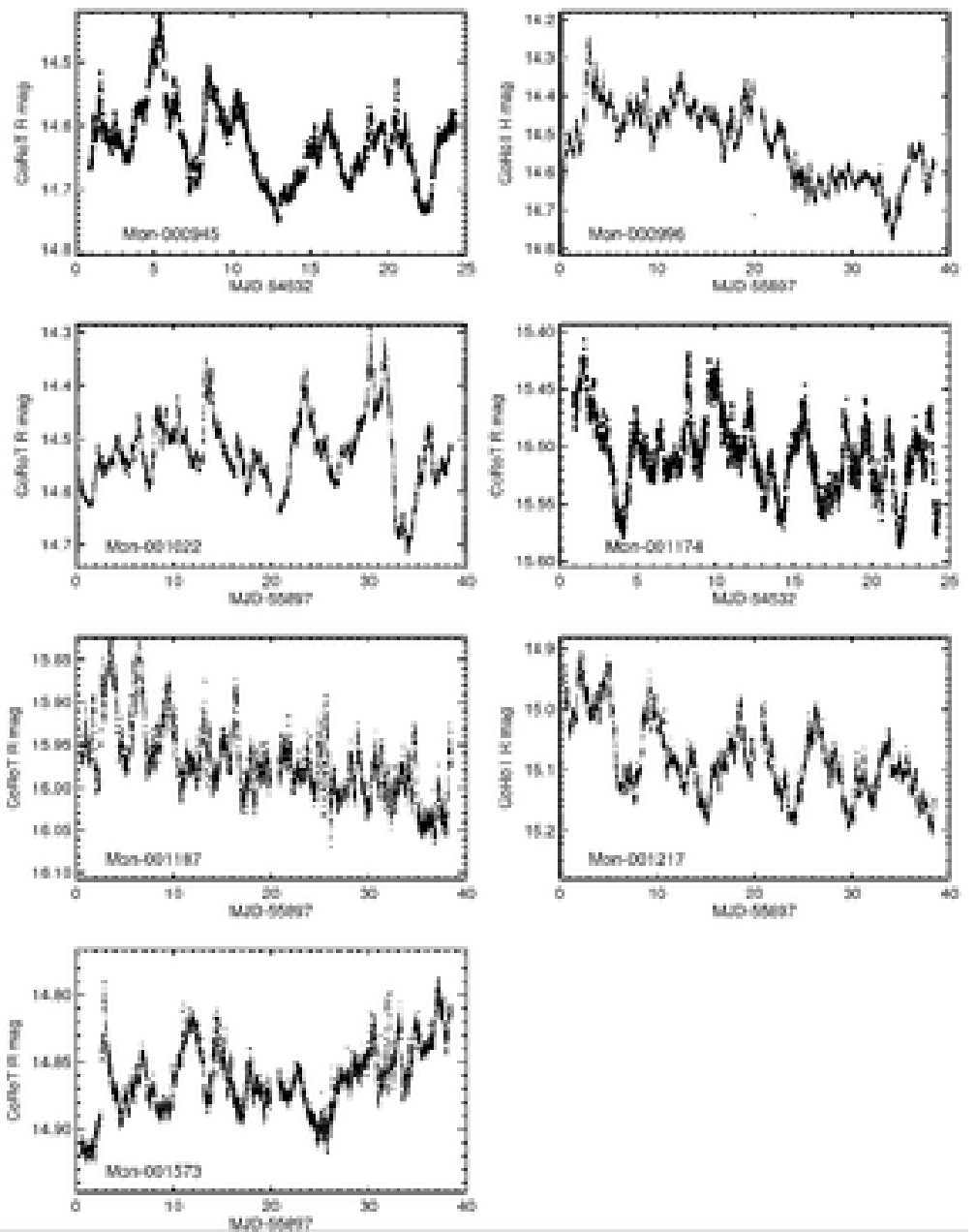}
\end{center}
\caption{{\em CoRoT} light curves, cont'd.
\label{fig:allcorotlcs3}}
\end{figure*}

\begin{figure*}
\begin{center}
\includegraphics[scale=1.2]{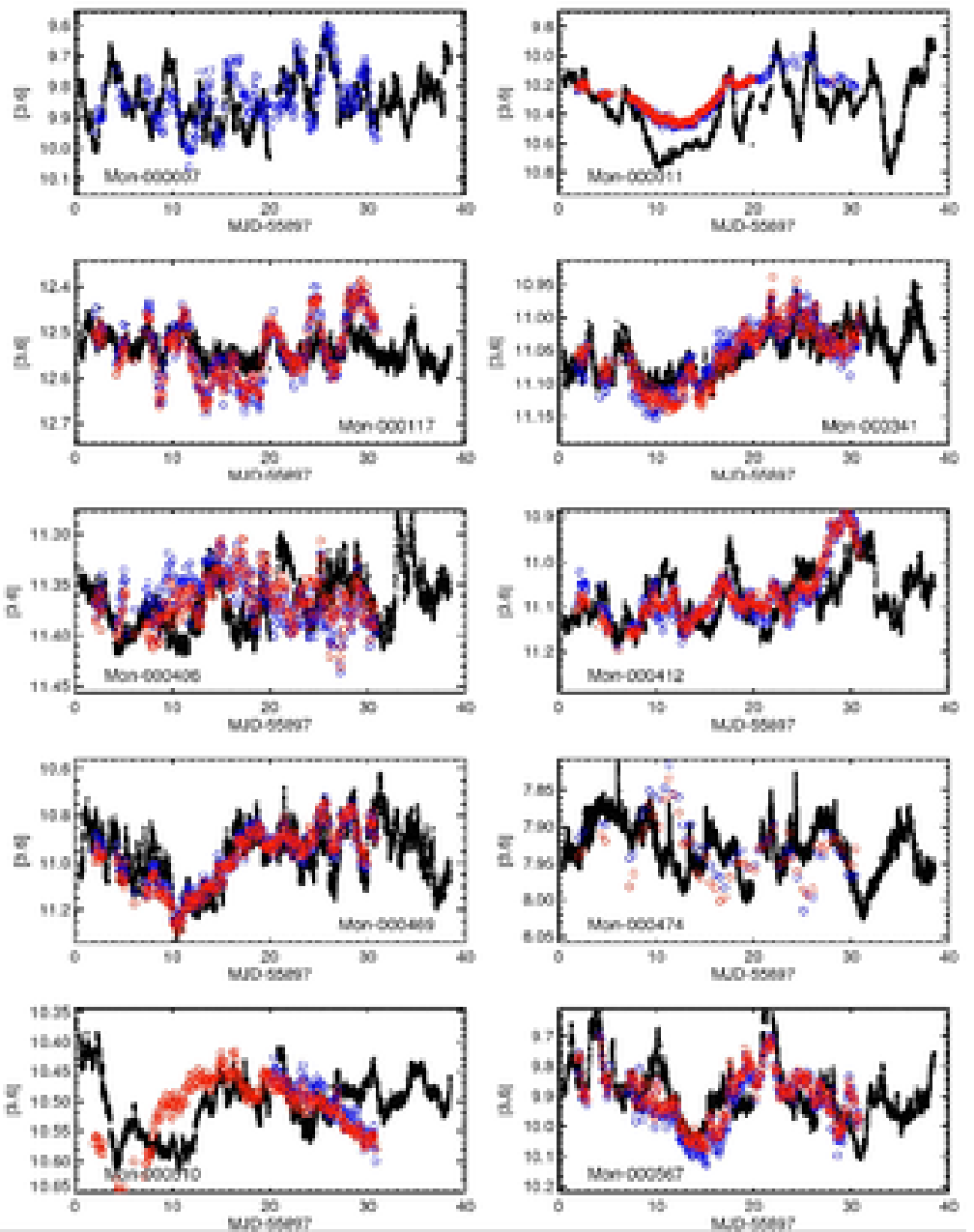}
\end{center}
\caption{2011 light curves for stars in Table~1 with both {\em Spitzer} and {\em CoRoT} data.
Blue circles are {\em Spitzer} 3.6 $\mu$m, red circles are {\em Spitzer} 4.5 $\mu$m, and
black dots are CoRoT data.
See discussion in \S 5.4.  In most cases, there is a fair correlation between the
shape and amplitudes of the light curves at the two wavelengths.
\label{fig:corot2011}}
\end{figure*}

Figure~\ref{fig:corot2011} shows the 2011 light curves for which we
have both {\em CoRoT} and {\em Spitzer} data.  The center of the $y$-axis scale is
set to the median magnitude for each band; the range in $y$ is the
same for each of the bands that are shown (for example, if the {\em CoRoT}
$y$-axis scale was from 14.6 to 14.0, and the [3.6] median magnitude
were 10.7, then the {\em Spitzer} $y$-axis scale would run from 11.0 to
10.4). 

\addtocounter{figure}{-1}
\begin{figure*}
\begin{center}
\includegraphics[scale=1.2]{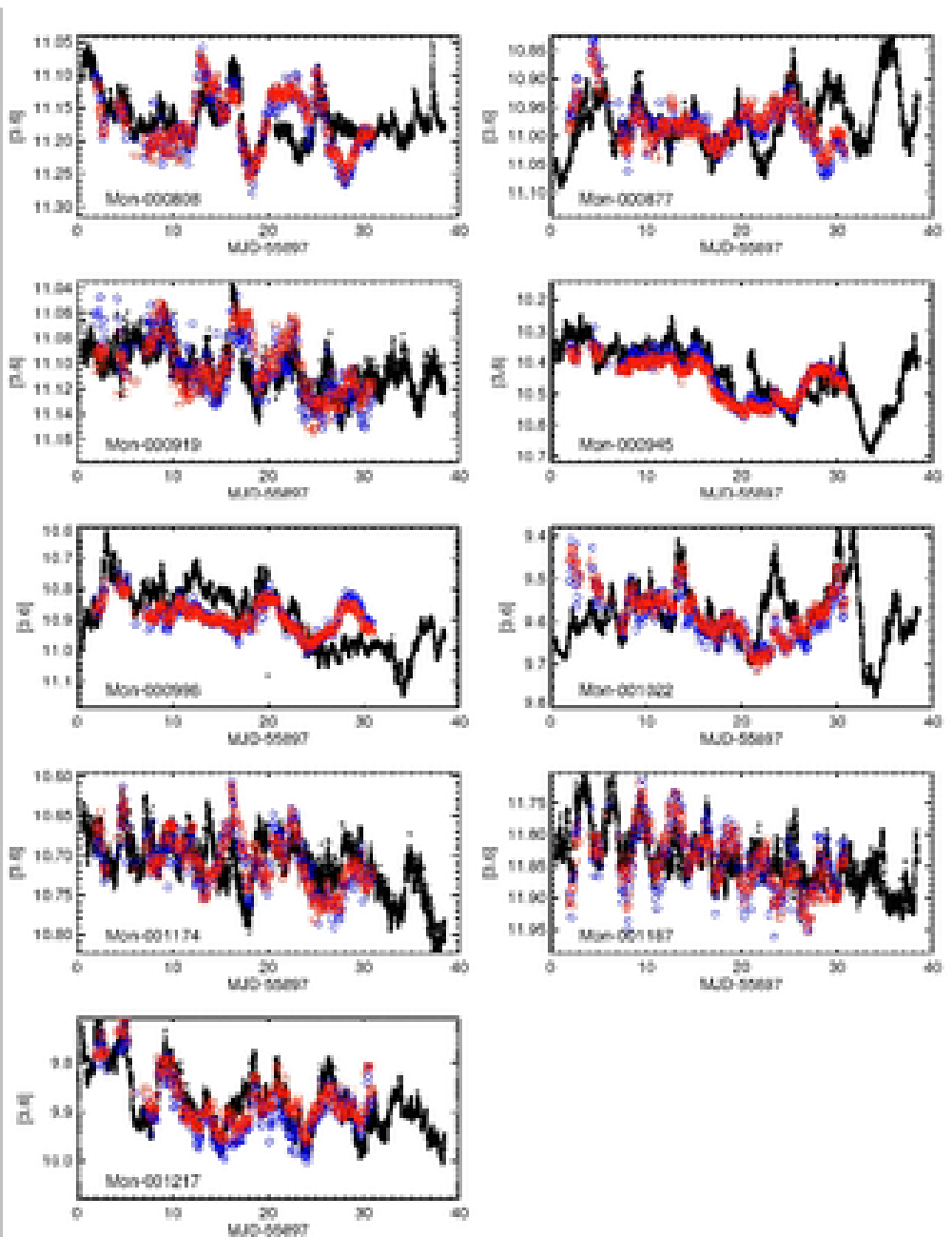}
\end{center}
\caption{2011 light curves, cont'd \label{fig:corot2011p2}}
\end{figure*}

Figure~\ref{fig:bigexcesses} shows the {\em CoRoT} light curves for the twelve
stars with very large UV excess that were not identified  as having
accretion burst dominated light curves.

\begin{figure*}
\begin{center}
\includegraphics[scale=0.8]{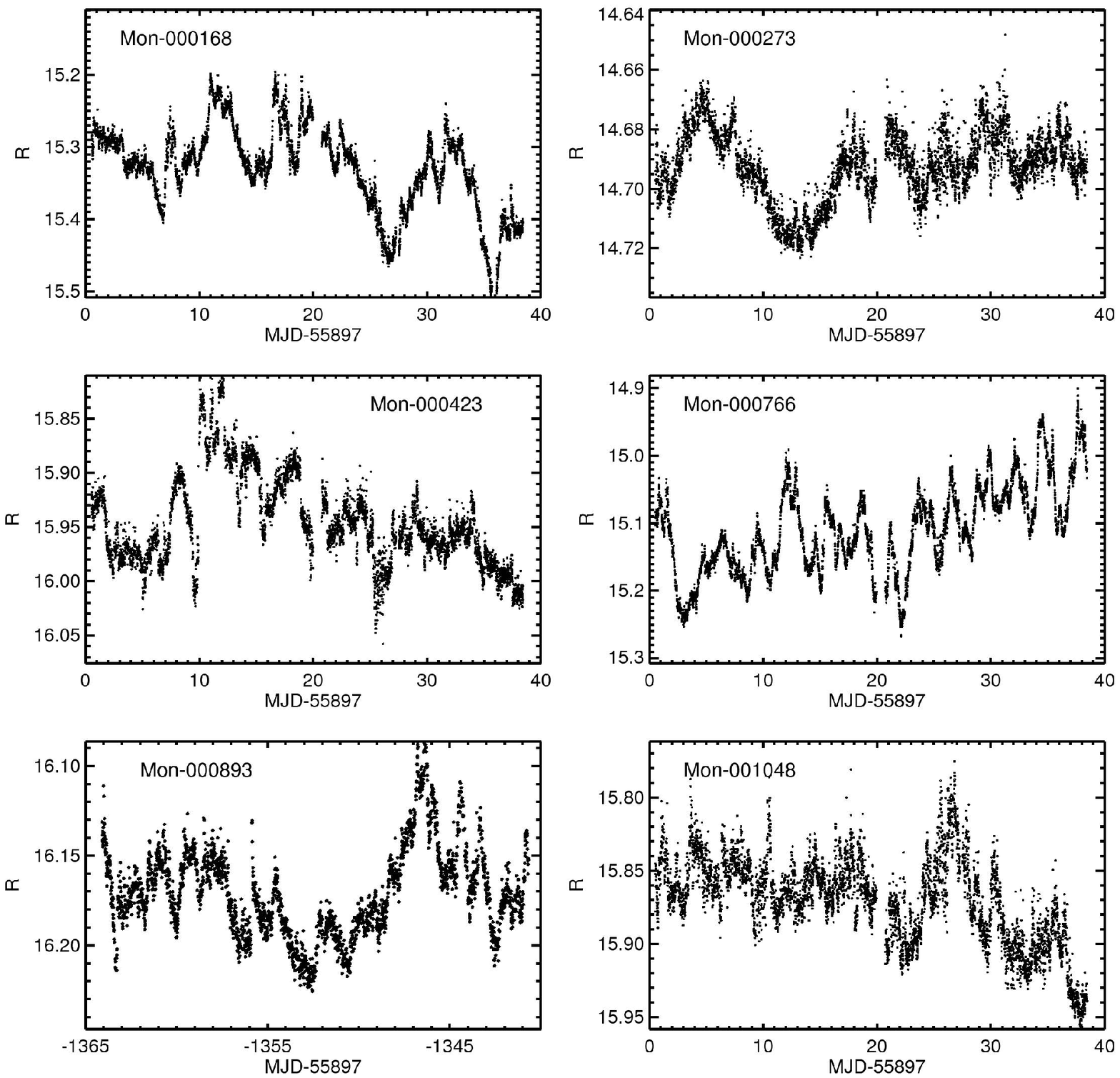}
\end{center}
\caption{{\em CoRoT} light curves for all 12 additional stars with extreme
UV excesses as defined in Figure~\ref{fig:colorcolorf4}, but which are
not included in Table~\ref{tab:basicinformation}.  Some of these light curves
do show flux bursts (e.g. Mon-000168) but were not included in the main sample
due to their having too much additional structure or too high a noise level.
See discussion in \S 4.
\label{fig:bigexcesses}}
\end{figure*}

\addtocounter{figure}{-1}
\begin{figure*}
\begin{center}
\includegraphics[scale=0.8]{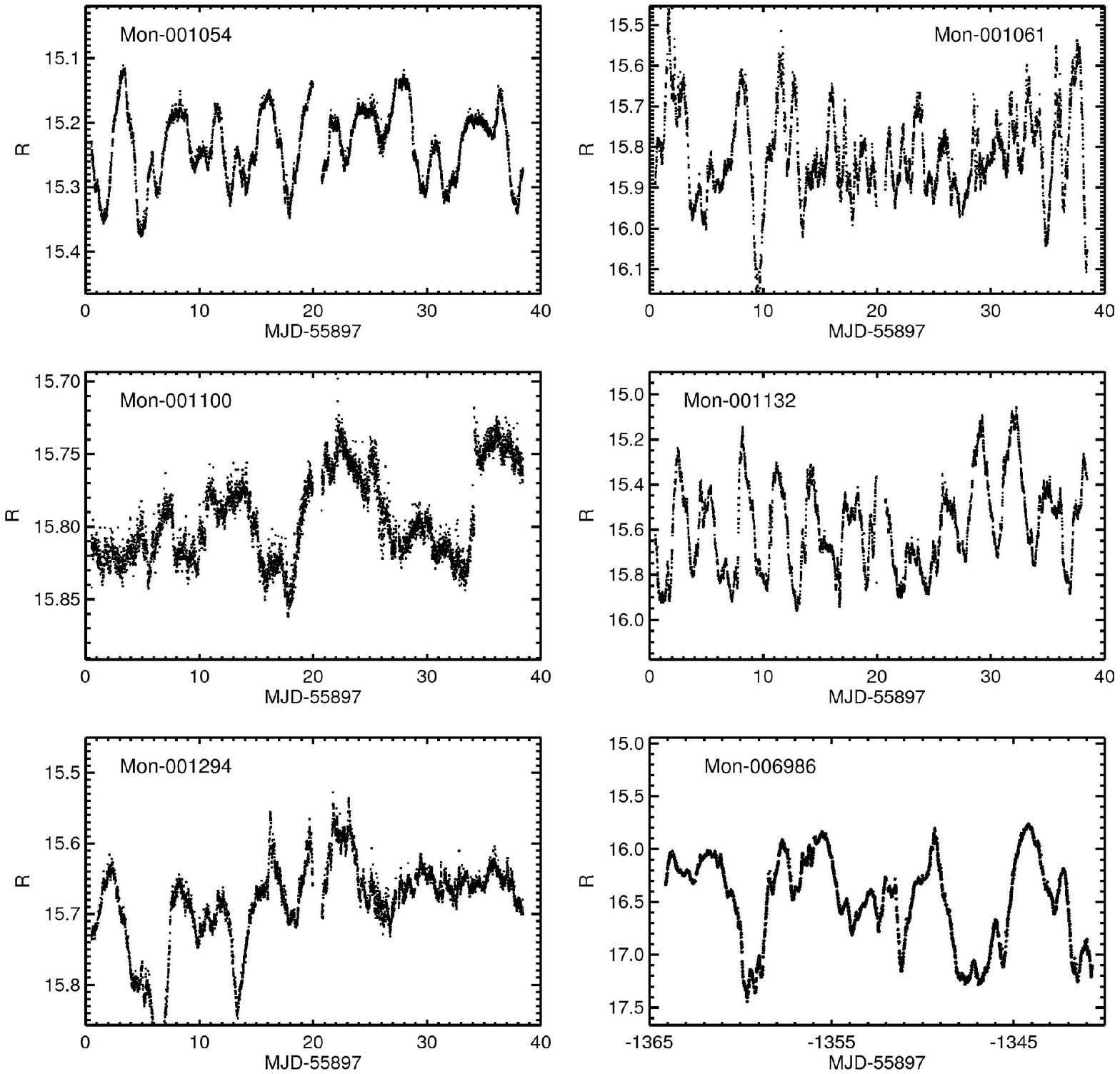}
\end{center}
\caption{--Cont.
\label{fig:bigexcessesp2}}
\end{figure*}

\vfill
\eject
\subsection*{Short and Long-Term Variability of H$\alpha$\ Profiles
for the Stars of Table 1}
We have MMT Hectochelle spectra for nearly all of the stars showing 
accretion burst light curves.   For that reason, we primarily show
Hectochelle data in Figure~\ref{fig:halphaprofiles}.   The MMT data
were obtained six or seven years prior to our light curves; we would
prefer to be able to show H$\alpha$ profiles obtained during our
photometric monitoring campaign.  For six of the stars of
Table~\ref{tab:basicinformation}, we have such spectra from the
VLT/FLAMES spectrograph (generally, six of the VLT spectra were
obtained during the {\em CoRoT} monitoring campaign, with the other fourteen
spectra obtained during January and February 2012).   Five of those
six stars also have MMT H$\alpha$\ profiles.   We can use those five
stars to address the extent to which the H$\alpha$\ profiles change on
both long and short-term timescales, and therefore the utility of
using the profile shapes from 2004/2005 to help interpret the 2011
light curves. Figure~\ref{fig:halpha2004-2012} provides this comparison.
The conclusion we draw from this comparison is that, in general, the profile shapes
do not seem to change their shapes drastically over either short (days to weeks)
or long (years) timescales.

\begin{figure*}
\begin{center}
\epsscale{1.0}
\includegraphics[scale=0.8]{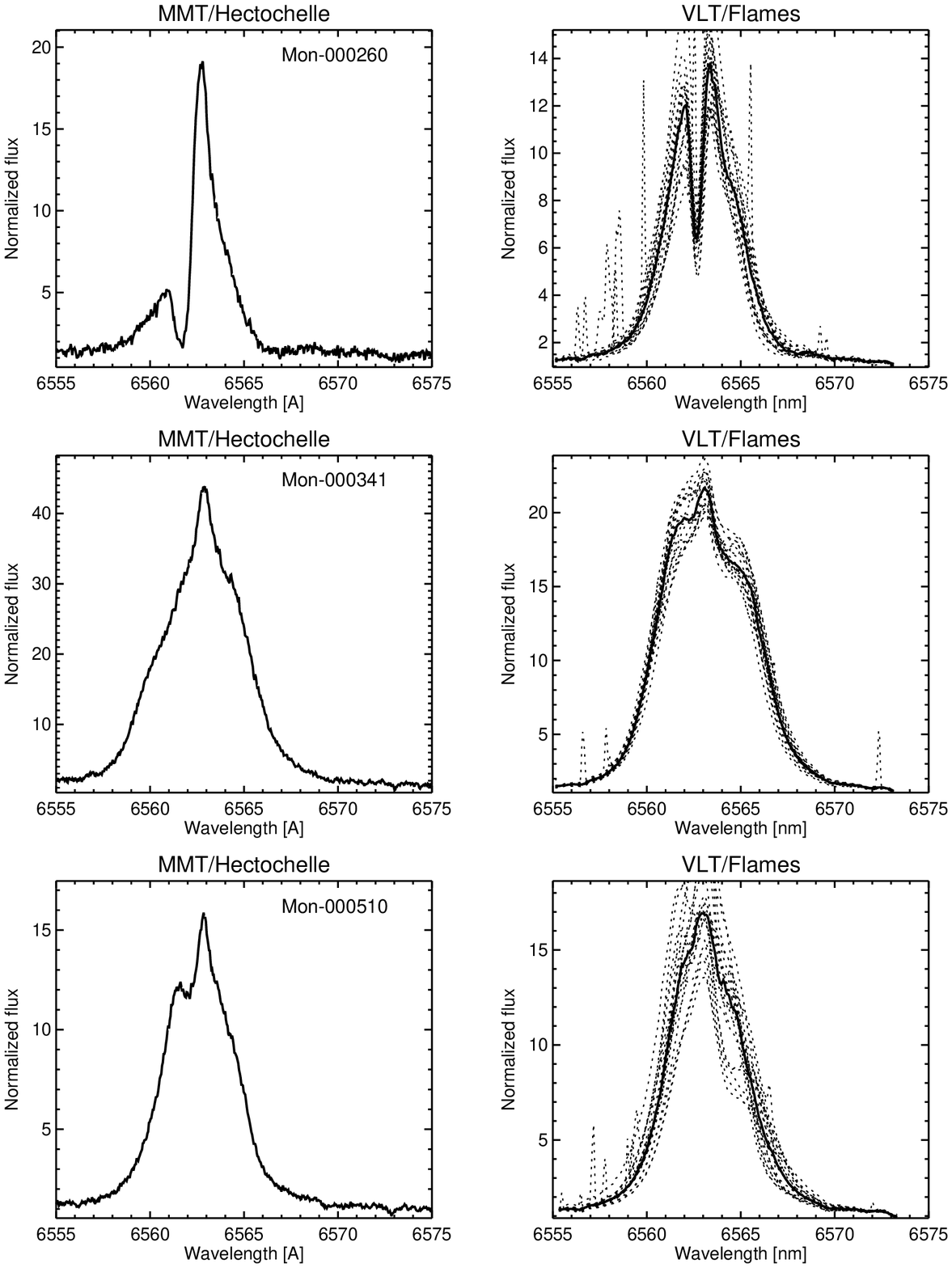}
\end{center}
\caption{Comparison of H$\alpha$\ profiles for objects with both 
MMT/Hectochelle data from 2004 or 2005 and VLT/Flames data from
2011/2012. For the VLT profiles, the dashed lines correspond to individual
epoch spectra, whereas the solid line corresponds to the average profile. \label{fig:halpha2004-2012}}
\end{figure*}

\addtocounter{figure}{-1}
\begin{figure*}
\begin{center}
\epsscale{1.0}
\includegraphics[scale=0.8]{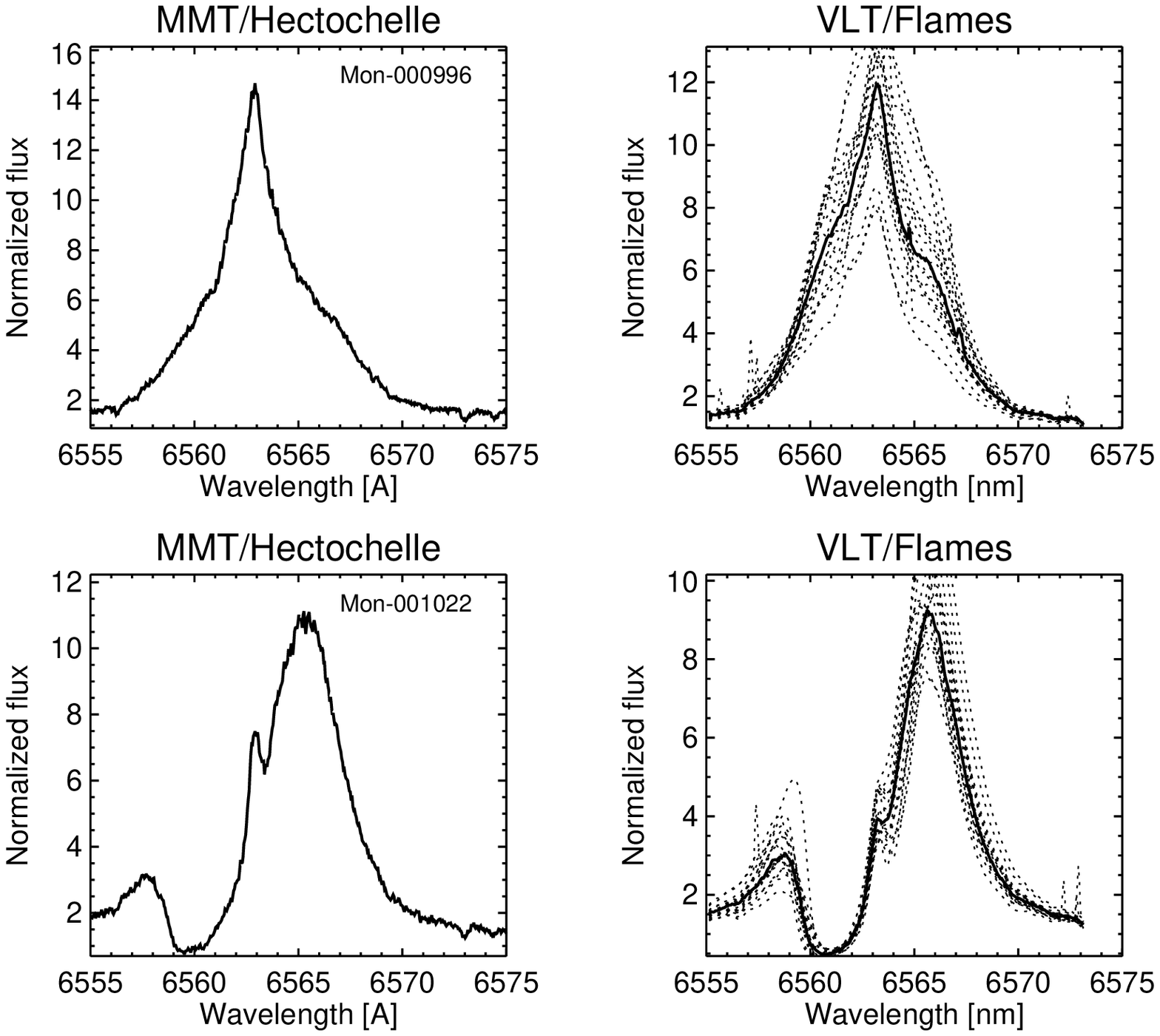}
\end{center}
\caption{cont.}
\end{figure*}

\subsection*{H$\alpha$\ Profiles for Stars with Variable Extinction Events 
in their {\em CoRoT} Lightcurves}

In several of the figures in the main body of the text, we have
adopted a set of stars to use as a comparison group to the accretion
burst group.   This comparison set is composed of stars whose {\em CoRoT}
light curves have prominent flux dips in either 2008 or 2011.  In some
cases, these are AA Tau analogs, where the flux dips are generally
broad and structured and occur approximately periodically.   In other
cases, the flux dips do not seem to show any obvious periodicity, and
we suspect the variable extinction arises from some type of disk
instability. In most cases, the light curves outside  the dips shows a
relatively constant continuum level.  Assuming these flux dips are due
to portions of the circumstellar disk passing through our line of
sight, which is the standard model, these stars must be viewed at
relatively high inclination.

\begin{figure*}
\begin{center}
\epsscale{1.0}
\includegraphics[scale=0.9]{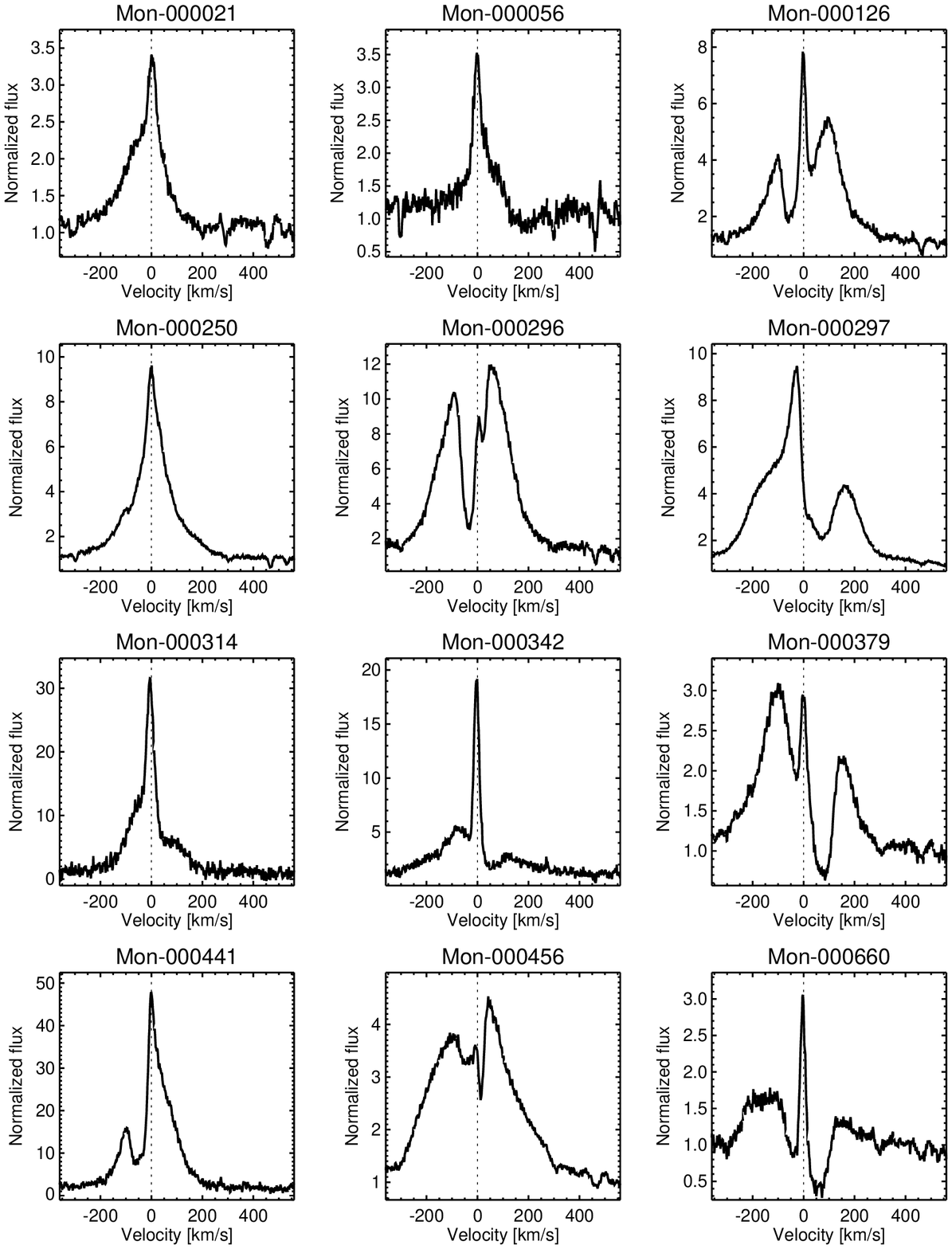}
\end{center}
\caption{H$\alpha$\ profiles of stars whose {\em CoRoT} lightcurves indicate the
presence of prominent extinction dips.  The narrow H$\alpha$\ emission peaks at rest
velocity in Mon 126, 314, 342, 372, 660, 811 and 824 may be spurious because of the
lack of sky subtraction in the MMT spectra. \label{fig:halphaextinction}}
\end{figure*}

We have either MMT Hectochelle or VLT spectra of most of the stars
that fall into this category.  Figure~\ref{fig:halphaextinction}
provides the H$\alpha$ profiles for these stars.   Compared to the
H$\alpha$ profiles of the  accretion burst group shown in
Figure~\ref{fig:halphaprofiles}, it is apparent that the variable
extinction group often have red-shifted absorption dips in their profiles,
which are completely absent from the accretion burst group.

\addtocounter{figure}{-1}
\begin{figure*}
\begin{center}
\epsscale{1.0}
\includegraphics[scale=0.9]{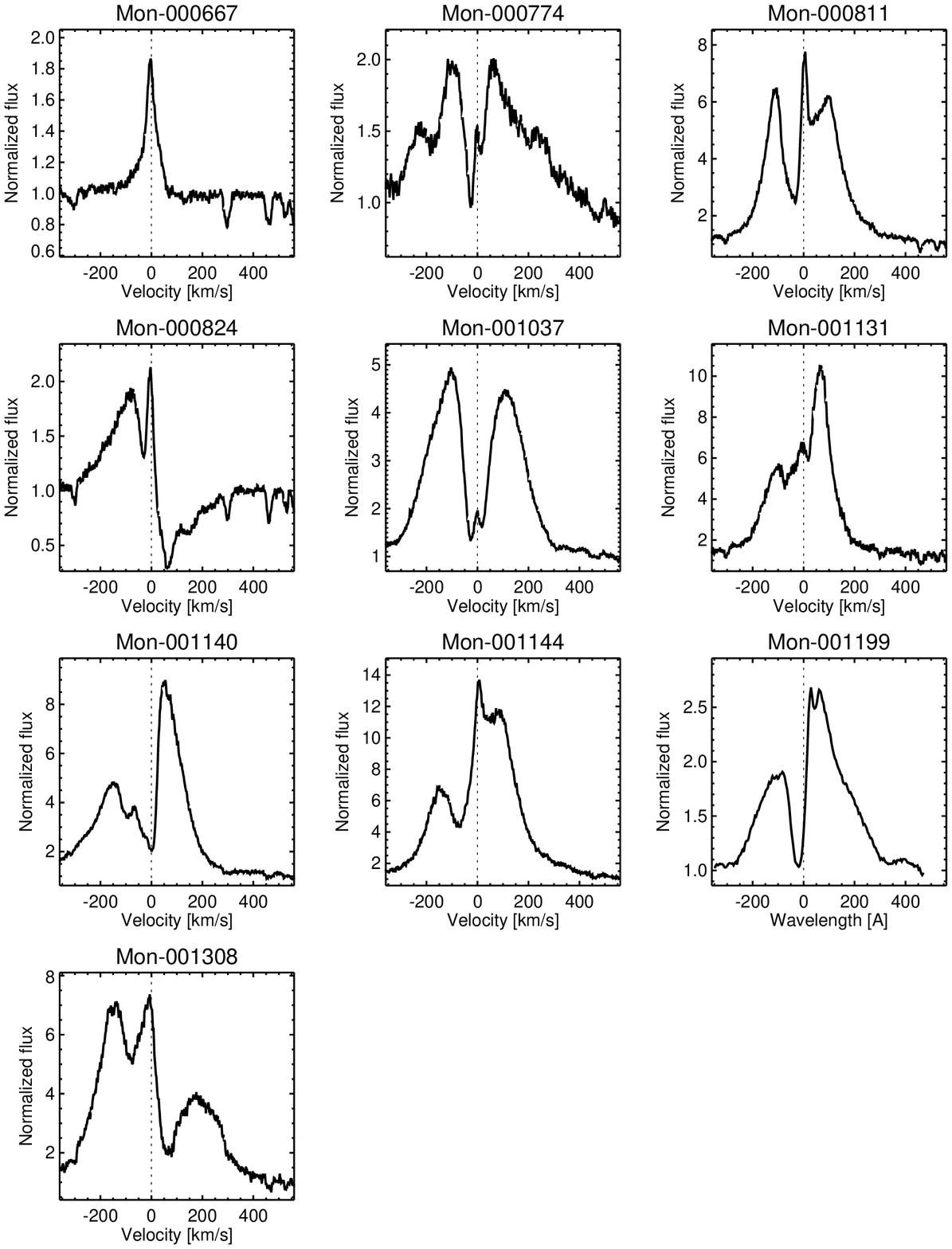}
\end{center}
\caption{H$\alpha$\ profiles, cont'd. \label{fig:halphaextinctionp2}}
\end{figure*}

\subsection*{Hot Spot Candidates}

A set of stars whose light curves may be dominated by relatively
stable hot spots located at high-latitude was identified in \S 7.
Table~\ref{tab:hotspotinfo}
provides basic information for these stars.

\begin{deluxetable*}{lccccccc}
\tabletypesize{\scriptsize}
\tablecolumns{8}
\tablewidth{0pt}
\tablecaption{Basic Information for YSOs Whose Light Curves May be Dominated
by Stable, High-Latitude Hot Spots\label{tab:hotspotinfo}}
\tablehead{
\colhead{Mon ID\tablenotemark{a}}  & \colhead{RA} & \colhead{DEC} &
\colhead{2MASS ID} &
\colhead{{\em CoRoT} 2008} & \colhead{{\em CoRoT} 2011} &
\colhead{SpT\tablenotemark{a}} & \colhead{H$\alpha$ EW
(\AA)\tablenotemark{a}}  
}
\startdata
Mon-000058 & 100.53625 & 9.689221 & 06420870+0941212 &      null & 616895632 & ... & 94.0 \\
Mon-000153 & 100.24962 & 9.784611 & 06405990+0947044 &      null & 400007889 &  M3 & 39.9 \\
Mon-000225 & 100.27596 & 9.417666 & 06410622+0925036 & 500007896 &      null &  M5 & 34.7 \\
Mon-000326 & 100.24508 & 9.655193 & 06405882+0939187 & 223980258 & 223980258 &  M0 & 27.9 \\
Mon-000448 & 100.26500 & 9.508055 & 06410360+0930290 & 400007803 & 602083897 & ... & 20.4 \\
Mon-000926 & 100.27679 & 9.477443 & 06410642+0928388 & 400007686 & 400007687 & M1.5 & 56.1 \\
Mon-001011 & 100.16887 & 9.583666 & 06404052+0935011 & 500008145 &      null & ... & ... \\
Mon-001054 & 100.15221 & 9.845999 & 06403652+0950456 & 400007538 & 400007538 &  M2 & 21.1 \\
Mon-001132 & 100.10779 & 9.849332 & 06402587+0950576 &      null & 602095740 & M2.5 & 166.0 \\
\enddata
\tablenotetext{a}{See Cody et al.\ (2014) for the sources of the
spectral type and H$\alpha$ equivalent width (EW) data. All of these
are in emission.}
\end{deluxetable*}

\acknowledgements{This work is based on observations made with
the Spitzer Space Telescope, which is operated by the
Jet Propulsion Laboratory, California Institute of Technology,
under a contract with NASA. Support for this
work was provided by NASA through an award issued
by JPL/Caltech.}


\end{document}